\documentclass{SSBAtran}

\usepackage{graphicx}
\usepackage[cmex10]{amsmath}
\newcommand{\orth}{\bot}

\begin{document}
%
\title{Normative theory of visual receptive fields}

%
%
 \author{\IEEEauthorblockN{%
  Tony Lindeberg}
  \IEEEauthorblockA{%
    Computational Brain Science Lab, Department of Computational Science and Technology,\\
    KTH Royal Institute of Technology, SE-100 44 Stockholm, Sweden.
  Email: tony@kth.se}
  }
%
%

\maketitle

\begin{abstract}
This article gives an overview of a normative computational theory of
visual receptive fields. It is described how idealized functional models of early spatial, spatio-chromatic and
spatio-temporal receptive fields can be derived in an axiomatic way
based on structural properties of the environment in combination with
assumptions about the internal structure of a vision system to
guarantee consistent handling of image representations over multiple
spatial and temporal scales.
Interestingly, this theory leads to predictions about visual receptive
field shapes with qualitatively very good similarity to biological receptive
fields measured in the retina, the LGN and the primary visual cortex (V1)
of mammals.

\end{abstract}


\section{Introduction}

\begin{figure}[hbt]
   \begin{center}
      \includegraphics[width=0.30\textwidth]{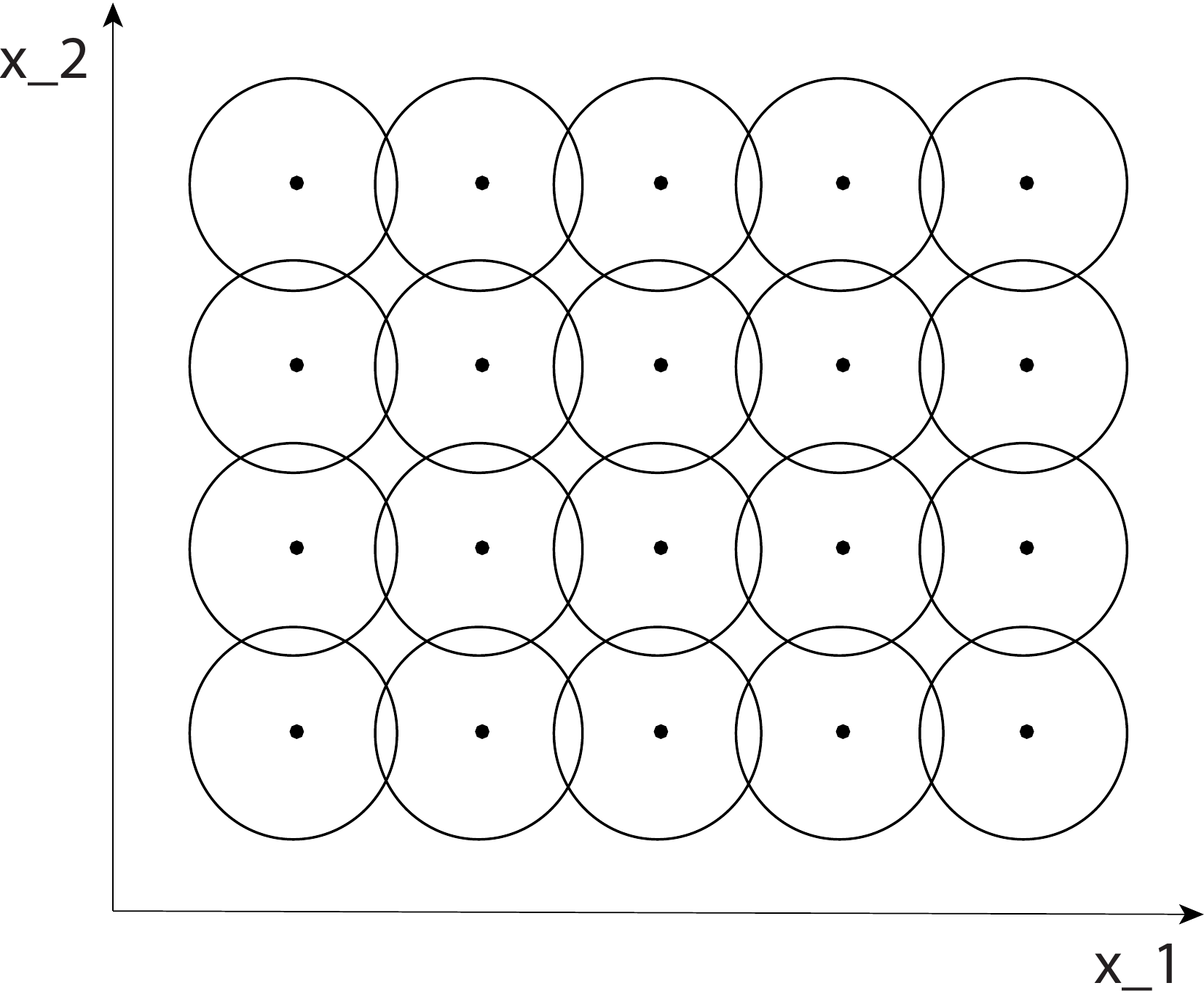}
   \end{center}
  \caption{A receptive field is traditionally defined as a region in the visual field for which
    a visual sensor/neuron/operator responds to visual stimuli.
    This figure shows a set of partially overlapping receptive fields
    over the spatial domain with all the receptive fields having the
    same spatial extent. More generally, one can conceive
    distributions of receptive fields over space or space-time with
    the receptive fields of different size, different shape and
    orientation in space as well as different directions in
    space-time, where adjacent receptive fields may also have significantly
  larger relative overlap than shown in this schematic illustration.
 In this work, we focus on a functional description of linear receptive
 fields, concerning how a neuron responds to visual stimuli over image space regarding
spatial receptive fields or over joint space-time regarding
spatio-temporal receptive fields.
}
  \label{fig-rec-fields}
\end{figure}

When light reaches a visual sensor such as the retina, 
the information necessary to infer properties about the surrounding
world is not contained in the measurement of image intensity
at a single point, but from the {\em relations\/} between intensity 
values at different points.
A main reason for this is that the incoming light constitutes 
an {\em indirect\/} source of information depending on the interaction
between geometric and material properties of objects in the 
surrounding world and on external illumination sources.
Another fundamental reason why cues to the surrounding world
need to be collected over {\em regions\/} in the visual field as 
opposed to at single image points is that the measurement process 
by itself requires the accumulation of energy over 
non-infinitesimal support regions over space and time.
Such a region in the visual field for which a neuron responds 
to visual stimuli is traditionally referred to as a {\em receptive field\/}
(Hubel and Wiesel 
\cite{HubWie59-Phys,HubWie62-Phys,HubWie05-book})
(see Figure~\ref{fig-rec-fields}).
In this work, we focus on a functional description of receptive
fields, regarding how a neuron with a purely spatial receptive field responds to visual stimuli over
image space, and regarding how a neuron with a spatio-temporal
receptive field responds to visual stimuli over space and time
(DeAngelis {\em et al.\/}\
\cite{DeAngOhzFre95-TINS,deAngAnz04-VisNeuroSci}).

If one considers the theoretical and computational problem of designing a vision
system that is going to make use of incoming reflected light to infer
properties of the surrounding world, one may ask what types of
image operations should be performed on the image data. 
Would any type of image operation be reasonable?
Specifically regarding the notion of receptive fields one may ask
what types of receptive field profiles would be reasonable.
Is it possible to derive a theoretical model of how
receptive fields ``ought to'' respond to visual data?

Initially, such a problem might be regarded as intractable
unless the question can be further specified.
It is, however, possible to address this problem systematically
using approaches that have been developed in the area of 
computer vision known as {\em scale-space theory\/}
(Iijima \cite{Iij62};
 Witkin \cite{Wit83};
 Koenderink \cite{Koe84};
 Koenderink and van Doorn \cite{KoeDoo87-BC,KoeDoo92-PAMI};
 Lindeberg \cite{Lin93-Dis,Lin94-SI,Lin10-JMIV,Lin13-ImPhys};
 Florack \cite{Flo97-book};
 Sporring {\em et al.\/}\ \cite{SpoNieFloJoh96-SCSPTH};
 Weickert {\em et al.\/}\ \cite{WeiIshImi99-JMIV};
 ter Haar Romeny \cite{Haa04-book}).
A paradigm that has been developed in this field is to impose 
{\em structural constraints\/} on the first stages of visual processing that
reflect {\em symmetry properties\/} of the environment.
Interestingly, it turns out to be possible to substantially reduce
the class of permissible image operations from such arguments.

The subject of this article is to describe how structural
requirements on the first stages of visual processing
as formulated in scale-space theory can be used for 
deriving idealized functional models of visual receptive fields and implications of how these
theoretical results can be used when modelling biological vision.
A main theoretical argument is that idealized functional models for linear receptive fields can be derived {\em by necessity\/}
given a small set of symmetry requirements that reflect properties 
of the world that one may naturally require an idealized vision system
to be adapted to.
In this respect, the treatment bears similarities to 
approaches in theoretical physics, where symmetry
properties are often used as main arguments
in the formulation of physical theories of the world.
The treatment that will follow will be general in the sense that 
{\em spatial,
spatio-chromatic and spatio-temporal receptive fields are
encompassed by the same unified theory\/}.

\begin{figure}[hbt]
  \begin{center}
    \begin{tabular}{c}
      \includegraphics[width=0.50\textwidth]{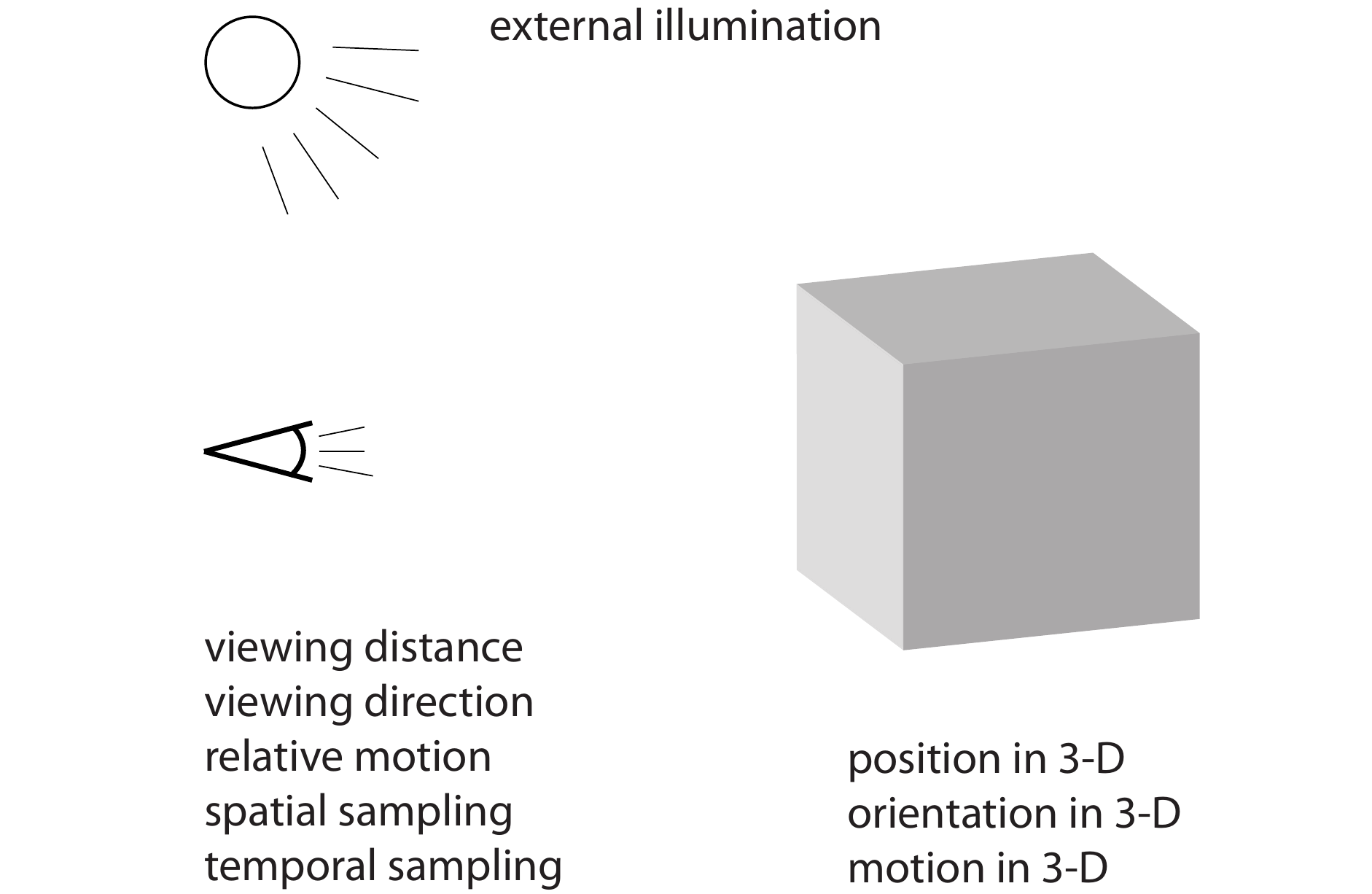} 
    \end{tabular}
  \end{center}
  \caption{Basic factors that influence the formation of
    images for an eye with a two-dimensional retina that observes objects in the
    three-dimensional world. In addition to the position, the orientation
    and the motion of the object in 3-D, the perspective projection onto
    the retina is affected by the viewing distance, the viewing
    direction and the relative motion of the eye in relation to the object, the spatial
  and the temporal sampling characteristics of the neurons in the retina as well the
  usually unknown external illumination field in relation to the
  geometry of the scene and the observer.}
  \label{fig-images-from-3D-world}
\end{figure}

This paper gives a condensed summary of a more general theoretical
framework for receptive fields derived and presented in
\cite{Lin10-JMIV,Lin13-BICY,Lin13-PONE,Lin16-JMIV}
and in turn developed to enable a consistent handling of receptive field
responses in terms of provable
covariance or invariance properties under natural image
transformations
(see Figure~\ref{fig-images-from-3D-world}).
In relation to the early publications on this topic
\cite{Lin10-JMIV,Lin13-BICY,Lin13-PONE}, this paper presents
an improved version of that theory leading to an improved
model for the temporal smoothing operation for the specific case of a
time-causal image domain \cite{Lin16-JMIV}, where the future cannot be accessed
and the receptive fields have to be solely based on information from
the present moment and a compact buffer of the past.
Specifically, this paper presents the improved axiomatic structure 
on a compact form more easy to access compared to the original
publications and also encompassing the better time-causal model.

It will be shown that the presented framework leads to predictions of {\em receptive field profiles\/}
in good agreement with receptive measurements reported in the literature
(Hubel and Wiesel \cite{HubWie59-Phys,HubWie62-Phys,HubWie05-book};
DeAngelis {\em et al.\/}\
\cite{DeAngOhzFre95-TINS,deAngAnz04-VisNeuroSci};
Conway and Livingstone \cite{ConLiv06-JNeurSci};
Johnson {\em et al.\/}\ \cite{JohHawSha08-JNeuroSci}). 
Specifically, explicit phenomenological models will be given of LGN
neurons and simple cells in V1 and will be compared to related models in terms
of Gabor functions (Mar\v{c}elja \cite{Mar80-JOSA}; 
Jones and Palmer \cite{JonPal87a,JonPal87b}; Ringach \cite{Rin01-JNeuroPhys,Rin04-JPhys}),
differences of Gaussians (Rodieck \cite{Rod65-VisRes}) and
Gaussian derivatives (Koenderink and van Doorn \cite{KoeDoo87-BC}; 
Young \cite{You87-SV}; Young {\em et al.\/}\ \cite{YouLesMey01-SV,YouLes01-SV}).
Notably, the evolution properties of the receptive field profiles in
this model can be described by diffusion equations and are therefore
suitable for implementation on a biological architecture,
since the computations can be expressed in terms of communications
between neighbouring computational units, where either a single
computational unit or a group of
computational units may be interpreted as corresponding to a neuron
or a group of neurons.
Specifically, computational models involving diffusion equations arise in mean field theory for
approximating the computations that are performed by populations of neurons
(Omurtag {\em et al.\/}\ \cite{OmuKniSir00-CompNeuro};
Mattia and Guidic \cite{MatGui02-PhysRevE};
Faugeras {\em et al.\/}\ \cite{FauTouCes09-FrontCompNeuroSci}).

\subsection{Structure of this article}

This paper is organized as follows: 
Section~\ref{sec-struc-req} gives an overview of and motivation to the assumptions that
the theory is based on. A set of structural requirements is formulated
to capture the effect of natural image transformations onto the
illumination field that reaches the retina and to
guarantee internal consistency between image representations that are computed
from receptive field responses over multiple spatial and temporal scales.

Section~\ref{sec-ideal-rec-field-fam} describes linear receptive families
that arise as consequences of these assumptions for the cases of
either a purely spatial domain or a joint spatio-temporal domain.
The issue of how to perform relative normalization between receptive
field responses over multiple spatial and temporal scales is treated,
so as to enable comparisons between receptive field responses at
different spatial and temporal scales.
We also show how the influence of illumination transformations and exposure control
mechanisms on the receptive field responses can be handled,
by describing invariance properties obtained by applying the derived linear receptive fields
over a logarithmically transformed intensity domain.

Section~\ref{eq-rel-biol-vision} shows examples of how spatial, spatio-chromatic
and spatio-temporal receptive fields in the retina, the LGN and the
primary visual cortex can be well modelled by the derived receptive
field families.

Section~\ref{sec-rel-prev-work} gives relations to previous work,
including conceptual and theoretical comparisons to previous use of
Gabor models of receptive fields, approaches for learning receptive
fields from image data and previous applications of a
logarithmic transformation of the image intensities.
Finally, Section~\ref{sec-summ} summarizes some of the main results.

\section{Assumptions underlying the theory: Structural requirements}
\label{sec-struc-req}

In the following, we shall describe a set of structural requirements that
can be stated concerning: (i)~spatial geometry, (ii)~spatio-temporal geometry,
(iii)~the image measurement process with its close relationship to the notion of
scale, (iv)~internal representations of image data that are to be
computed by a general purpose vision system and (v)~the parameterization of image intensity with regard 
to the influence of illumination variations.

For modelling the image formation
process, we will at any point on the retina approximate the spherical
retina by a perspective projection onto the tangent plane of the
retinal surface at that image point, below represented as the image plane.
Additionally, we will approximate the possibly non-linear geometric
transformations regarding spatial and spatio-temporal geometry by
local linearizations at every image point, and corresponding to the
derivative of the possibly non-linear transformation.
In these ways, the theoretical analysis can be substantially simplified, while still
enabling accurate modelling of essential functional properties of receptive fields in
relation to the effects of natural image transformations as arising
from interactions with the environment.

\subsection{Static image data over a spatial domain}
\label{sec-assum-spat-domain}

In the following, we will describe a theoretical model for the
computational function of 
applying visual receptive fields to local image patterns.

For time-independent data $f$ over a two-dimensional spatial image
domain, we would like to define a family of image representations
$L(\cdot;\; s)$ over a possibly multi-dimensional scale parameter $s$,
where the internal image representations $L(\cdot;\; s)$ are computed by applying some
parameterized family of image operators ${\cal T}_s$ to the image data $f$:
\begin{equation}
  \label{eq-transf-kernel-def}
  L(\cdot;\; s) = {\cal T}_s \, f(\cdot).
\end{equation}
Specifically, we will assume that the family of
image operators ${\cal T}_s$ should satisfy:

\paragraph{Linearity}

For the earliest processing stages to make as few irreversible
decisions as possible, we assume that they should be linear
\begin{equation}
   {\cal T}_s(a_1 f_1 + a_2 f_2) = a_1 {\cal T}_s f_1 + b_1 {\cal T}_s f_2.
\end{equation}
Specifically, linearity implies that any particular
scale-space properties (to be detailed below) that we derive for the
zero-order image representation $L$ will transfer to any spatial
derivative $L_{x_1^{\alpha_1} x_2^{\alpha_2}}$ of $L$, so that
\begin{equation}
  L_{x_1^{\alpha_1} x_2^{\alpha_2}}(\cdot;\; s) 
  = \partial_{x_1^{\alpha_1} x_2^{\alpha_2}} ({\cal T}_s \, f)
  = {\cal T}_s (\partial_{x_1^{\alpha_1} x_2^{\alpha_2}} f).
\end{equation}
In this sense, the assumption of linearity reflects the requirement of
a lack of bias to particular types of image structures, with the
underlying aim that the processing performed in the first
processing stages should be {\em generic\/}, to be used as input for a large
variety of visual tasks. By the assumption of linearity, local image
structures that are captured by {\em e.g.\/}\ first- or second-order
derivatives will be treated in a structurally similar manner, which
would not necessarily be the case if the first local neighbourhood
processing stage of the first layer of receptive fields would instead be
genuinely non-linear.%
\footnote{Note, however, that the assumption about linearity of some first layers of
receptive fields does, however, not
exclude the possibility of defining later stage non-linear receptive fields that
operate on the output from the linear receptive fields, such as the
computations performed by complex cells in the primary visual cortex.
Neither does this assumption of linearity exclude the possibility of
transforming the raw image intensities by a pointwise non-linear
mapping function prior to the application of linear receptive fields
based on processing over local neighbourhoods. 
In Section~\ref{sec-intens-var} it will be specifically shown that a
pointwise logarithmic transformation of the image intensities prior to
the application of linear receptive fields has theoretical advantages
in terms of invariance properties of derivative-based receptive field responses under local multiplicative
illumination transformations.}

This genericity property is closely related to the basic property of the mammalian vision system,
that the computations performed in the retina, the LGN and the primary
visual cortex provide general purpose output that is used as input to higher-level visual areas.

\paragraph{Shift invariance}

To ensure that the visual interpretation of an object should be the same
irrespective of its position in the image plane, we assume that the
first processing stages should be shift invariant, so that if an
object is moved a distance $\Delta x = (\Delta x_1, \Delta x_2)$ in the image plane,
the receptive field response should remain on a similar form while shifted with the
same distance. Formally, this requirement can be stated that
the family of image operators ${\cal T}_s$ should commute
with the shift operator defined  by $S_{\Delta x}(f)(x) = f(x - \Delta x)$:
\begin{equation}
   {\cal T}_s \left( S_{\Delta x} f \right) = S_{\Delta x} \left( {\cal T}_s f \right).
\end{equation}
In other words, if we shift the input by a translation and then apply
the receptive field operator ${\cal T}_s$, the result should be
similar as applying the receptive field operator to the original input
and then shifting the result.

\paragraph{Convolution structure}

Together, the assumptions about linearity and shift-invariance imply that
${\cal T}_s$ will correspond to a {\em convolution\/} operator \cite{HirWid55}.
This implies that the representation $L$ can be computed from the image data $f$ by
convolution with some parameterized family of convolution kernels $T(\cdot;\; s)$:
\begin{equation}
  \label{eq-conv-op-spat}
  L(\cdot;\; s) = T(\cdot;\; s) * f.
\end{equation}

\paragraph{Semi-group structure over spatial scales}

To ensure that the transformation from any finer scale $s_1$ to any
coarser scale $s_1 + s_2$ should be of the same form for any $s_2 > 0$
(a requirement of algebraic closedness), we assume that the
result of convolving two kernels $T(\cdot;\; s_1)$ and $T(\cdot;\; s_2)$
from the family with each other should be a kernel within the same family of kernels
and with added parameter values $T(\cdot;\; s_1+s_2)$:
\begin{equation}
  T(\cdot;\; s_1) * T(\cdot;\; s_2) = T(\cdot;\; s_1+s_2).
\end{equation}
This assumption specifically implies that the representation 
$L(\cdot;\; s_2)$ at a coarse scale $s_2$ can be computed from the
representation $L(\cdot;\; s_1)$ at a finer scale $s_1 < s_2$
by a convolution operation of the same form (\ref{eq-conv-op-spat}) as the transformation from
the original image data $f$ while using the difference in scale levels
$s_2 - s_1$ as the parameter
\begin{equation}
  \label{eq-casc-smooth-spat}
  L(\cdot;\; s_2) = T(\cdot;\; s_2 - s_1) * L(\cdot;\; s_1).
\end{equation}
This property does in turn imply that if we are able to derive
specific properties of the family of transformations ${\cal T}_s$ (to
be detailed below),
then these properties will not only hold for the transformation from
the original image data $f$ to the representations $L(\cdot;\; s)$ at
coarser scales, but also between any pair of scale levels $s_2 > s_1$,
with the aim that image representations at coarser scales
should be possible to regard as simplifications of corresponding
image representations at finer scales.

In terms of mathematical concepts, this form of algebraic structure is
referred to as a semi-group structure over spatial scales
\begin{equation}
  \label{eq-semi-group-ops-spat}
    {\cal T}_{s_1}  {\cal T}_{s_2} =  {\cal T}_{s_1 + s_2}.
\end{equation}

\paragraph{Scale covariance under spatial scaling transformations}

If a visual observer looks at the same object from different distances, we would like the
internal representations derived from the receptive field responses to be sufficiently similar, so that the object
can be recognized as the same object while appearing with a different size on
the retina. Specifically, it is thereby natural to require that the
receptive field responses should be of a similar form while resized in the image
plane. 

This corresponds to a requirement of spatial scale covariance 
under uniform scaling transformations of the spatial domain 
$x' = S_s \, x$: 
\begin{equation}
    L'(x';\; s') = L(x;\; s) \quad \Leftrightarrow \quad
         {\cal T}_{S_s(s)} \, {\cal S}_s \, f = {\cal S}_s \, {\cal T}_s \, f
 \end{equation}
to hold for some transformation $s' = S_s(s)$ of the scale parameter
$s$.

\paragraph{Affine covariance under affine transformations}

If a visual observer looks at the same local surface patch from two
different viewing directions,
then the local surface patch may be deformed in different ways onto
the different views and
with different amounts of perspective foreshortening from the
different viewing directions. If we approximate the local deformations
caused by the perspective mapping by local affine transformations,
then the transformation between the two differently deformed views of
the local surface patch can in turn be described by a composed local affine
transformation $x' = A \, x$.
If we are to use receptive field responses as a basis for higher level
visual operations, it is natural to require that the receptive field
response of an affine deformed image patch should remain on a similar
form while
being reshaped by a corresponding affine transformation.

This corresponds to a requirement of affine covariance under general
affine transformations $x' = A \; x$: 
\begin{equation}
    L'(x';\; s') = L(x;\; s) \quad \Leftrightarrow \quad
         {\cal T}_{A(s)} \, {\cal A} \, f = {\cal A} \, {\cal T}_s \, f
 \end{equation}
to hold for some transformation $s' = A(s)$ of the scale parameter.

\paragraph{Non-creation of new structure with increasing scale}

If we apply the family of transformations ${\cal T}_s$ for computing
representations at coarser scales from representations at finer scales
according to (\ref{eq-transf-kernel-def}) and
(\ref{eq-casc-smooth-spat}),
there could be a potential risk that the family of transformations
could amplify spurious
structures in the input to produce macroscopic amplifications
in the representations at coarser scales that do not directly correspond to
simplifications of corresponding structures in the original image data.
To prevent such undesirable phenomena from occurring, we require that
local spurious structures must not be amplified and express this
condition in terms of the evolution properties over scales at local
maxima and minima in the image intensities as smoothed by the family
of convolution kernels $T(\cdot;\; s)$:
If a point $x_0$ for some scale $s_0$ is a local maximum point in the
image plane, then the value at this maximum point $L(x_0;\; s_0)$ must
not increase to coarser scales $s > s_0$.
Similarly, if a point is a local minimum point in the image plane,
then the value at this minimum point $L(x_0;\; s_0)$ must
not decrease to coarser scales $s > s_0$.

\begin{figure}[hbt]
  \begin{center}
      \includegraphics[width=0.40\textwidth]{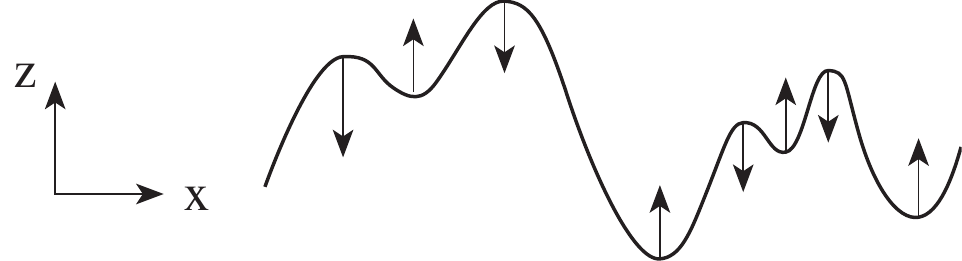}
  \end{center}
  \caption{The requirement of non-enhancement of local extrema is a
    way of restricting the class of possible image operations by
    formalizing the notion that new image structures must not
    be created with increasing scale, by requiring that the value at a
    local maximum must not increase and that the value at a local
    minimum must not decrease from finer to coarser scales $s$.}
  \label{fig-non-enh-extr}
\end{figure}

Formally, this requirement that new structures should not be created
from finer to coarser scales, can be formalized into the requirement
of  {\em non-enhancement of local extrema\/}, which implies that
if at some scale $s_0$ a point $x_0$ is a local
maximum (minimum) for the mapping from $x$ to $L(x;\; s_0)$, then
(see Figure~\ref{fig-non-enh-extr}):
  \begin{itemize}
    \item
      $(\partial_s L)(x;\; s) \leq 0$ at any spatial maximum,
   \item
      $(\partial_s L)(x;\; s) \geq 0$ at any spatial minimum.
    \end{itemize}
This condition implies a strong condition on the class of possible smoothing kernels $T(\cdot;\, s)$.

\subsection{Time-dependent image data over space-time 
}

To model the computational function of spatio-temporal receptive
fields in time-dependent image patterns, we do for 
a time-dependent spatio-temporal domain first inherit the structural
requirements regarding a spatial domain and complement the spatial scale
parameter $s$ by a temporal scale parameter $\tau$. In addition, we assume:

\paragraph{Scale covariance under temporal scaling transformations}

If a similar type of spatio-temporal event $f(x, t)$ occurs at
different speeds, faster or
slower, it is natural to require that the receptive field responses
should be of a similar form, while occurring correspondingly faster or
slower.

This corresponds to a requirement of temporal scale covariance 
under a temporal scaling transformation of the temporal domain $t' = S_{\tau} t$:
\begin{multline}
    L'(x', t';\; s', \tau') = L(x, t;\; s, \tau) \quad \Leftrightarrow \quad\\
         {\cal T}_{S_{\tau}(s, \tau)} \, {\cal S}_{\tau} \, f = {\cal S}_{\tau} \, {\cal T}_{s, \tau} \, f
 \end{multline}
to hold for some transformation $(s', \tau') = S_{\tau}(s, \tau)$ of the spatio-temporal scale
parameters $(s, \tau)$.

\paragraph{Galilean covariance under Galilean transformations}

If an observer looks at the same object in the world for different relative
motions $v = (v_1, v_2)$ between the object and the observer, it is natural to require
that the internal representations of the object should be sufficiently
similar so as to enable a coherent perception of the object under
different relative motions relative to the observer.
Specifically, we may require that the receptive field responses under
relative motions should remain on the same form while being
transformed in a corresponding way as the relative motion pattern.

If we at any point in space-time locally linearize the possibly
non-linear motion
pattern $x(t) = (x_1(t), x_2(t))$ by a local Galilean transformation 
$x' = x + v \, t$ over space-time
\begin{equation}
    f' = {\cal G}_v \, f  \quad \Leftrightarrow \quad
         f'(x', t') = f(x, t) \quad \mbox{with} \quad x' = x + v \, t,
\end{equation}
then the requirement of guaranteeing a consistent visual interpretation under different relative
motions between the object and the observer can be stated as a
requirement of Galilean covariance:
\begin{multline}
    L'(x', t';\; s', \tau') = L(x, t;\; s, \tau) \quad \Leftrightarrow \quad\\
    {\cal T}_{G_v(s,\tau)} \, {\cal G}_v \, f = {\cal G}_v \, {\cal T}_{s,\tau} \, f
\end{multline}
to hold for some transformation $G_v(s,\tau)$ of the 
spatio-temporal scale parameters $(s, \tau)$.

\paragraph{Semi-group structure over temporal scales in the case of a non-causal
  temporal domain}

To ensure that the representations between different spatio-temporal
scale levels $(s_1, \tau_1)$ and $(s_2, \tau_2)$ should be sufficiently
well-behaved internally, we will make use of different types of assumptions
depending on whether the temporal domain is regarded as time-causal or
non-causal. Over a time-causal temporal domain, the future cannot be
accessed, which is the basic condition for real-time visual perception by a
biological organism. Over a non-causal temporal domain, the temporal
kernels may extend to the relative future in relation to any
pre-recorded time moment, which is sometimes used as a conceptual
simplification when analysing pre-recorded time-dependent data although not at all
realistic in a real-world setting.

For the case of a non-causal temporal domain, we make use of a similar
type of semi-group property (\ref{eq-semi-group-ops-spat}) as
formulated over a purely spatial domain, while extending the
semi-group property over both the spatial scale parameter $s$ and the
temporal scale parameter $\tau$:
\begin{equation}
    {\cal T}_{s_1, \tau_1}  {\cal T}_{s_2, \tau_2} 
    =  {\cal T}_{s_1 + s_2, \tau_1 + \tau_2}.
\end{equation}
In analogy with the case of a purely spatial domain, this requirement
guarantees that the transformation from any finer spatio-temporal
scale level $(s_1, \tau_1)$
to any coarser spatio-temporal scale level $(s_2, \tau_2) \geq (s_1, \tau_1)$ 
will always be of the same form 
(algebraic closedness)
\begin{equation}
   L(\cdot, \cdot;\; s_2, \tau_2) 
  = {\cal T}_{s_2 - s_1, \tau_2 - \tau_1} \, L(\cdot, \cdot;\; s_1, \tau_1).
\end{equation}
Specifically, this assumption implies that if we are able to establish
desirable properties of the family of transformations ${\cal T}_{s, \tau}$ (to
be detailed below),
then these relations hold between any pair of spatio-temporal scale
levels $(s_1, \tau_1)$ and $(s_2, \tau_2)$ with 
$(s_2, \tau_2) \geq (s_1, \tau_1)$.

\paragraph{Cascade structure over temporal scales in the case of a time-causal
  temporal domain}

Since it can be shown that the assumption of a semi-group structure
over temporal scales leads to undesirable temporal dynamics in terms
of {\em e.g.\/}\ longer temporal delays for a time-causal temporal
domain \cite[Appendix~A]{Lin17-JMIV},
we do for a time-causal temporal domain instead assume a weaker
cascade smoothing property over temporal scales for the temporal
smoothing kernel over temporal scales 
\begin{equation}
  \label{eq-casc-smooth-prop-in-proof-temp-dyn}
  L(\cdot;\; \tau_2) 
  = h(\cdot;\; \tau_1 \mapsto \tau_2)
     * L(\cdot;\; \tau_1),
\end{equation}
where the temporal kernels $h(t;\; \tau)$ should for any triplets of temporal scale
values and temporal delays $\tau_1$, $\tau_2$
and $\tau_3$ obey the transitive property
\begin{equation}
   h(\cdot;\; \tau_1 \mapsto \tau_2) * h(\cdot;\; \tau_2 \mapsto \tau_3) = h(\cdot;\; \tau_1 \mapsto \tau_3).
\end{equation}
This weaker assumption of a cascade smoothing property
(\ref{eq-casc-smooth-prop-in-proof-temp-dyn}) still ensures that a
representation at a coarser temporal scale $\tau_2$ should with a
corresponding requirement of an accompanying simplifying condition on the family of
kernels $h$ (to be detailed below) constitute a simplification of the representation at a
finer temporal scale $\tau_1$, while not implying as hard constraints
as a semi-group structure.

\paragraph{Non-enhancement of local space-time extrema in the case of
  a non-causal temporal domain}

In the case of a non-causal temporal domain, we again build on the
notion of non-enhancement of local extrema to guarantee that the
representations at coarser spatio-temporal scales should constitute true
simplifications of corresponding representations at finer scales
Over a spatio-temporal domain, we do, however, state the requirement in
terms of local extrema over joint space-time instead of over local extrema
over image space.
If a point $(x_0, t_0)$ for some scale $(s_0, \tau_0)$ is a local maximum point over
space-time, then the value at this maximum point $L(x_0, t_0;\; s_0, \tau_0)$ must
not increase to coarser scales $(s, \tau) \geq (s_0, \tau_0)$.
Similarly, if a point is a local minimum point over space-time,
then the value at this minimum point $L(x_0, t_0;\; s_0, \tau_0)$ must
not decrease to coarser scales $(s, \tau) \geq (s_0, \tau_0)$.

Formally, this requirement of non-creation of new structure from finer
to coarser spatio-temporal scales, can be stated as follows:
If at some scale $(s_0, \tau_0)$ a point $(x_0, t_0)$ is a local
maximum (minimum) for the mapping from $(x, t)$ to $L(x, t;\; s_0, \tau_0)$, then
  \begin{itemize}
    \item
      $\alpha \, (\partial_s L)(x, t;\; s, \tau) + \beta \, (\partial_{\tau} L)(x, t;\; s, \tau) \leq 0$ 
      at any spatio-temporal maximum
   \item
       $\alpha \, (\partial_s L)(x, t;\; s, \tau) + \beta \, (\partial_{\tau} L)(x, t;\; s, \tau) \geq 0$ 
      at any spatio-temporal minimum
   \end{itemize}
should hold in any positive spatio-temporal direction defined from any
non-negative linear combinations of $\alpha$ and $\beta$.
This condition implies a strong condition on the class of possible
smoothing kernels 
$T(\cdot, \cdot;\, s, \tau)$.

\paragraph{Non-creation of new local extrema or zero-crossings for a
  purely temporal signal in the case of
  a non-causal temporal domain}

In the case of a time-causal temporal domain, we do instead state a
requirement for purely temporal signals, based on the cascade
smoothing property (\ref{eq-casc-smooth-prop-in-proof-temp-dyn}).
We require
that for a purely temporal signal $f(t)$, the transformation from a
finer temporal scale $\tau_1$ to a coarser temporal scale $\tau_2$
must not increase the number of local extrema or the number of zero-crossings in the
signal.

\begin{figure}[hbtp]
  \begin{center}
    \begin{tabular}{c}
      \hspace{-2mm} {\footnotesize $g(x_1, x_2;\; s)$} \hspace{-2mm} \\
      \hspace{-2mm} \includegraphics[width=0.14\textwidth]{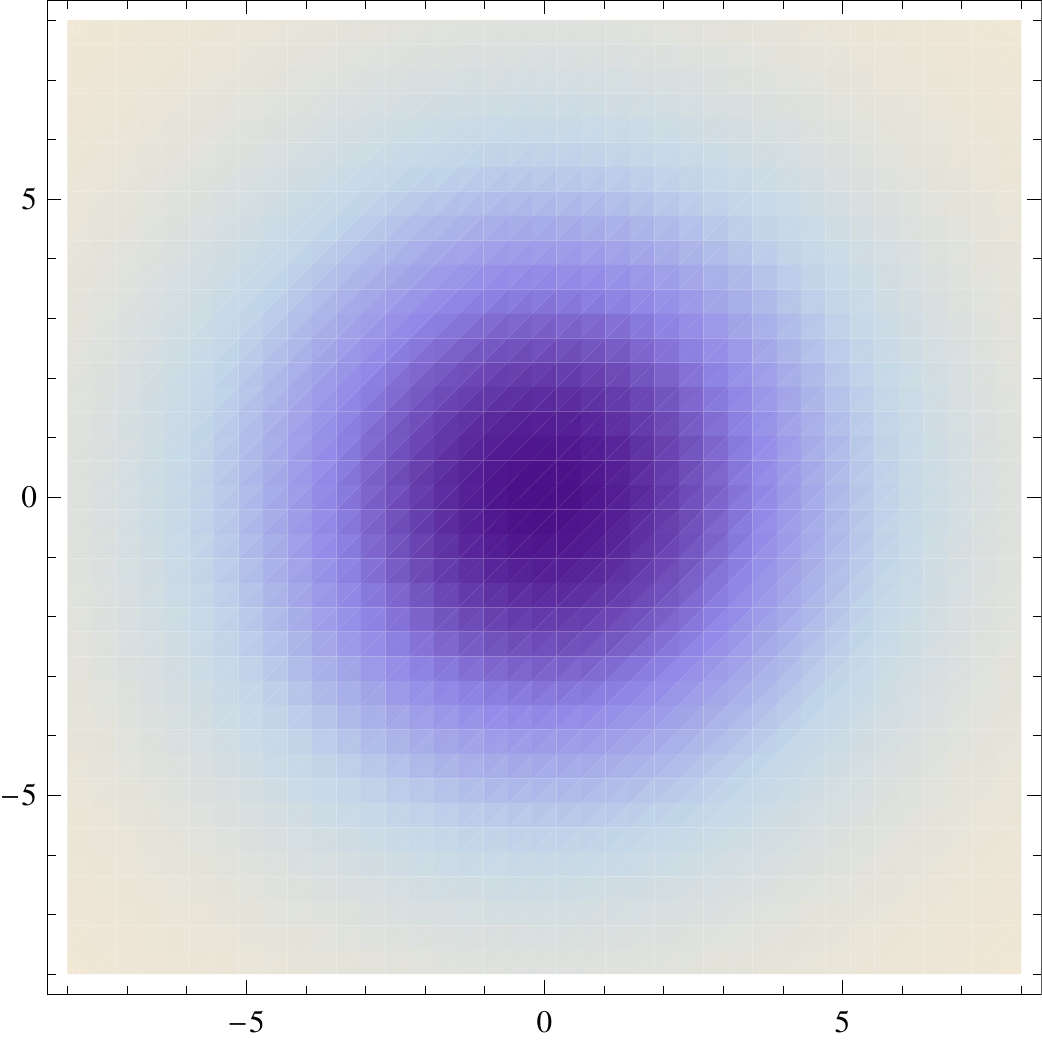} \hspace{-2mm} \\
    \end{tabular} 
  \end{center}
  \vspace{-6mm}
  \begin{center}
    \begin{tabular}{cc}
      \hspace{-2mm}  {\footnotesize  $g_{x_1}(x_1, x_2;\; s)$} \hspace{-2mm} 
      & \hspace{-2mm} {\footnotesize $g_{x_2}(x_1, x_2;\; s)$} \hspace{-2mm} \\
      \hspace{-2mm} \includegraphics[width=0.14\textwidth]{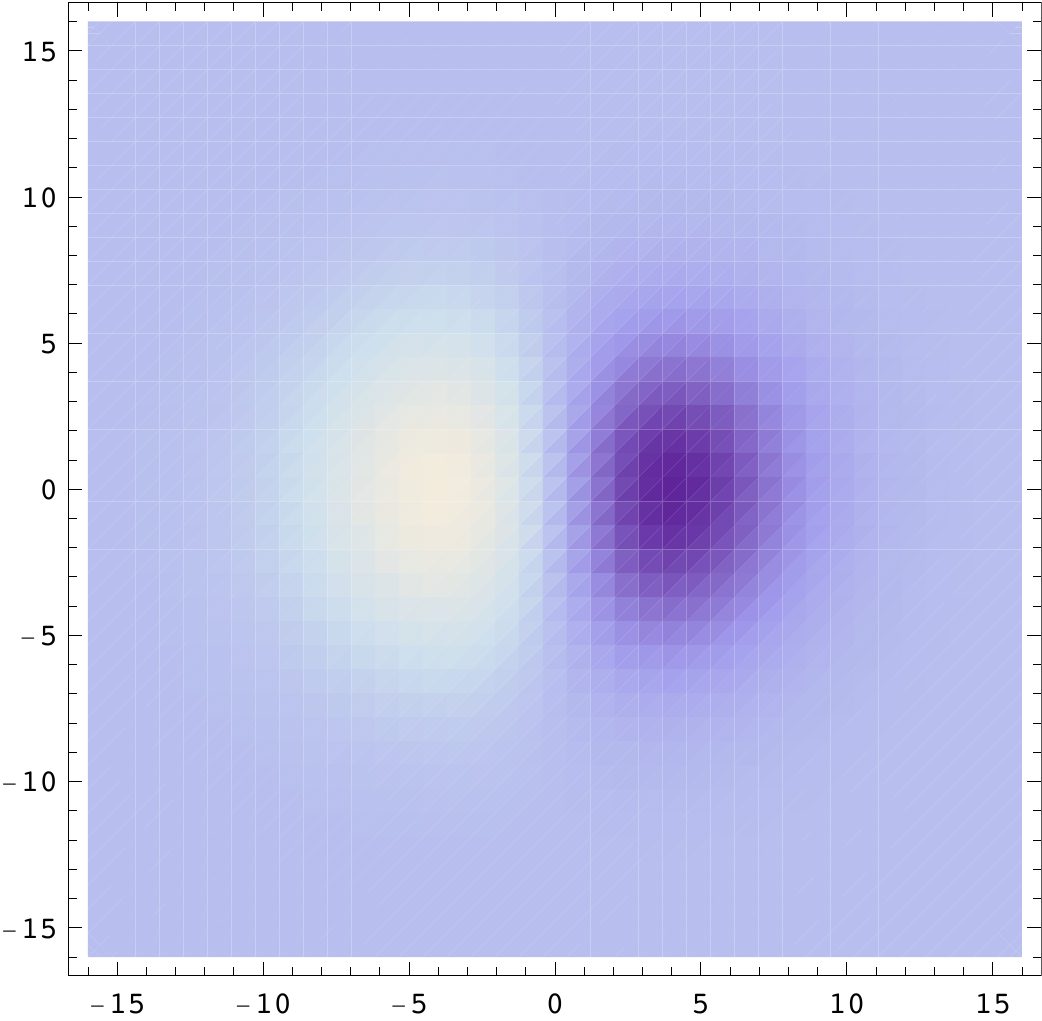} \hspace{-2mm} &
      \hspace{-2mm} \includegraphics[width=0.14\textwidth]{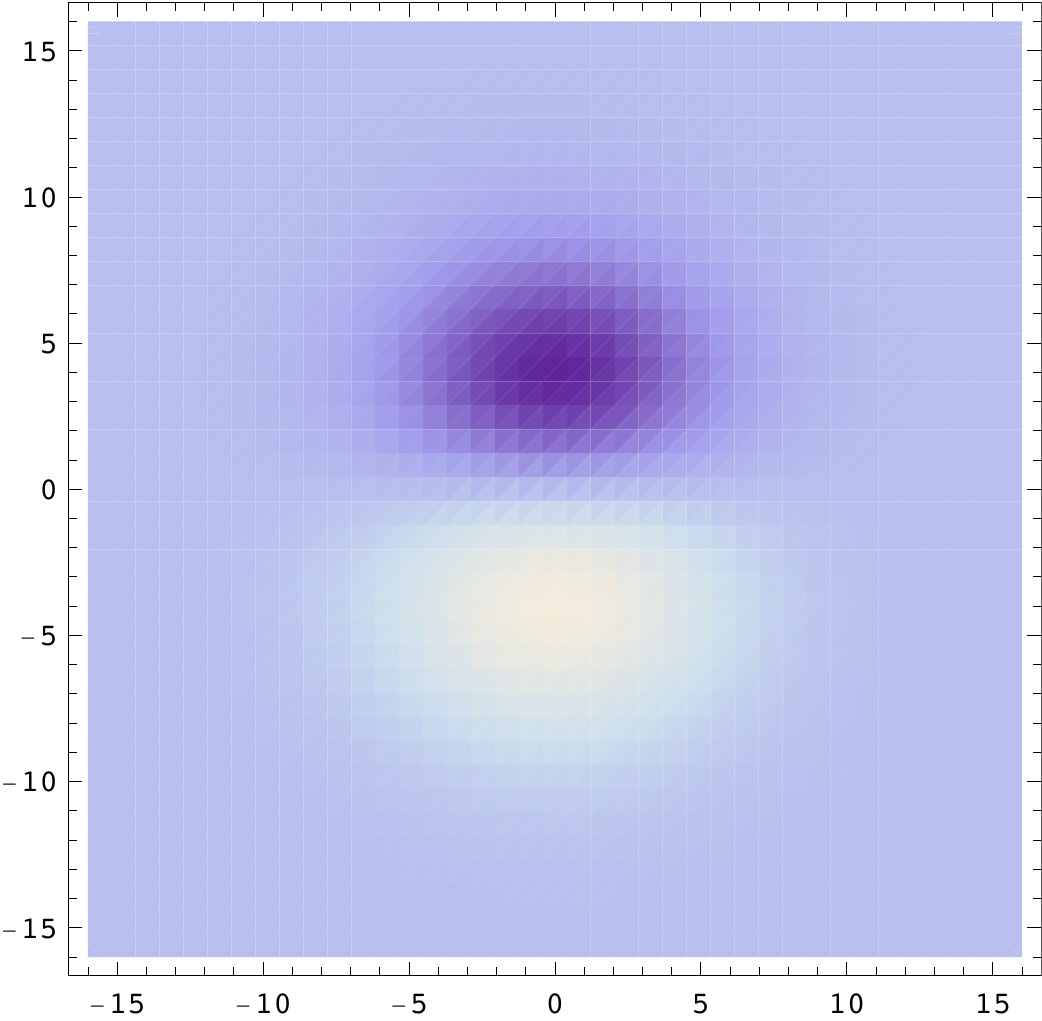} \hspace{-2mm} \\
    \end{tabular} 
  \end{center}
  \vspace{-6mm}
  \begin{center}
    \begin{tabular}{ccc}
     \hspace{-2mm} {\footnotesize $g_{x_1 x_1}(x_1, x_2;\; s)$} \hspace{-2mm} 
     & \hspace{-2mm} {\footnotesize $g_{x_1 x_2}(x_1, x_2;\; s)$} \hspace{-2mm} 
     & \hspace{-2mm} {\footnotesize $g_{x_2 x_2}(x_1, x_2;\; s)$} \hspace{-2mm} \\
      \hspace{-2mm} \includegraphics[width=0.14\textwidth]{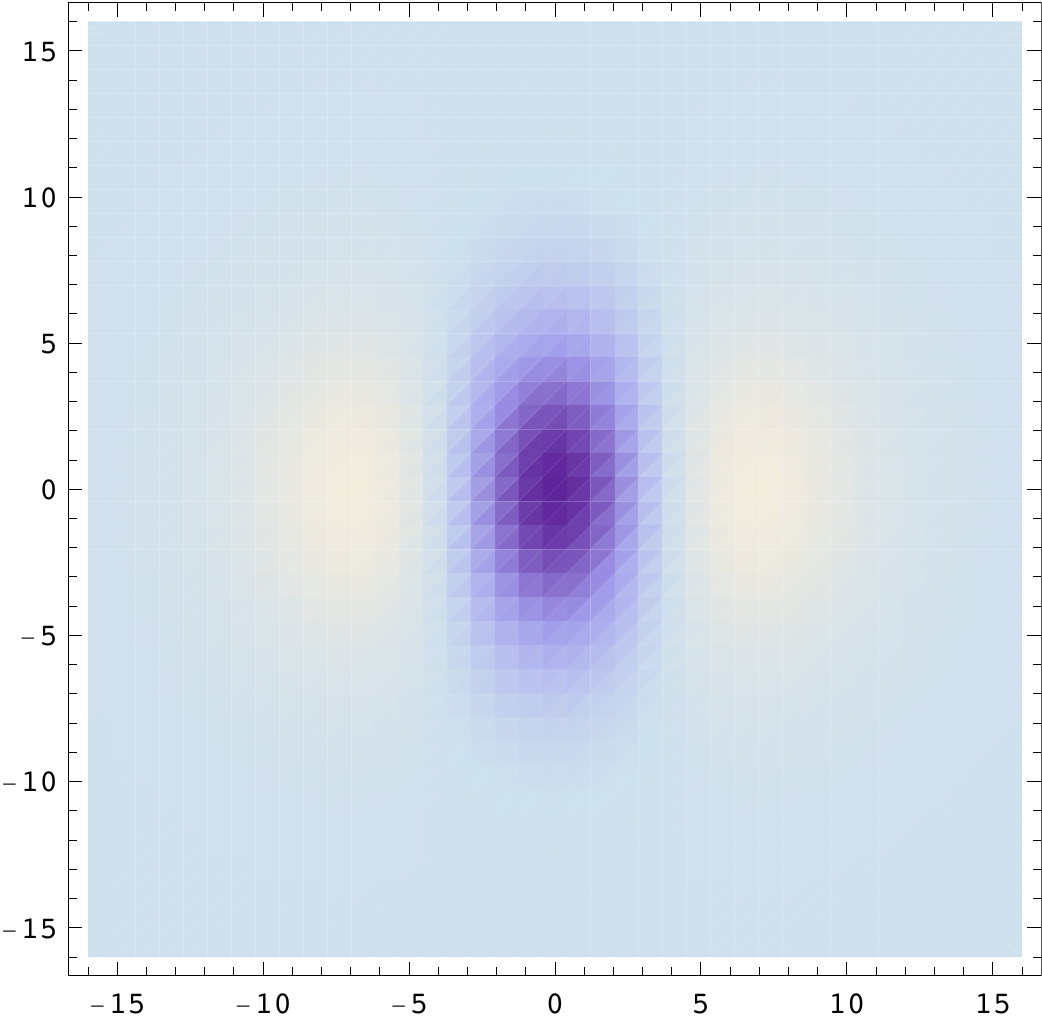} \hspace{-2mm} &
      \hspace{-2mm} \includegraphics[width=0.14\textwidth]{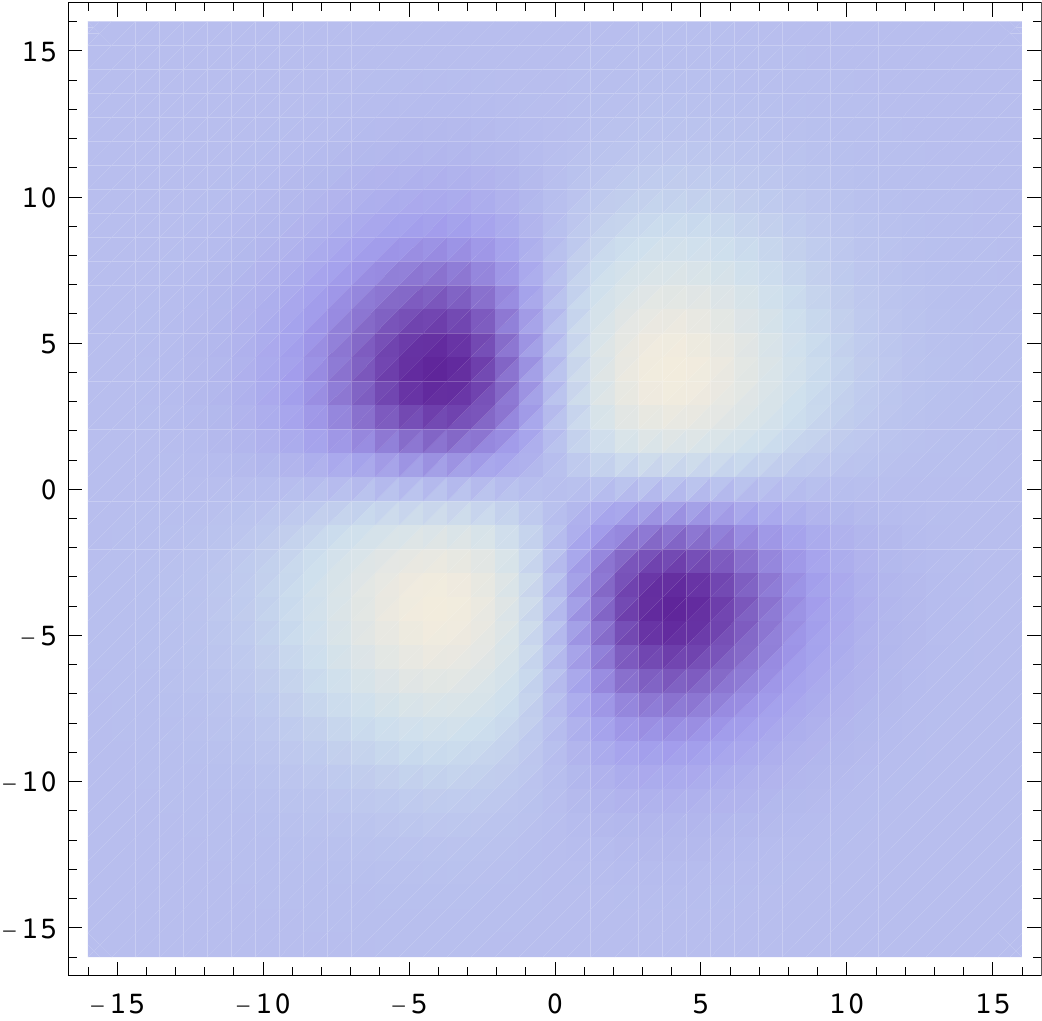} \hspace{-2mm} &
      \hspace{-2mm} \includegraphics[width=0.14\textwidth]{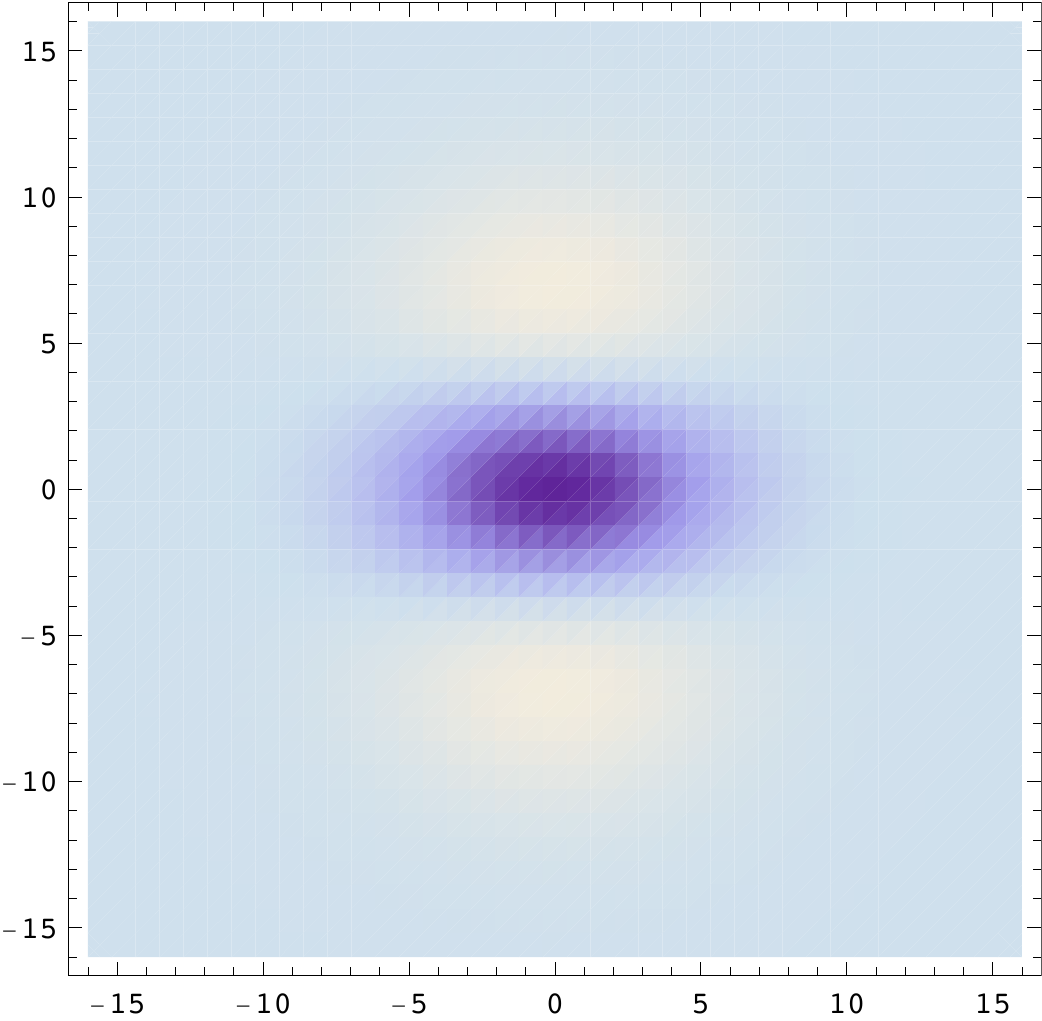} \hspace{-2mm} \\
    \end{tabular} 
  \end{center}
  \vspace{-3mm}
  \caption{Spatial receptive fields formed by the 2-D Gaussian kernel with its partial derivatives up to
    order two.    The corresponding family of receptive fields is closed under
    translations, rotations and scaling transformations, meaning that
    if the underlying image is subject to a set of such image
    transformations then it will always be possible to find some
    possibly other receptive field such that the receptive field
    responses of the original image and the transformed image can be
    matched.}
  \label{fig-Gauss-ders-2D}

  \medskip

  \begin{center}
    \begin{tabular}{ccc}
    \hspace{-2mm} {\footnotesize $g(x;\; \Sigma_1)$} \hspace{-2mm} 
    & \hspace{-2mm} {\footnotesize $g(x;\; \Sigma_2)$} \hspace{-2mm} 
    & \hspace{-2mm} {\footnotesize $g(x;\; \Sigma_3)$} \hspace{-2mm} \\
     \hspace{-2mm} \includegraphics[width=0.14\textwidth]{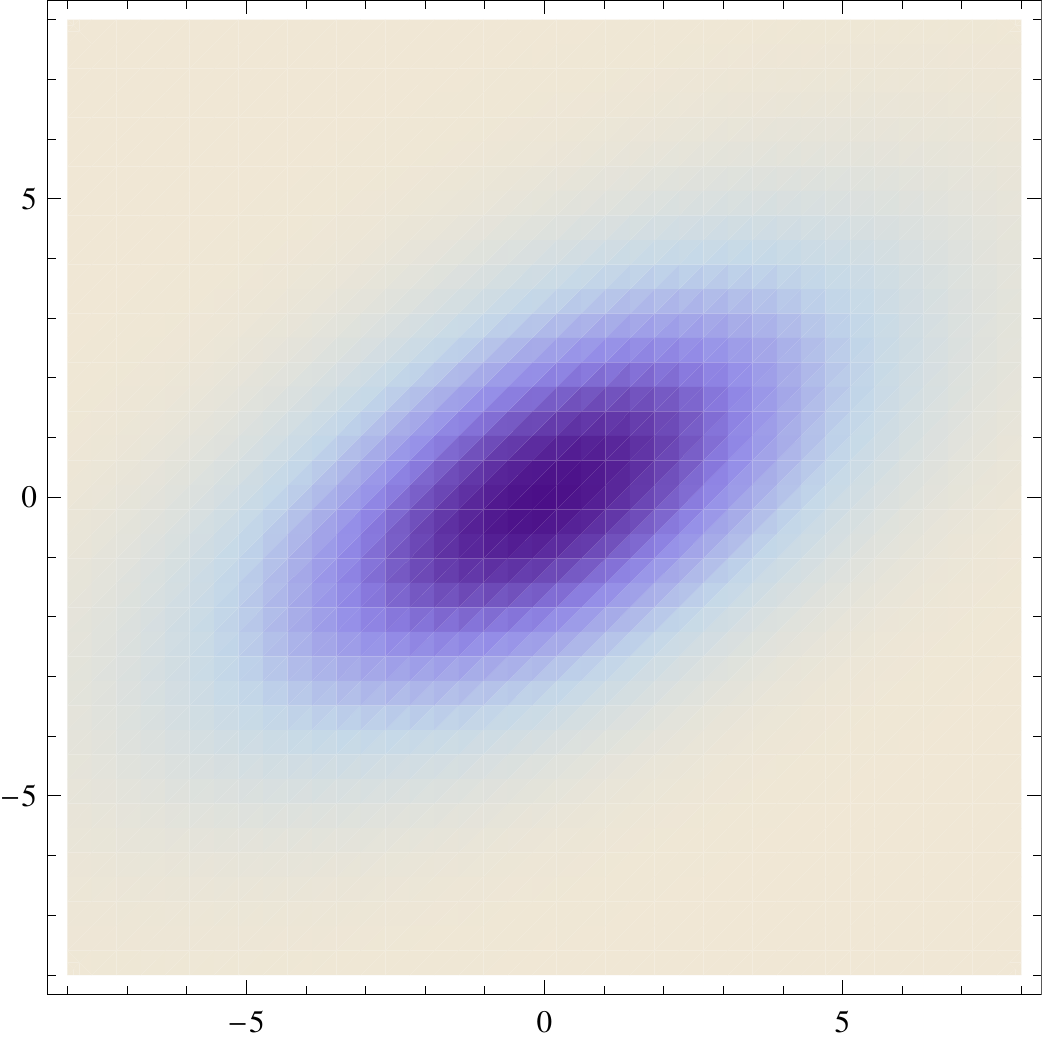} \hspace{-2mm} &
     \hspace{-2mm} \includegraphics[width=0.14\textwidth]{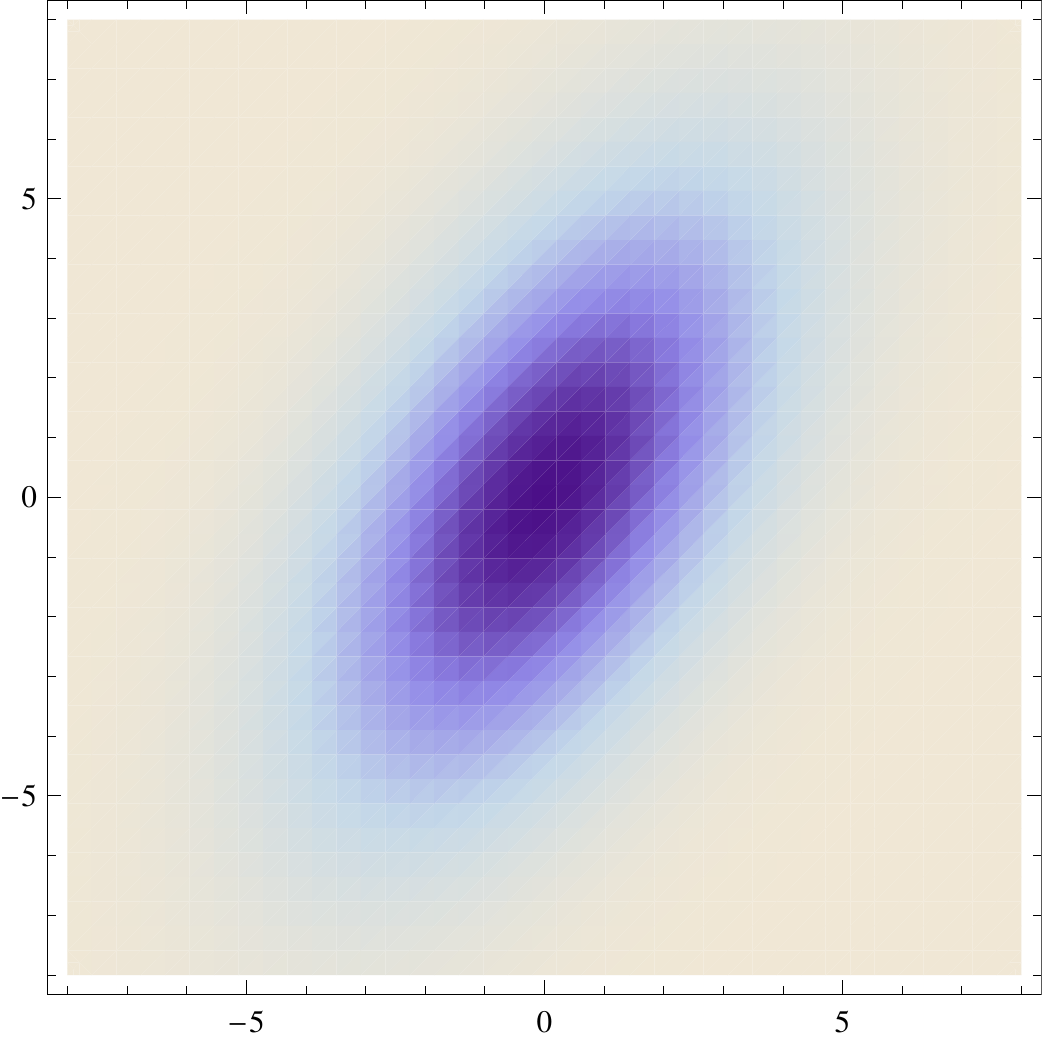} \hspace{-2mm} &
     \hspace{-2mm} \includegraphics[width=0.14\textwidth]{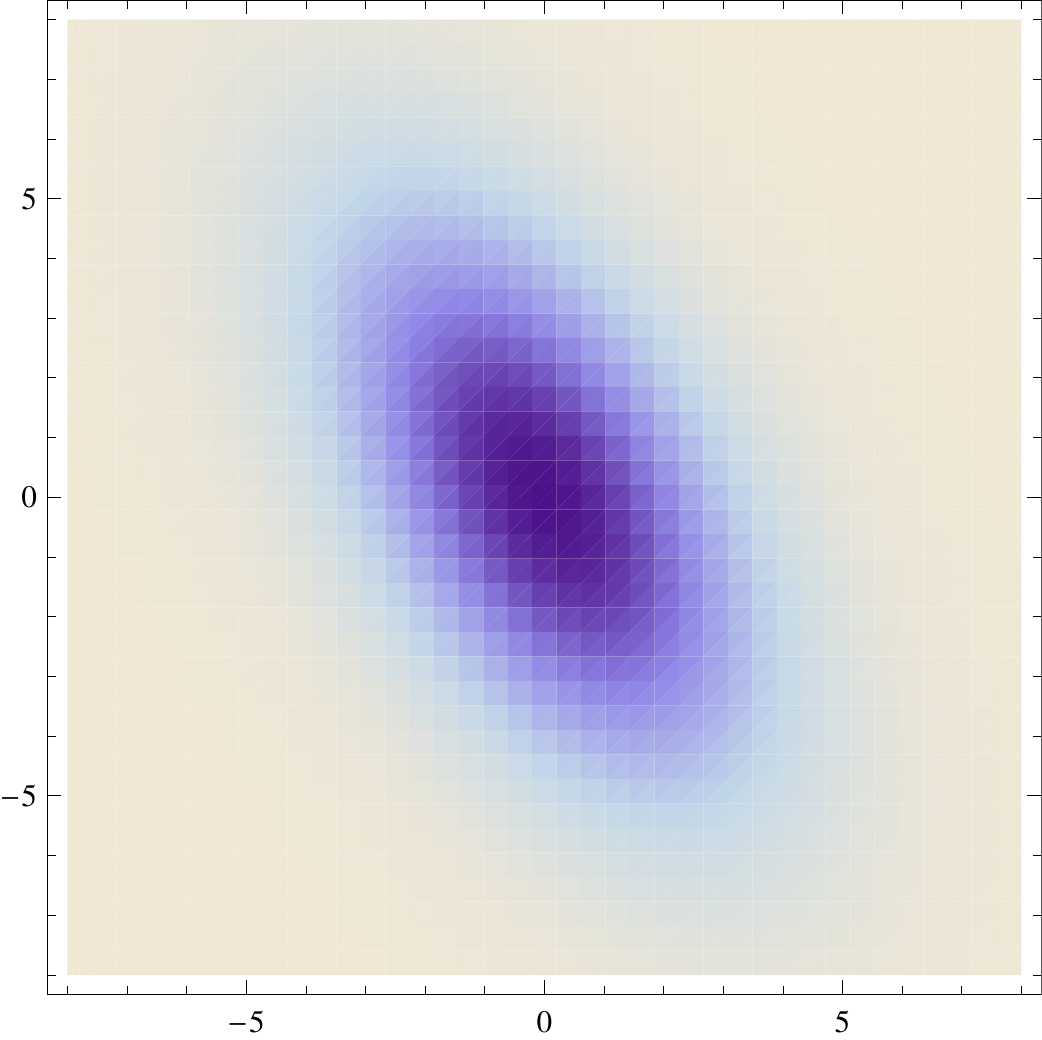} \hspace{-2mm} \\
   \end{tabular} 
  \end{center}
  \vspace{-6mm}
   \begin{center}
    \begin{tabular}{ccc}
    \hspace{-2mm} {\footnotesize $g_{\varphi_1}(x;\; \Sigma_1)$} \hspace{-2mm} 
    & \hspace{-2mm} {\footnotesize $g_{\varphi_2}(x;\; \Sigma_2)$} \hspace{-2mm} 
    & \hspace{-2mm} {\footnotesize $g_{\varphi_3}(x;\; \Sigma_3)$} \hspace{-2mm} \\
      \hspace{-2mm} \includegraphics[width=0.14\textwidth]{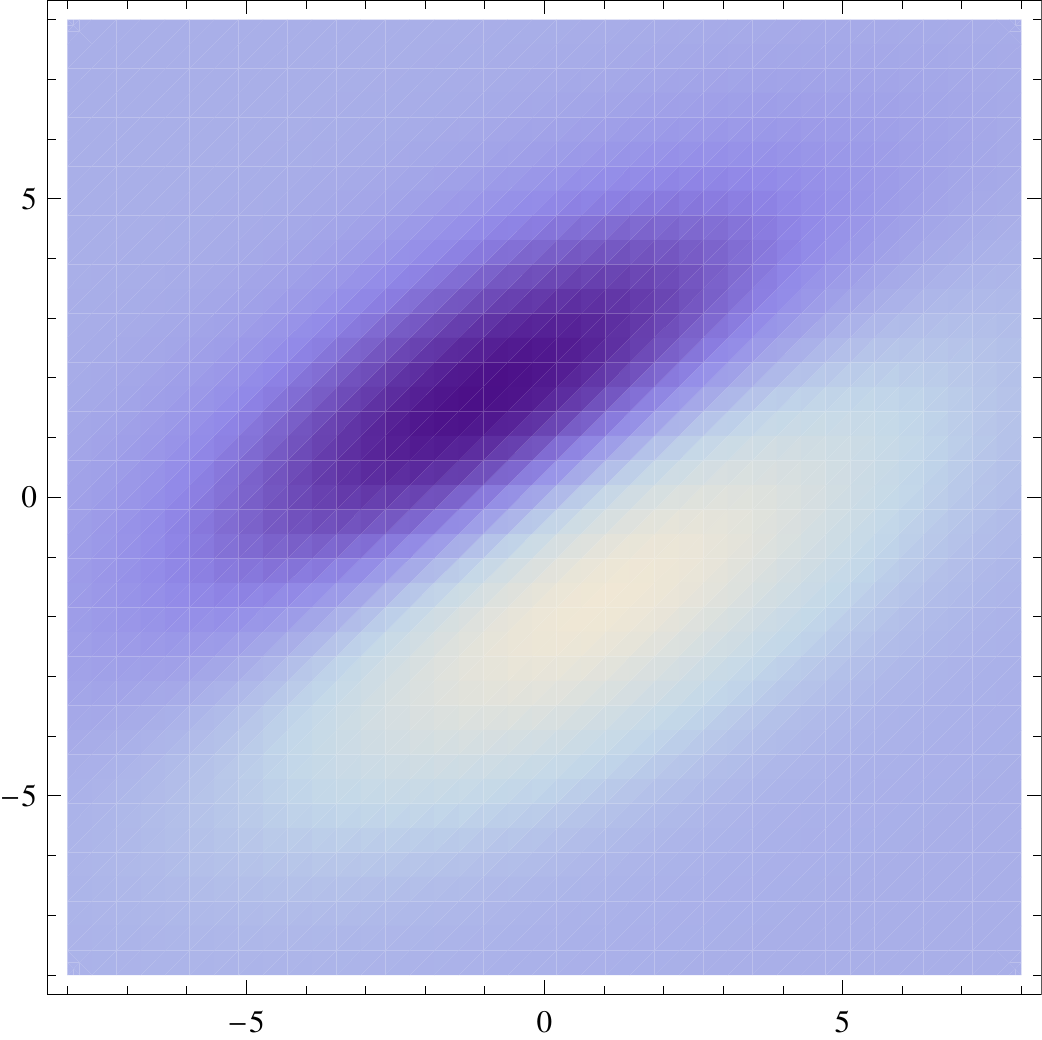} \hspace{-2mm} &
      \hspace{-2mm} \includegraphics[width=0.14\textwidth]{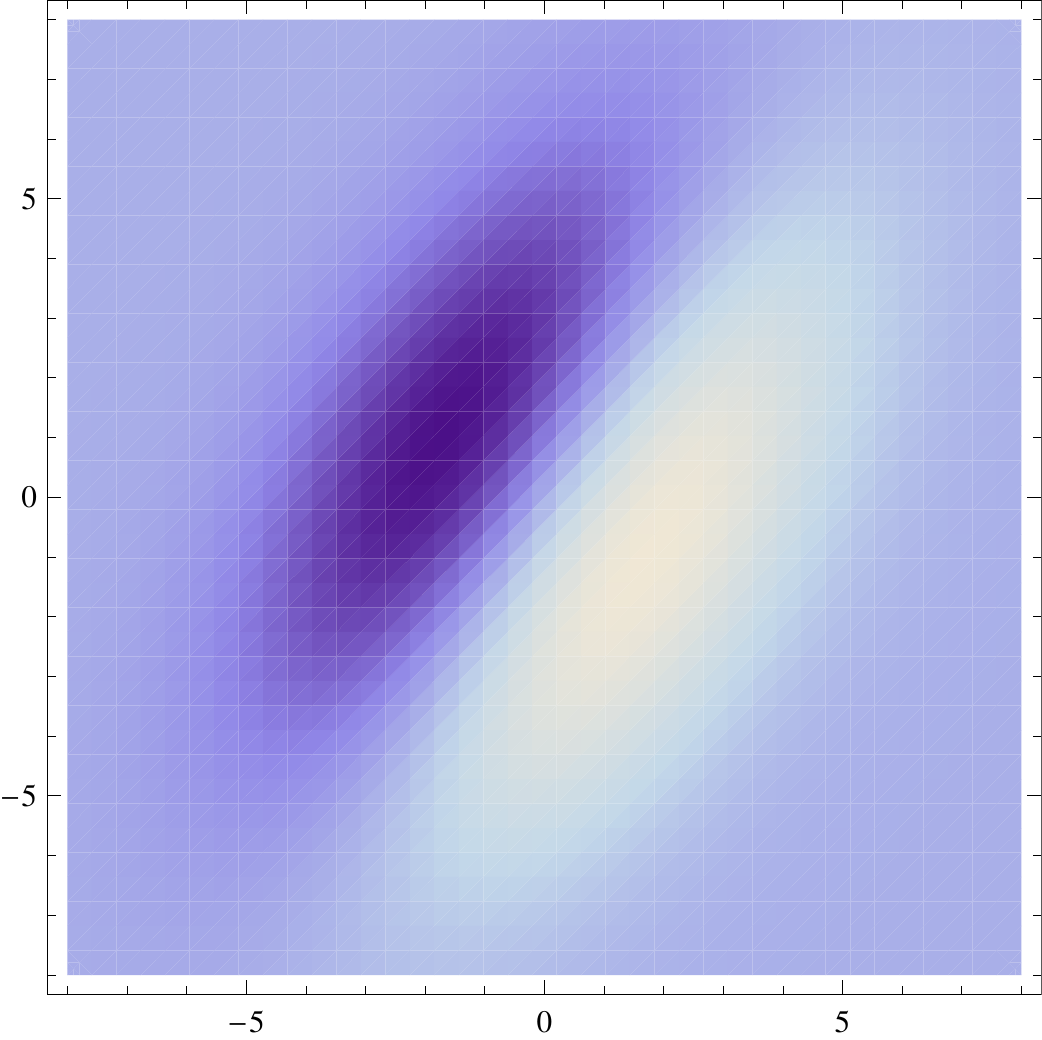} \hspace{-2mm} &
      \hspace{-2mm} \includegraphics[width=0.14\textwidth]{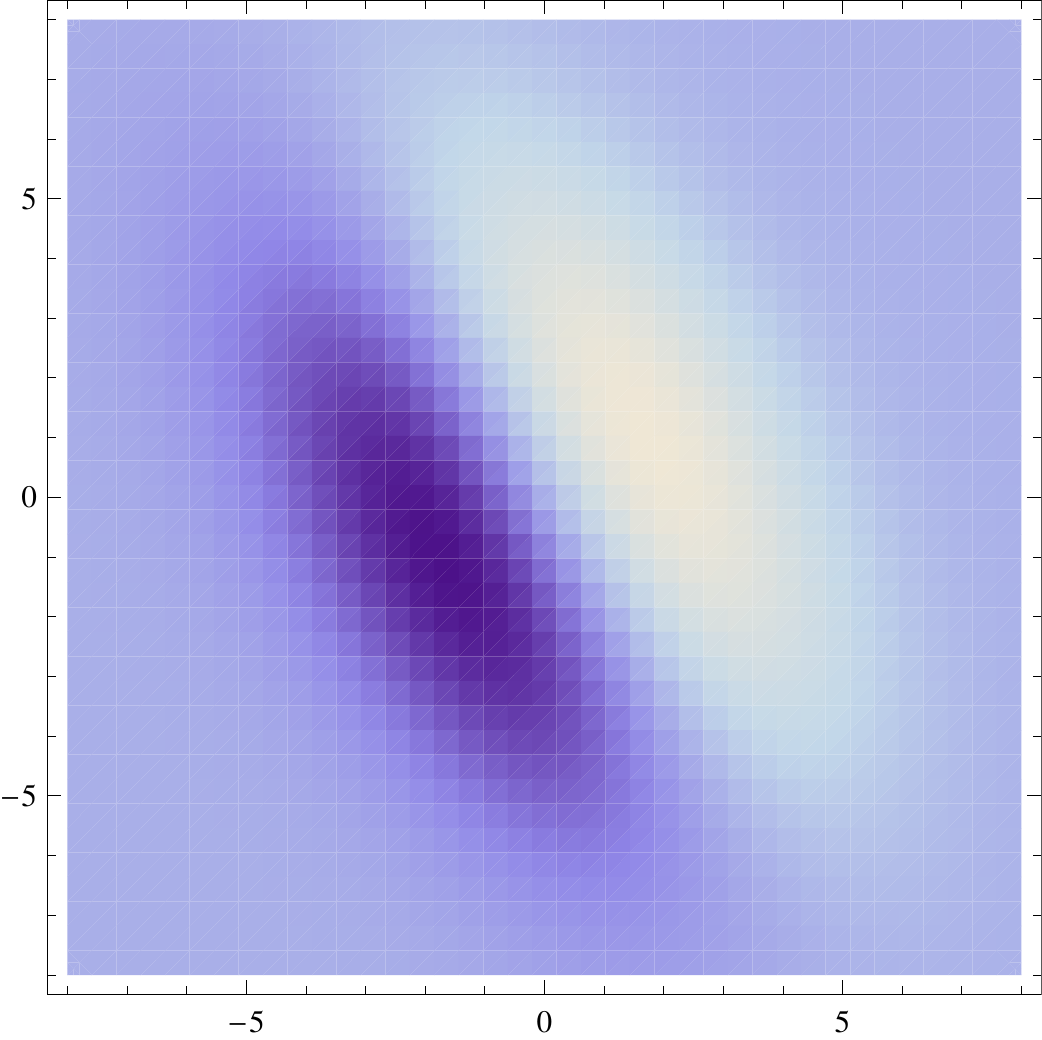} \hspace{-2mm} \\
    \end{tabular} 
  \end{center}
  \vspace{-6mm}
  \begin{center}
    \begin{tabular}{ccc}
   \hspace{-2mm} {\footnotesize $g_{\varphi_1\varphi_1}(x;\; \Sigma_1)$} \hspace{-2mm} 
    & \hspace{-2mm} {\footnotesize $g_{\varphi_2\varphi_2}(x;\; \Sigma_2)$} \hspace{-2mm} 
    & \hspace{-2mm} {\footnotesize $g_{\varphi_3\varphi_3}(x;\; \Sigma_3)$} \hspace{-2mm} \\
      \hspace{-2mm} \includegraphics[width=0.14\textwidth]{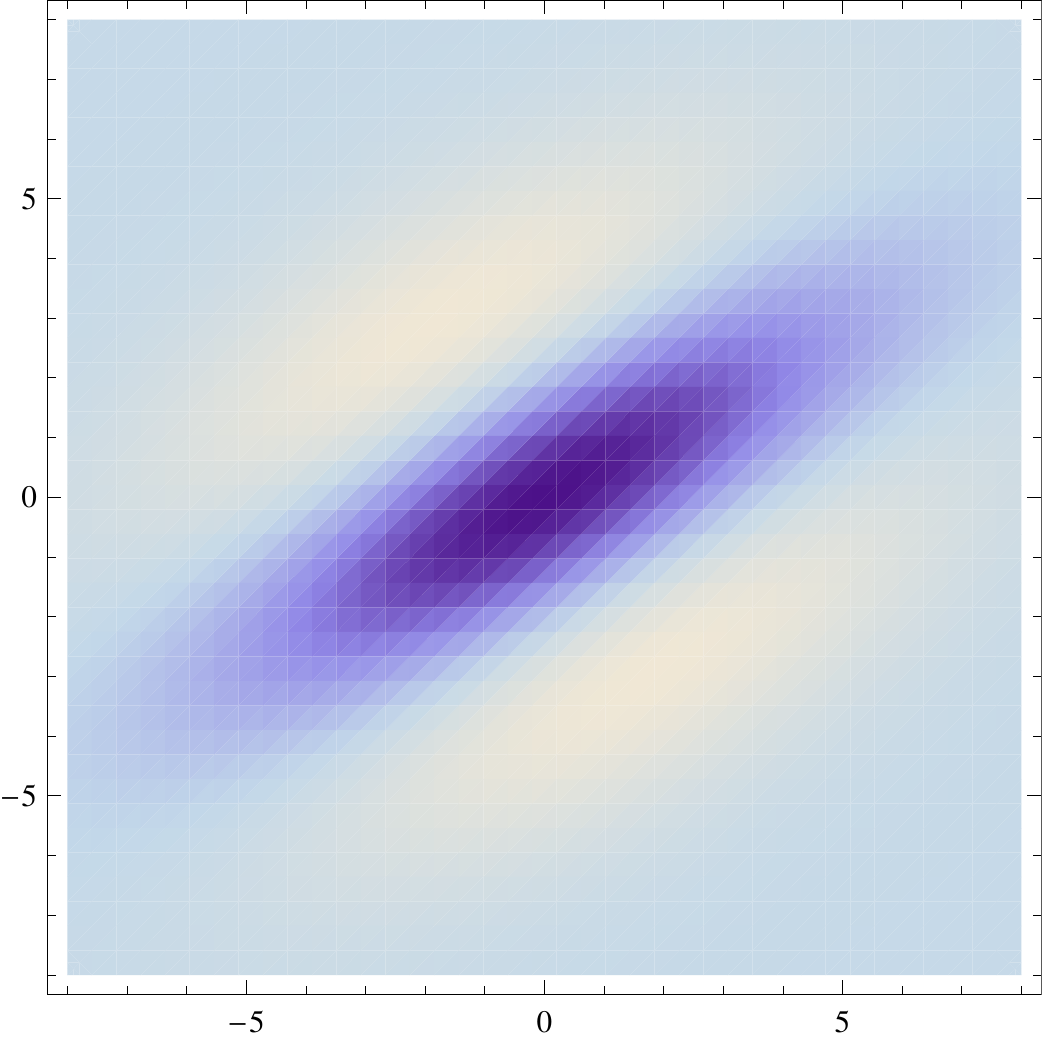} \hspace{-2mm} &
      \hspace{-2mm} \includegraphics[width=0.14\textwidth]{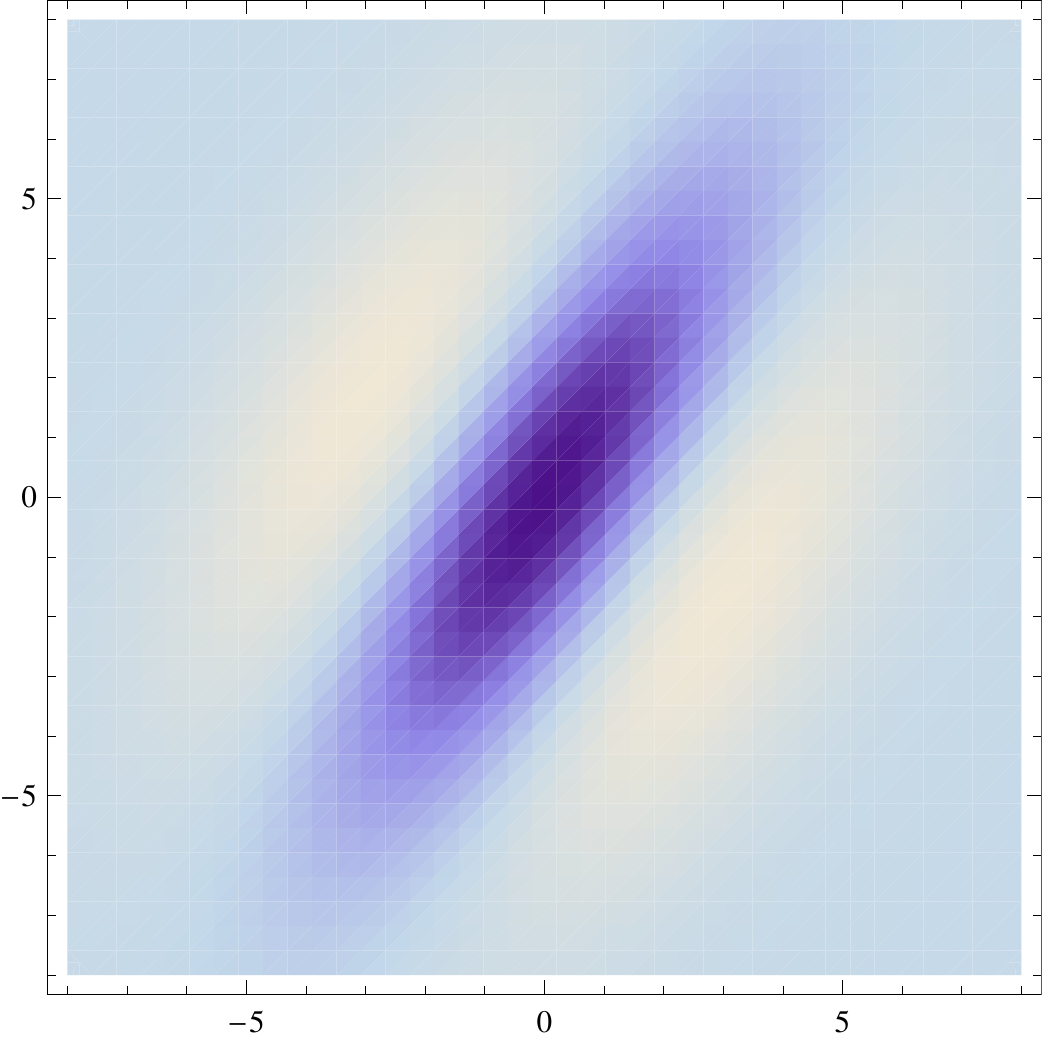} \hspace{-2mm} &
      \hspace{-2mm} \includegraphics[width=0.14\textwidth]{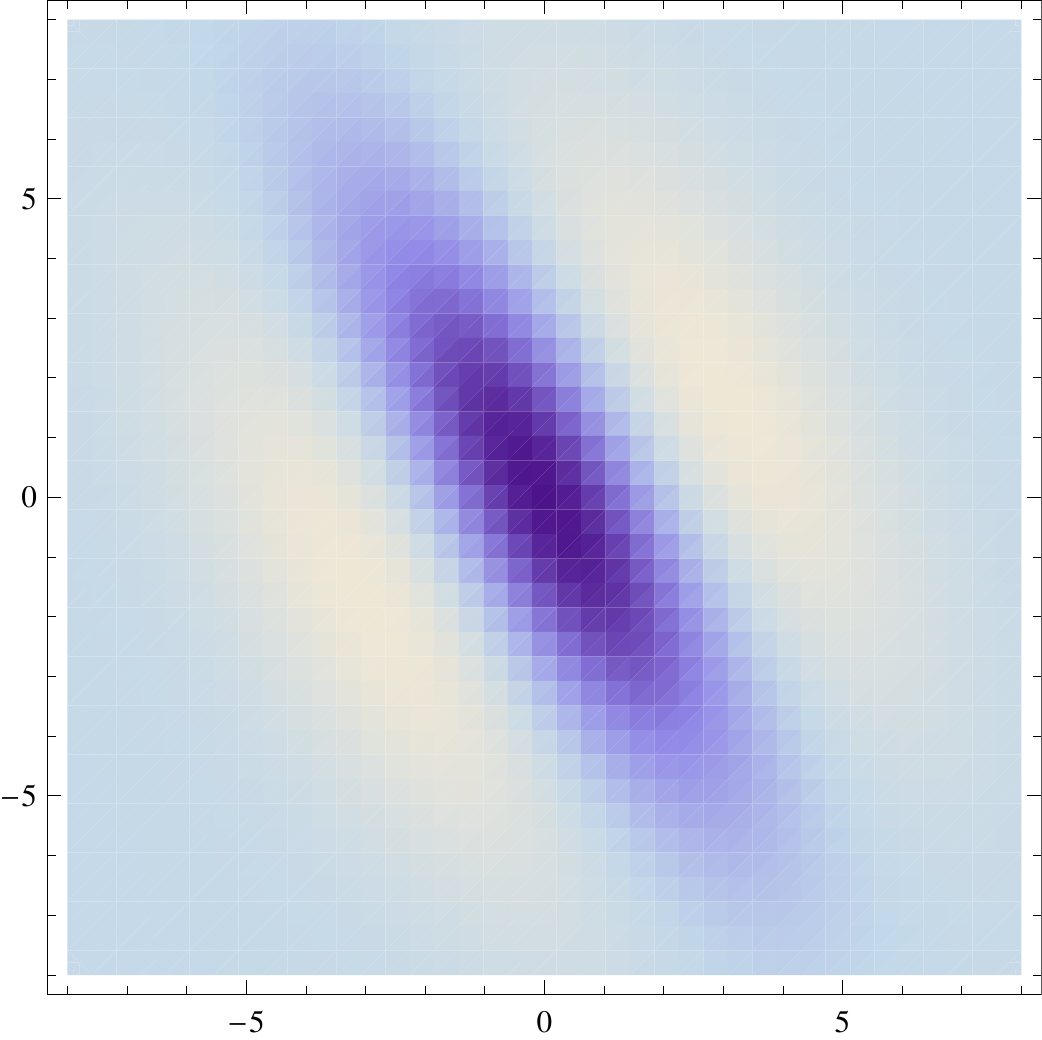} \hspace{-2mm} \\
    \end{tabular} 
  \end{center}
  \vspace{-3mm}
  \caption{Spatial receptive fields formed by affine Gaussian kernels and directional
    derivatives of these, here using three different covariance
    matrices $\Sigma_1$, $\Sigma_2$ and $\Sigma_3$ corresponding to
    the directions $\theta_1 = \pi/6$, $\theta_2 = \pi/3$ and
    $\theta_3 = 2\pi/3$ of the major eigendirection of the covariance matrix and
    with first- and second-order directional derivatives computed in the corresponding
    orthogonal directions $\varphi_1$, $\varphi_2$ and $\varphi_3$. The corresponding family of receptive fields
  is closed under general affine transformations of the spatial
  domain, including translations, rotations, scaling transformations and
  perspective foreshortening (although this figure only illustrates
  variabilities in the orientation of the filter, thereby disregarding
variations in both the size and the degree of elongation).
This closedness property implies that receptive field responses
computed from different views of a smooth local surface patch can be
perfectly matched, if the transformation between the two views can be
modelled as a local affine transformation.}
  \label{fig-aff-Gaussian-ders}
\end{figure}

\section{Idealized receptive field families}
\label{sec-ideal-rec-field-fam}

\subsection{Spatial image domain}
\label{sec-spat-RF}

Based on the above assumptions in Section~\ref{sec-assum-spat-domain},
it can be shown \cite{Lin10-JMIV} that when complemented with certain regularity
assumptions in terms of Sobolev norms, they imply that a spatial
scale-space representation $L$ as determined by these must satisfy a
diffusion equation of the form 
\begin{equation}
  \label{eq-vel-adapt-scsp-diff-eq}
  \partial_s L 
  = \frac{1}{2} \nabla^T \left( \Sigma_0 \nabla L \right) - \delta_0^T \nabla L
\end{equation}
for some positive semi-definite covariance matrix $\Sigma_0$ and
some translation vector $\delta_0$.
In terms of convolution kernels, this corresponds to Gaussian kernels
of the form
\begin{equation}
  \label{eq-gauss-gen-spattemp}
  g(x;\; \Sigma_s, \delta_s) 
   = \frac{1}{(2 \pi)^{N/2} \sqrt{\det \Sigma_s}} \,
      e^{- {(x - \delta_s)^T \Sigma_s^{-1} (x - \delta_s)}/{2}},
\end{equation}
which for a given $\Sigma_s = s \, \Sigma_0$ and a given 
$\delta_s = s \, \delta_0$ satisfy (\ref{eq-vel-adapt-scsp-diff-eq}).
If we additionally require these kernels to be mirror symmetric
through the origin, then we obtain
{\em affine Gaussian kernels\/}
\begin{equation}
  \label{eq-aff-gauss}
  g(x;\; \Sigma) 
   = \frac{1}{(2 \pi)^{N/2} \sqrt{\det \Sigma}} \,
      e^{-x^T \Sigma^{-1} x/2}.
\end{equation}
Their spatial derivatives constitute a canonical family for
expressing receptive fields over a spatial domain that can be
summarized on the form
\begin{equation}
  \label{eq-spat-RF-model}
   T(x;\; s, \Sigma)  = g(x;\; s \, \Sigma).
\end{equation}

\begin{figure}[!b]

   \begin{center}
     \begin{tabular}{c}
       \includegraphics[width=0.42\textwidth]{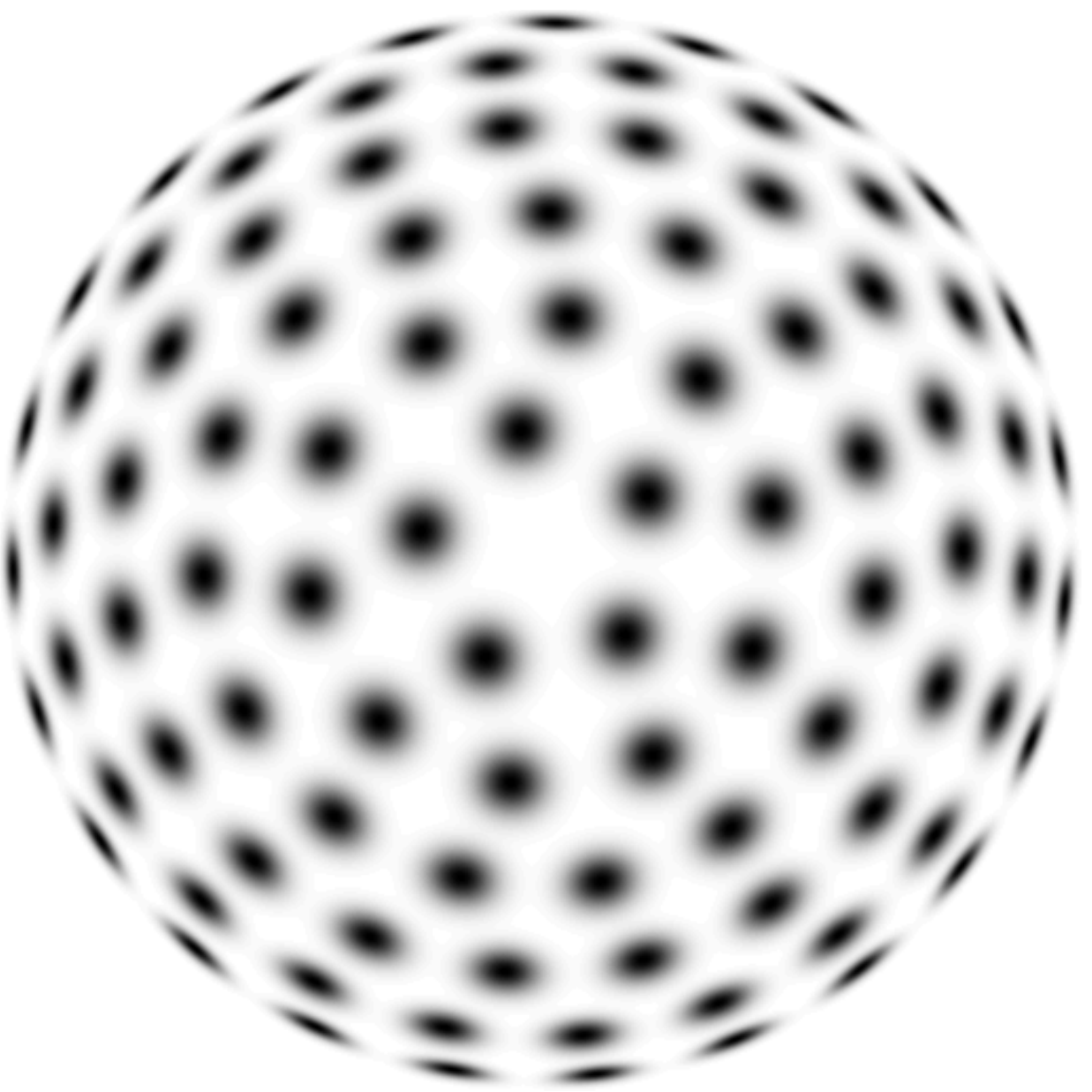} 
     \end{tabular}
   \end{center}

  \caption{Distribution of affine Gaussian receptive fields
    corresponding to a uniform distribution on a hemisphere regarding   
zero-order smoothing kernels. 
In the most idealized version of the theory, one can
    think of all affine receptive fields with their directional
    derivatives in preferred directions aligned to the eigendirections
    of the covariance matrix $\Sigma$ as being present at any
    position in the image domain. When restricted to a limited number
    of receptive fields in an actual implementation, there is also an
    issue of distributing a fixed number of receptive fields over the spatial
    coordinates $x = (x_1, x_2)$ and the filter parameters $s$ and $\Sigma$.}
  \label{fig-distr-aff-rec-fields}
\end{figure}

\noindent
Incorporating the fact that spatial derivatives of these kernels
are also compatible with the assumptions underlying this theory,
this does specifically for the case of a two-dimensional spatial image domain
lead to spatial receptive fields that can be compactly summarized
on the form
\begin{equation}
  \label{eq-spat-RF-model}
   T_{{\varphi}^{m_1} {\orth{\varphi}}^{m_2}}(x_1, x_2;\; s, \Sigma)  
  = \partial_{\varphi}^{m_1} \partial_{\bot \varphi}^{m_2} 
      \left( g(x_1, x_2;\; s \Sigma) \right),
\end{equation}
where 
\begin{itemize}
\item
  $x = (x_1, x_2)$ denote the spatial coordinates,
\item
  $s$ denotes the spatial scale,
\item
 $\Sigma$ denotes a spatial covariance matrix determining the shape of
 a spatial affine Gaussian kernel,
\item
  $m_1$ and $m_2$ denote orders of spatial differentiation,
\item
  $\partial_{\varphi} = \cos \varphi \, \partial_{x_1} + \sin \varphi \, \partial_{x_2}$,
  $\partial_{\bot \varphi} = \sin \varphi \, \partial_{x_1} - \cos \varphi \, \partial_{x_2}$
 denote spatial directional derivative operators in two orthogonal
 directions $\varphi$ and $\bot \varphi$ aligned with the eigenvectors
 of the covariance matrix $\Sigma$,
\item
  $g(x;\; s, \Sigma)  = \frac{1}{2 \pi s \sqrt{\det\Sigma}} e^{-x^T \Sigma^{-1} x/2s}$
  is an affine Gaussian kernel with its size determined by the spatial scale parameter 
  $s$ and its shape by the spatial covariance matrix $\Sigma$.
\end{itemize}
Figure~\ref{fig-Gauss-ders-2D} and
Figure~\ref{fig-aff-Gaussian-ders} show examples of spatial receptive
fields from this family up to second order of spatial
differentiation. 
Figure~\ref{fig-Gauss-ders-2D} shows partial derivatives of the
Gaussian kernel for the specific case when the covariance matrix
$\Sigma$ is restricted to a unit matrix and the Gaussian kernel thereby
becomes rotationally symmetric.
The resulting family of receptive fields is closed under scaling
transformations over the spatial domain, implying that
if an object is seen from different distances to the observer, then it will always be
possible to find a transformation of the scale parameter $s$ between
the two image domains so that the receptive field responses computed
from the two image domains can be matched.
Figure~\ref{fig-aff-Gaussian-ders} shows examples of affine Gaussian
receptive fields for covariance matrices $\Sigma$ that do not
correspond to rescaled copies of the unit matrix.
The resulting full family of affine Gaussian derivative kernels is
closed under general affine transformations, implying that for two
different perspective views of a local smooth surface patch, it will always
be possible to find a transformation of the covariance matrices
$\Sigma$ between the two domains so that the receptive field responses
can be matched, if the transformation between the two image domains is
approximated by a local affine transformation.

In the most idealized version of the theory, one should think of
receptive fields for all combinations of filter parameters as being present
at every image point, as illustrated in
Figure~\ref{fig-distr-aff-rec-fields} concerning affine Gaussian
receptive fields over different orientations in image space and
different eccentricities.

\subsection{Spatio-temporal image domain}
\label{sec-spat-temp-RF}

Over a non-causal spatio-temporal domain, corresponding arguments as
in Section~\ref{sec-spat-RF} lead to a similar form of diffusion
equation as in Equation~(\ref{eq-vel-adapt-scsp-diff-eq}), while
expressed over the joint space-time domain $p = (x, t)$ and
with $\delta_0$ interpreted as a local drift velocity.
After splitting the composed affine Gaussian
spatio-temporal smoothing kernel corresponding to
(\ref{eq-gauss-gen-spattemp}) while expressed over the joint space-time domain into
separate smoothing operations over space and time, this leads to
zero-order spatio-temporal receptive fields of the form \cite{Lin10-JMIV,Lin13-BICY}:
\begin{equation}
\label{eq-spat-temp-RF-model}
  T(x_1, x_2, t;\; s, \tau;\; v, \Sigma) 
  = g(x_1 - v_1 t, x_2 - v_2 t;\; s, \Sigma) \, h(t;\; \tau).
\end{equation}
After combining that
result with the results from corresponding theoretical analysis for a time-causal
spatio-temporal domain in \cite{Lin10-JMIV,Lin16-JMIV}, the resulting
spatio-temporal derivative kernels constituting the spatio-temporal
extension of the spatial receptive field model (\ref{eq-spat-RF-model}) can be
reparametrised and summarized on the following form
(see \cite{Lin10-JMIV,Lin13-BICY,Lin13-PONE,Lin16-JMIV}):
\begin{align}
  \begin{split}
  \label{eq-spat-temp-RF-model-der}
    & T_{{\varphi}^{m_1} {\bot \varphi}^{m_2} {\bar t}^n}(x_1, x_2, t;\; s, \tau;\; v, \Sigma)
  \end{split}\nonumber\\
  \begin{split}
    & = \partial_{\varphi}^{m_1} \partial_{\bot \varphi}^{m_2} \partial_{\bar t}^n 
           \left( g(x_1 - v_1 t, x_2 - v_2 t;\; s, \Sigma) \, h(t;\; \tau) \right),
  \end{split}
\end{align}
where 
\begin{itemize}
\item
  $x = (x_1, x_2)$ denote the spatial coordinates,
\item
  $t$ denotes time,
\item
  $s$ denotes the spatial scale,
\item
  $\tau$ denotes the temporal scale,
\item
  $v = (v_1, v_2)^T$ denotes a local image velocity,
\item
 $\Sigma$ denotes a spatial covariance matrix determining the shape of
 a spatial affine Gaussian kernel,
\item
  $m_1$ and $m_2$ denote orders of spatial differentiation,
\item
  $n$ denotes the order of temporal differentiation,
\item
  $\partial_{\varphi} = \cos \varphi \, \partial_{x_1} + \sin \varphi  \, \partial_{x_2}$ and
  $\partial_{\bot \varphi} = \sin \varphi \, \partial_{x_1} - \cos \varphi \, \partial_{x_2}$
 denote spatial directional derivative operators in two orthogonal
 directions $\varphi$ and $\bot \varphi$ aligned with the eigenvectors
 of the covariance matrix $\Sigma$,
\item
  $\partial_{\bar t} = v_1 \partial_{x_1} + v_2 \partial_{x_2} + \partial t$
  is a velocity-adapted temporal derivative operator aligned to the
  direction of the local image velocity $v = (v_1, v_2)^T$,
\item
  $g(x;\; s, \Sigma)  = \frac{1}{2 \pi s \sqrt{\det\Sigma}} e^{-x^T \Sigma^{-1} x/2s}$
  is an affine Gaussian kernel with its size determined by the spatial scale parameter 
  $s$ and its shape determined by the spatial covariance matrix $\Sigma$,
\item
  $g(x_1 - v_1 t, x_2 - v_2 t;\; s, \Sigma)$ denotes a spatial affine
  Gaussian kernel that moves with image velocity $v = (v_1, v_2)$ in space-time and
\item
  $h(t;\; \tau)$ is a temporal smoothing kernel over time
  corresponding to a Gaussian kernel 
  $h(t;\; \tau) = g(t;\; \tau) = 1/\sqrt{2 \pi \tau} \exp(-t^2/2\tau)$
  in the case of non-causal time or a cascade of first-order
  integrators or equivalently truncated exponential kernels coupled in
  cascade $h(t;\; \tau) = h_{composed}(\cdot;\; \mu)$ according to (\ref{eq-comp-trunc-exp-cascade}) over a time-causal temporal domain.
\end{itemize}
This family of spatio-temporal scale-space kernels can be seen as a
canonical family of linear receptive fields over a spatio-temporal
domain.

\begin{figure}[!h]
   \begin{center}
      \includegraphics[width=0.47\textwidth]{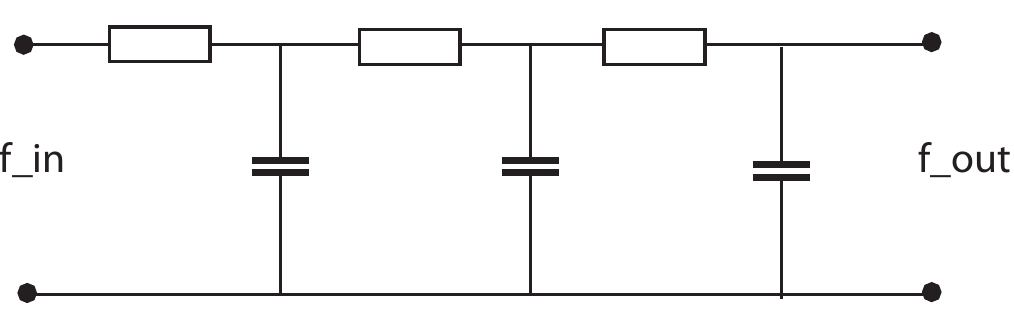}
   \end{center}
\caption{Electric wiring diagram consisting of a set of resistors and
  capacitors that emulate a series of first-order integrators coupled in
  cascade, if we regard the time-varying voltage $f_{in}$ as
  representing the time varying input signal and the resulting output
  voltage $f_{out}$ as representing the time varying output signal at a
  coarser temporal scale.
  According to the theory of temporal scale-space
  kernels for one-dimensional signals 
  (Lindeberg \protect\cite{Lin90-PAMI,Lin16-JMIV}; 
   Lindeberg and Fagerstr{\"o}m \protect\cite{LF96-ECCV}),
  the corresponding equivalent truncated exponential kernels are the only
  primitive temporal smoothing kernels that guarantee both temporal
  causality and non-creation of local extrema
  (alternatively zero-crossings) with increasing temporal scale.}
  \label{fig-first-order-integrators-electric}
\end{figure}

For the case of a time-causal temporal domain, the result states that
truncated exponential kernels of the form
 \begin{equation}
  \label{eq-trunc-exp-kern-prim}
    h_{exp}(t;\; \mu_k) 
    = \left\{
        \begin{array}{ll}
          \frac{1}{\mu_k} e^{-t/\mu_k} & t \geq 0 \\
          0         & t < 0
        \end{array}
      \right.
  \end{equation}
coupled in cascade constitute the natural temporal smoothing kernels.
These do in turn lead to a composed temporal convolution kernel of the form
\begin{equation}
  \label{eq-comp-trunc-exp-cascade}
  h_{composed}(\cdot;\; \mu) 
  = *_{k=1}^{K} h_{exp}(\cdot;\; \mu_k)
\end{equation}
and corresponding to a set of first-order integrators coupled in
cascade (see Figure~\ref{fig-first-order-integrators-electric}).

\begin{figure}[!p]
  \begin{center}
    \begin{tabular}{c}
     \hspace{-2mm} {\footnotesize $-T(x, t;\; s, \tau)$} \hspace{-2mm} \\
      \hspace{-2mm} \includegraphics[width=0.15\textwidth]{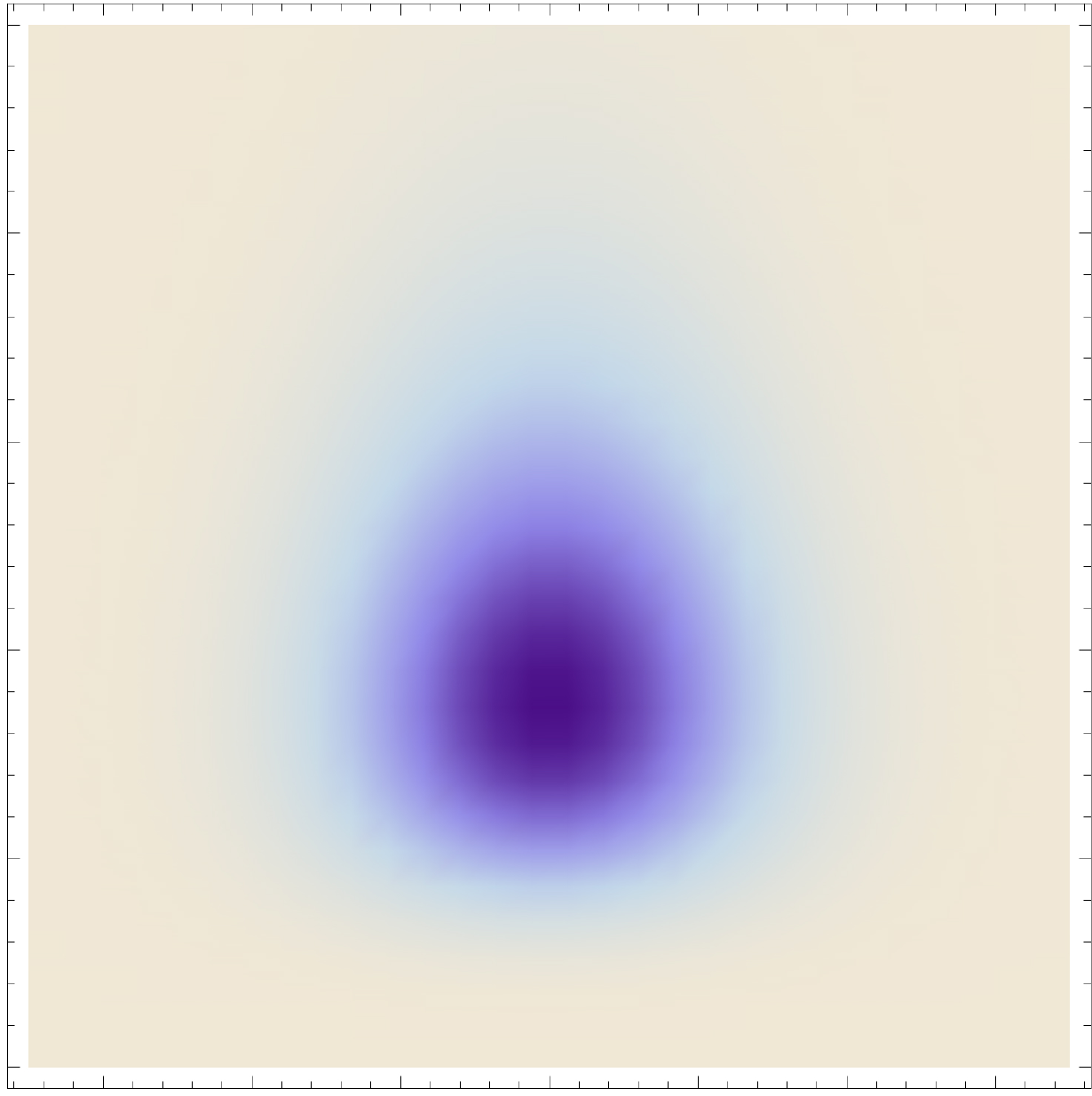} \hspace{-2mm} \\
    \end{tabular} 
  \end{center}
  \vspace{-6mm}
  \begin{center}
    \begin{tabular}{cc}
      \hspace{-2mm} {\footnotesize $T_x(x, t;\; s, \tau)$} \hspace{-2mm} 
      & \hspace{-2mm} {\footnotesize $T_t(x, t;\; s, \tau, \delta)$} \hspace{-2mm} \\
      \hspace{-2mm} \includegraphics[width=0.15\textwidth]{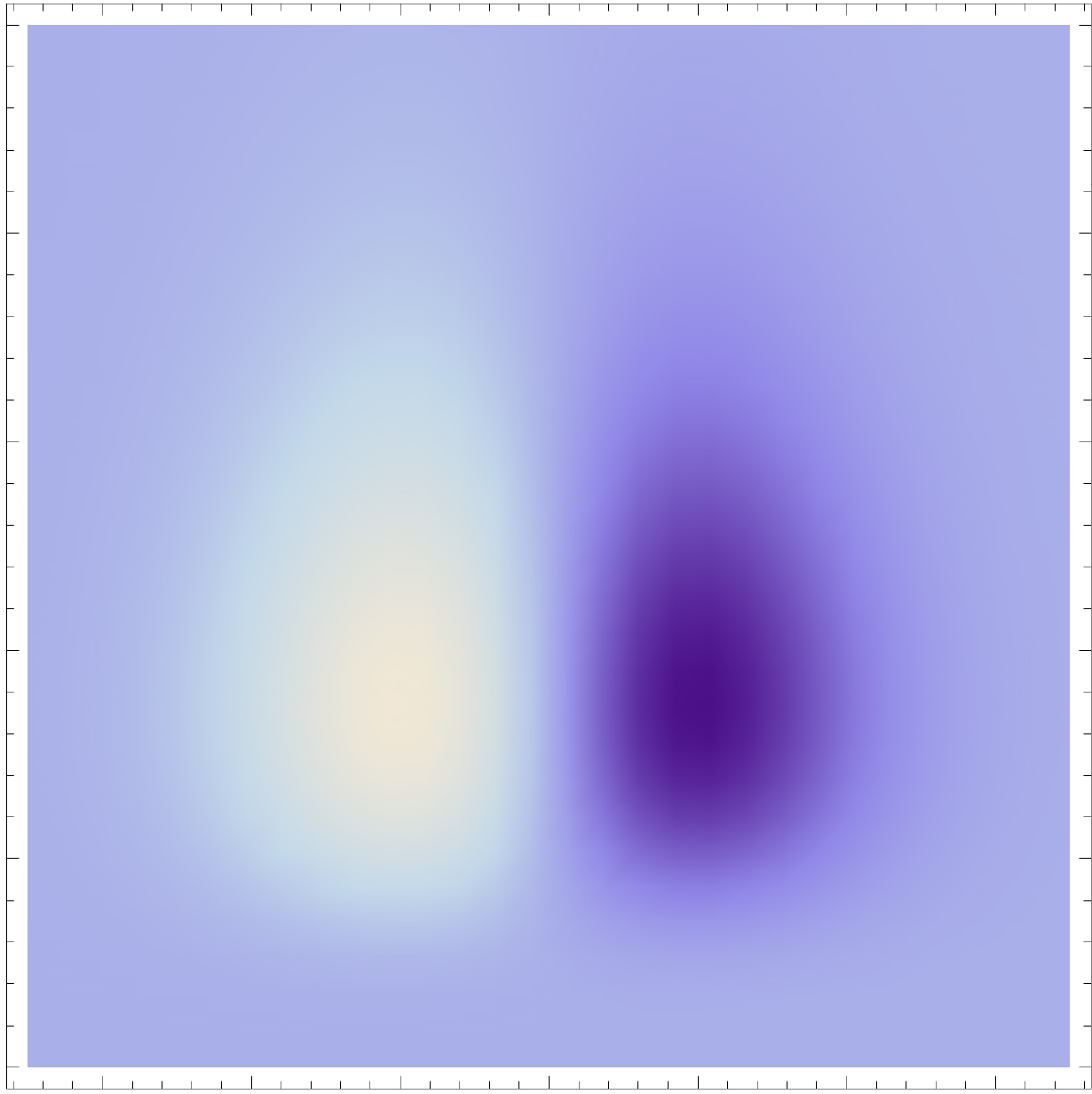} \hspace{-2mm} &
      \hspace{-2mm} \includegraphics[width=0.15\textwidth]{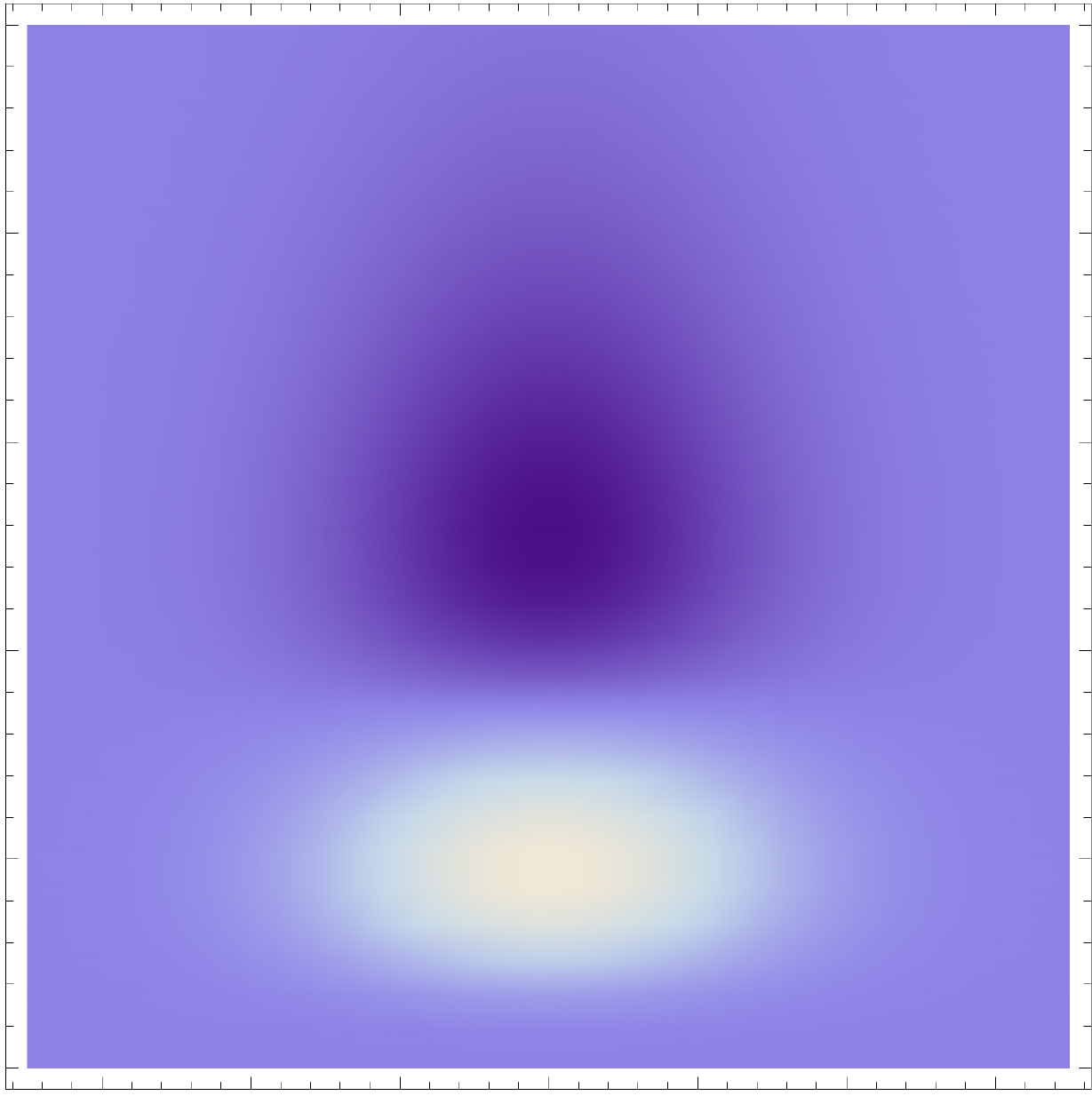} \hspace{-2mm} \\
    \end{tabular} 
  \end{center}
  \vspace{-6mm}
  \begin{center}
    \begin{tabular}{ccc}
      \hspace{-2mm} {\footnotesize $T_{xx}(x, t;\; s, \tau)$} \hspace{-2mm} 
      & \hspace{-2mm} {\footnotesize $T_{xt}(x, t;\; s, \tau)$} \hspace{-2mm} 
      & \hspace{-2mm} {\footnotesize $T_{tt}(x, t;\; s, \tau)$} \hspace{-2mm} \\
      \hspace{-2mm} \includegraphics[width=0.15\textwidth]{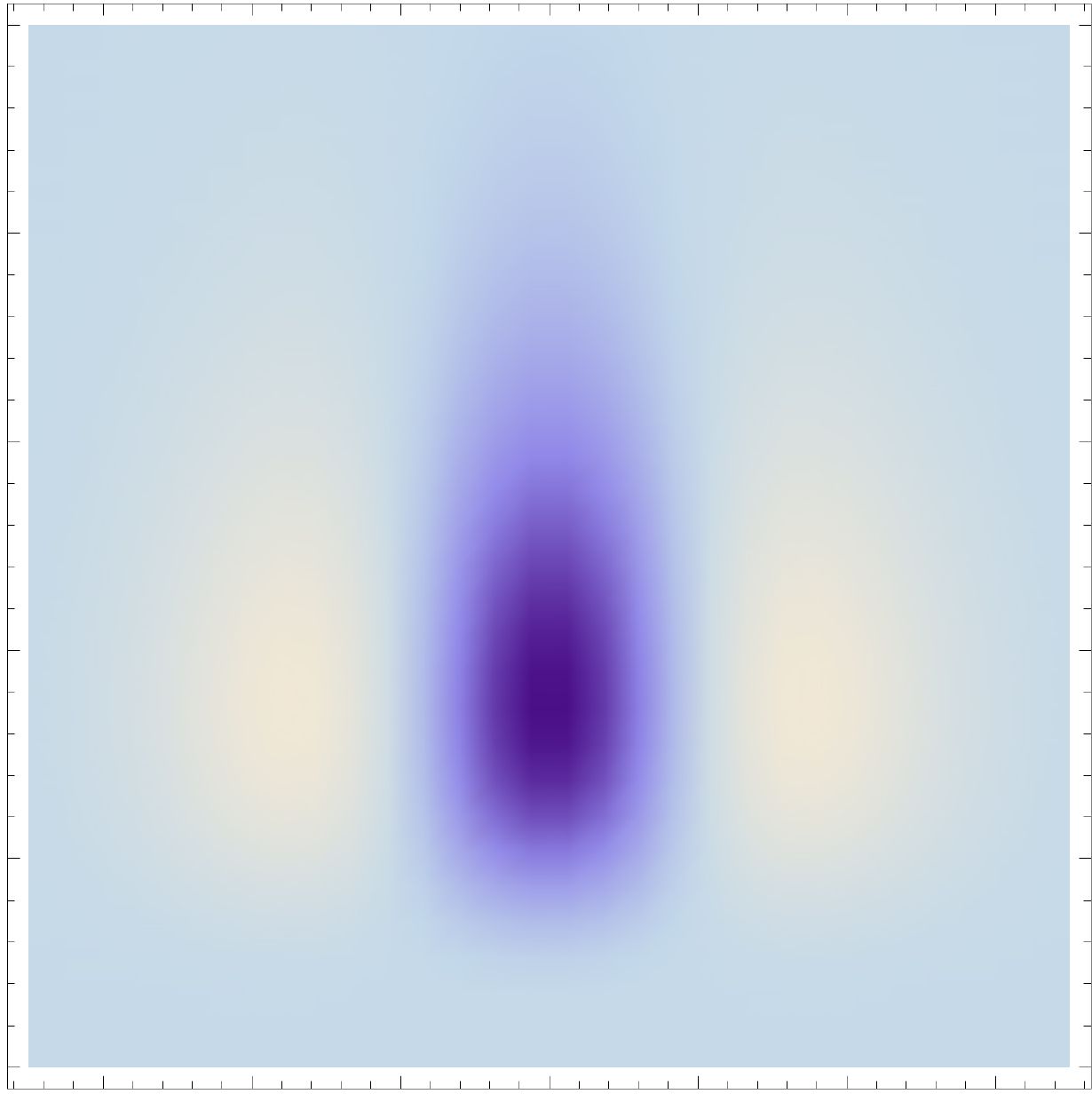} \hspace{-2mm} &
      \hspace{-2mm} \includegraphics[width=0.15\textwidth]{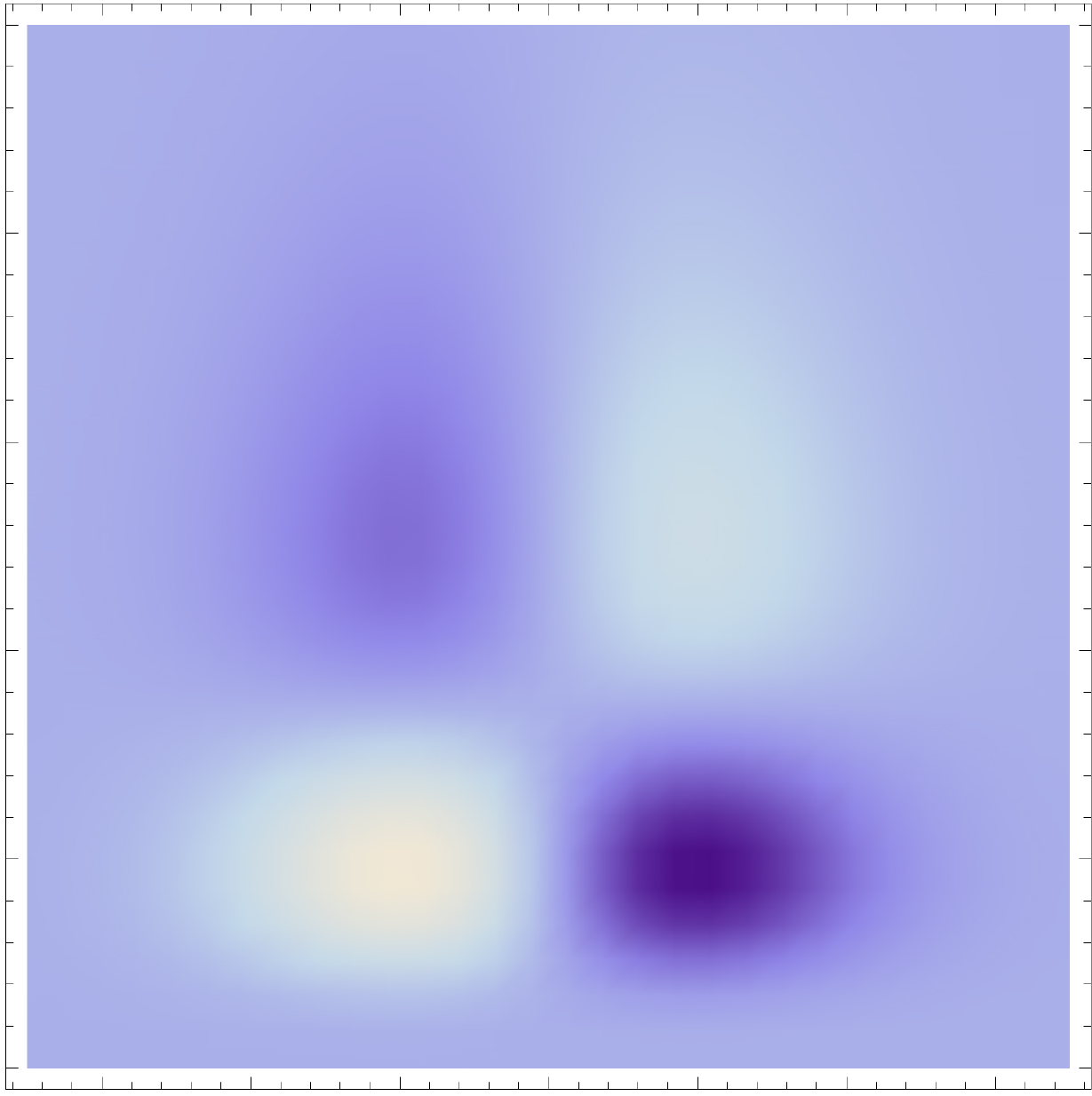} \hspace{-2mm} &
      \hspace{-2mm} \includegraphics[width=0.15\textwidth]{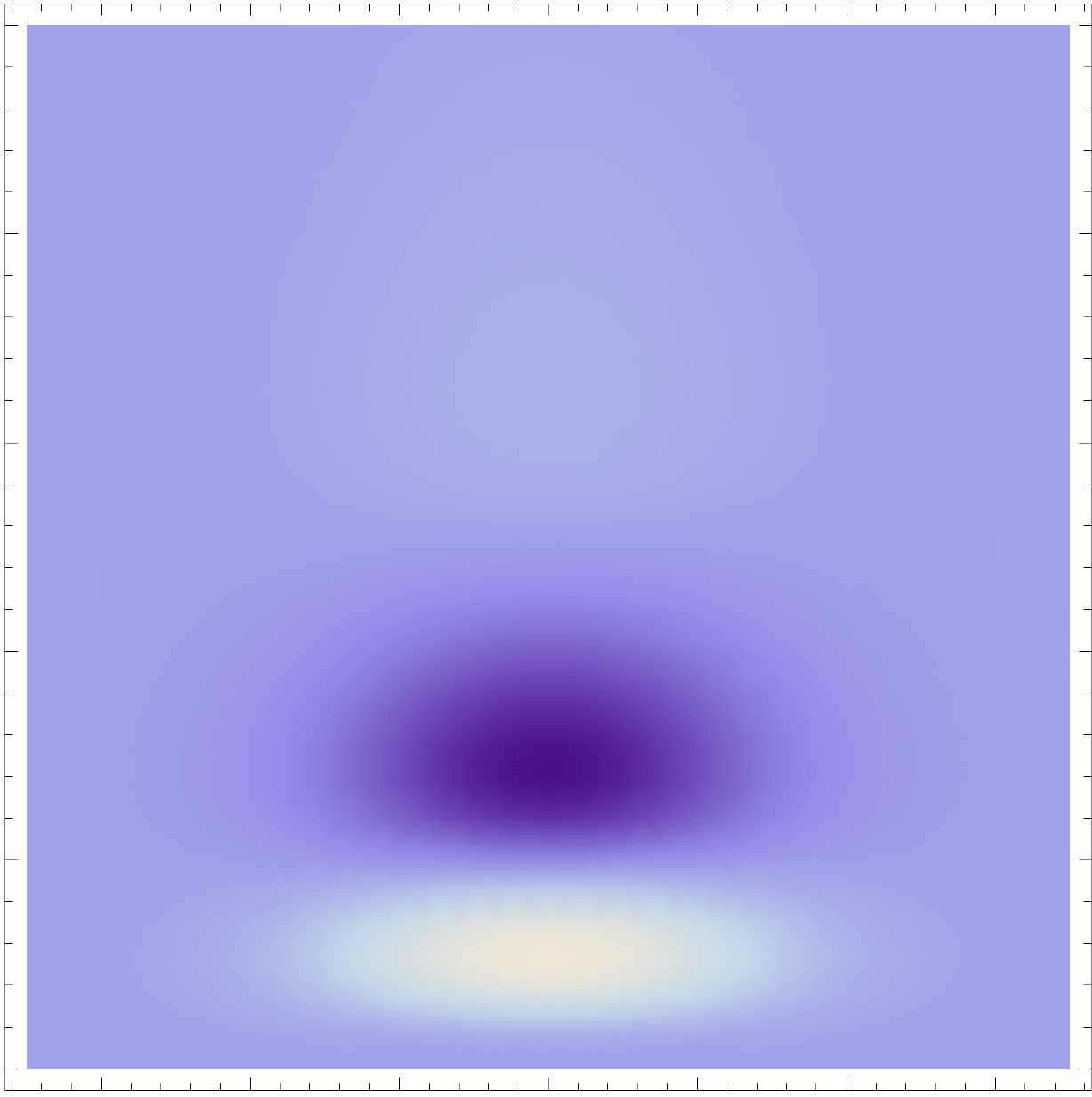} \hspace{-2mm} \\
    \end{tabular} 
  \end{center}
  \caption{{\em Space-time separable kernels\/}
           $T_{x^{m}t^{n}}(x, t;\; s, \tau) = \partial_{x^m t^n} (g(x;\; s) \, h(t;\; \tau))$ 
           up to order two obtained as the composition of Gaussian
           kernels over the spatial domain $x$ and a cascade of
           truncated exponential kernels over the temporal domain $t$
           with a logarithmic distribution of the intermediate
           temporal scale levels that approximates the time-causal
           limit kernel ($s = 1$, $\tau = 1$, $K = 7$, $c = \sqrt{2}$).
           The corresponding family of spatio-temporal receptive
           fields is closed under spatial scaling transformations as
           well as under temporal scaling transformations for temporal
           scaling factors that are integer powers of the distribution
           parameter $c$ of the temporal smoothing kernel.
           (Horizontal axis: space $x$. Vertical axis: time $t$.)}
  \label{fig-non-caus-sep-spat-temp-rec-fields}

  \medskip

  \begin{center}
    \begin{tabular}{c}
     \hspace{-2mm} {\footnotesize $-T(x, t;\; s, \tau, v)$} \hspace{-2mm} \\
      \hspace{-2mm} \includegraphics[width=0.15\textwidth]{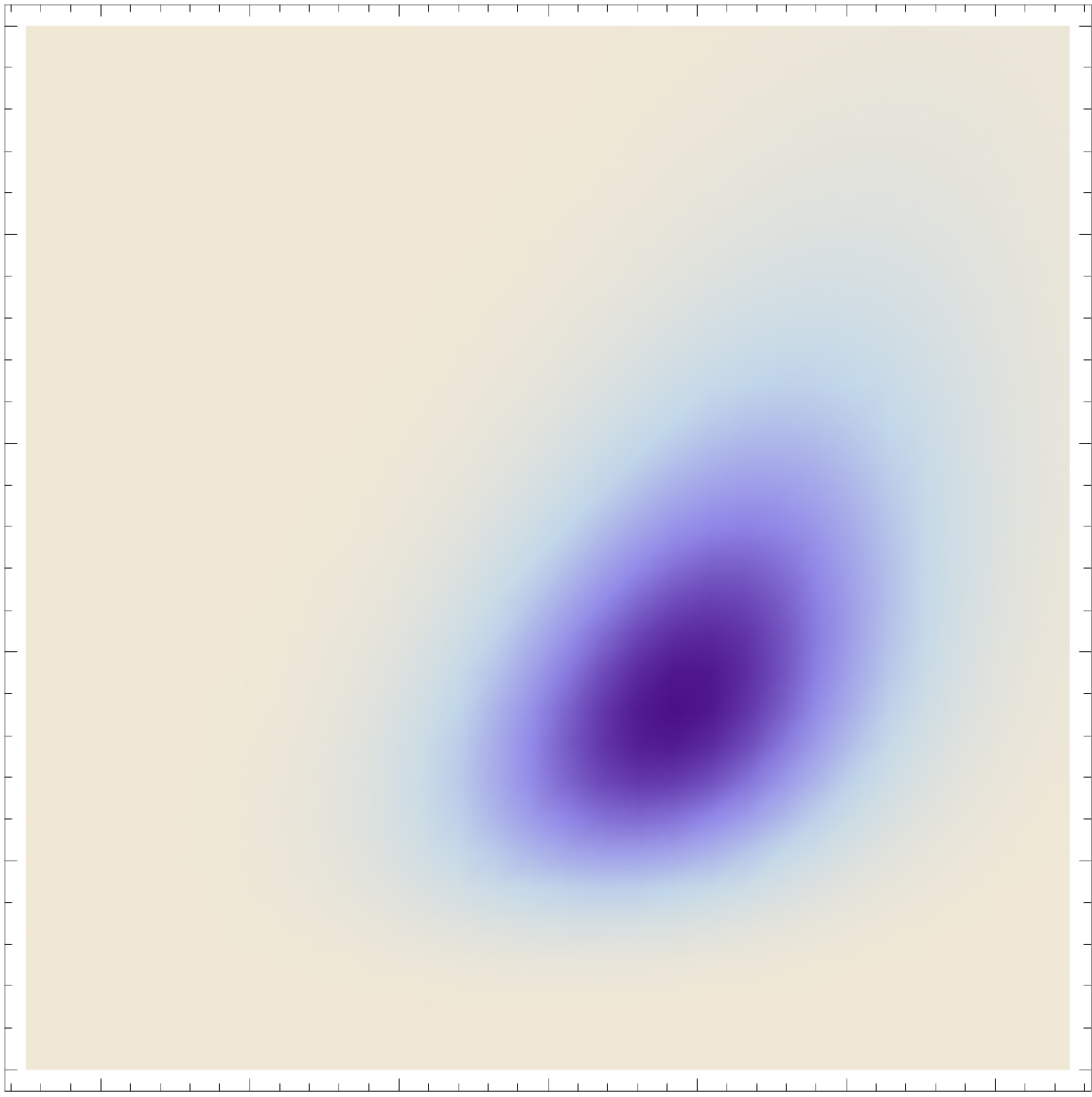} \hspace{-2mm} \\
    \end{tabular} 
  \end{center}
  \vspace{-6mm}
  \begin{center}
    \begin{tabular}{cc}
      \hspace{-2mm} {\footnotesize $T_x(x, t;\; s, \tau, v)$} \hspace{-2mm} 
      & \hspace{-2mm} {\footnotesize $T_t(x, t;\; s, \tau, \delta)$} \hspace{-2mm} \\
      \hspace{-2mm} \includegraphics[width=0.15\textwidth]{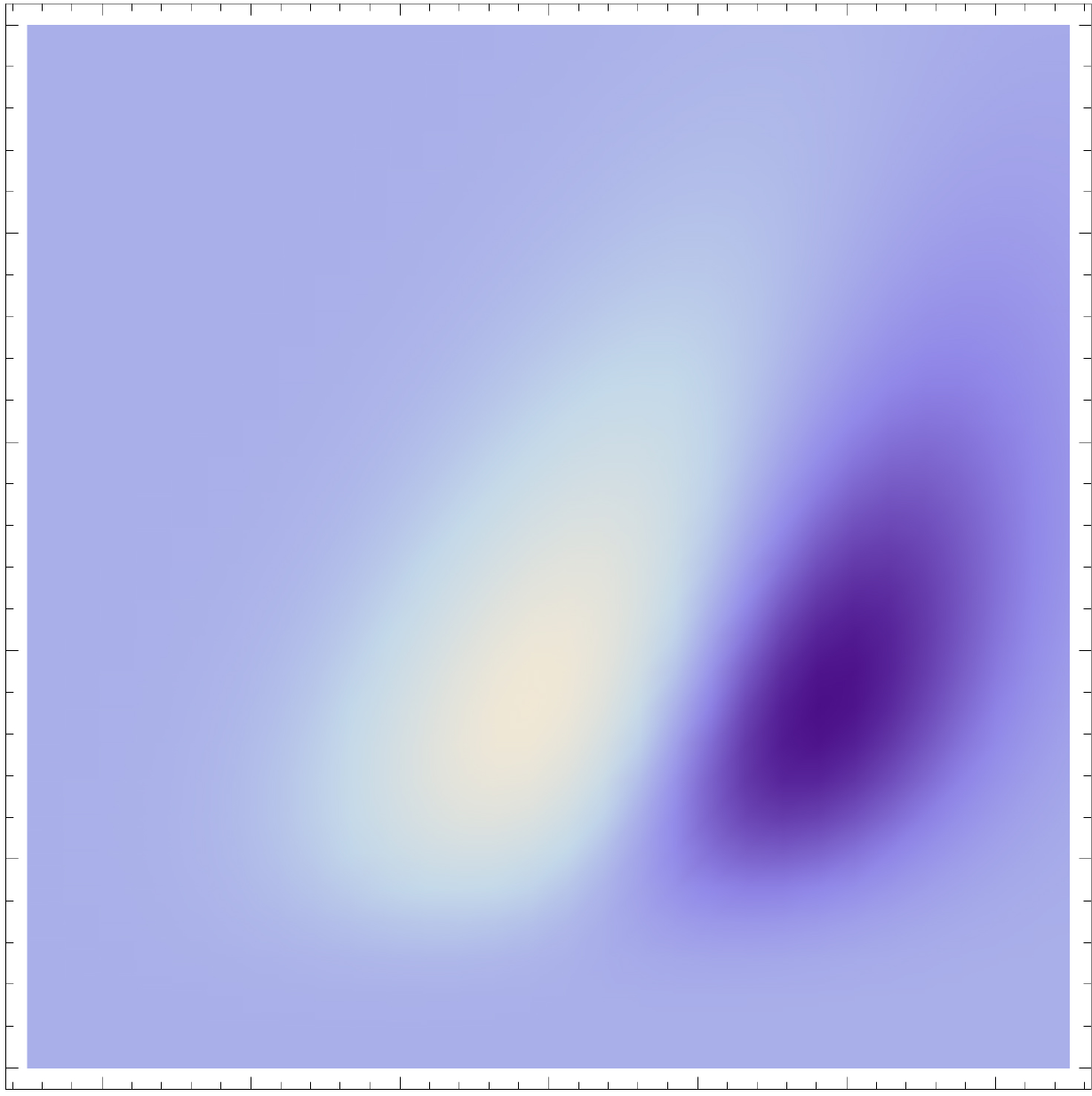} \hspace{-2mm} &
      \hspace{-2mm} \includegraphics[width=0.15\textwidth]{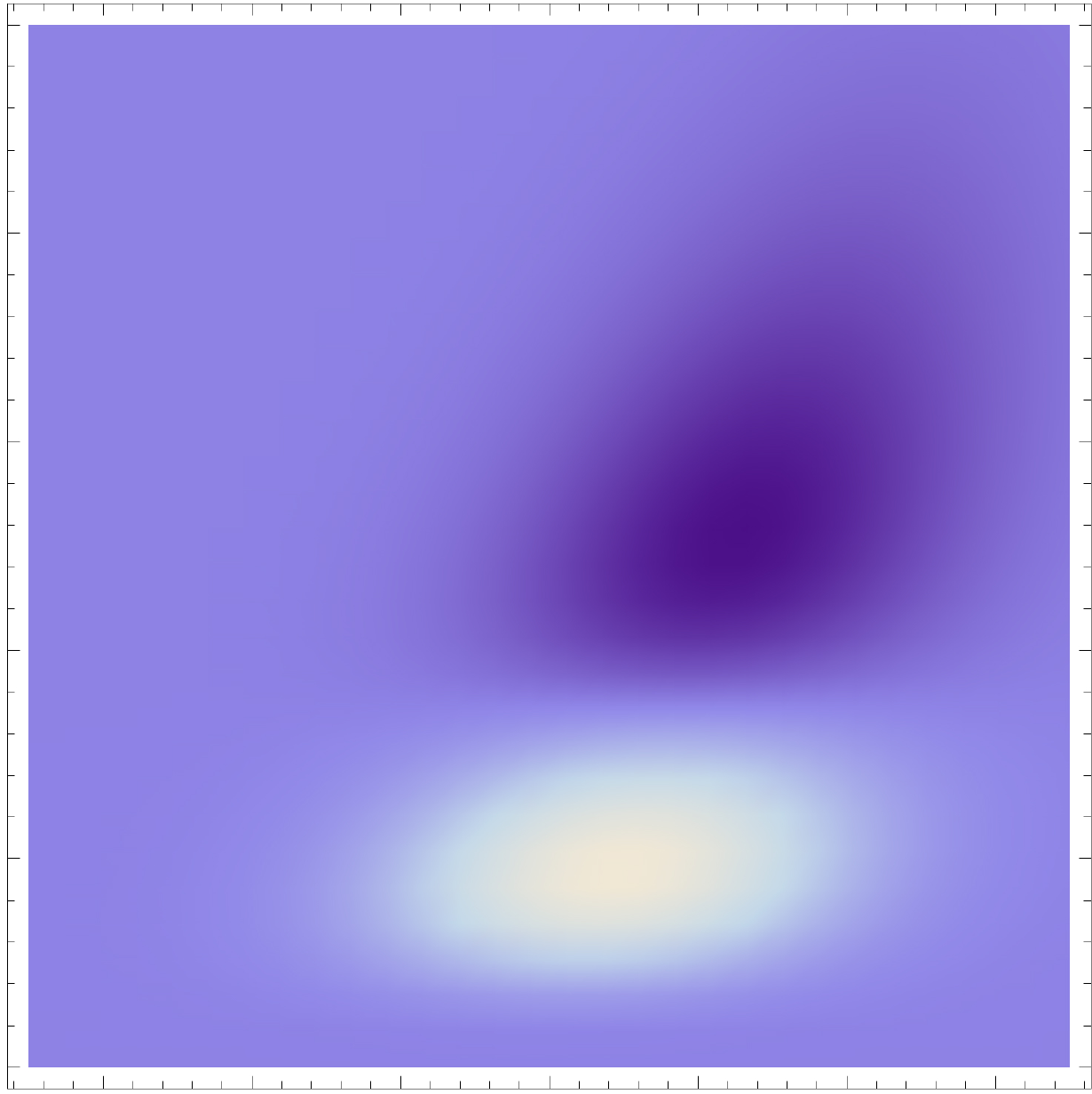} \hspace{-2mm} \\
    \end{tabular} 
  \end{center}
  \vspace{-6mm}
  \begin{center}
    \begin{tabular}{ccc}
      \hspace{-2mm} {\footnotesize $T_{xx}(x, t;\; s, \tau, v)$} \hspace{-2mm} 
      & \hspace{-2mm} {\footnotesize $T_{xt}(x, t;\; s, \tau, v)$} \hspace{-2mm} 
      & \hspace{-2mm} {\footnotesize $T_{tt}(x, t;\; s, \tau, v)$} \hspace{-2mm} \\
      \hspace{-2mm} \includegraphics[width=0.15\textwidth]{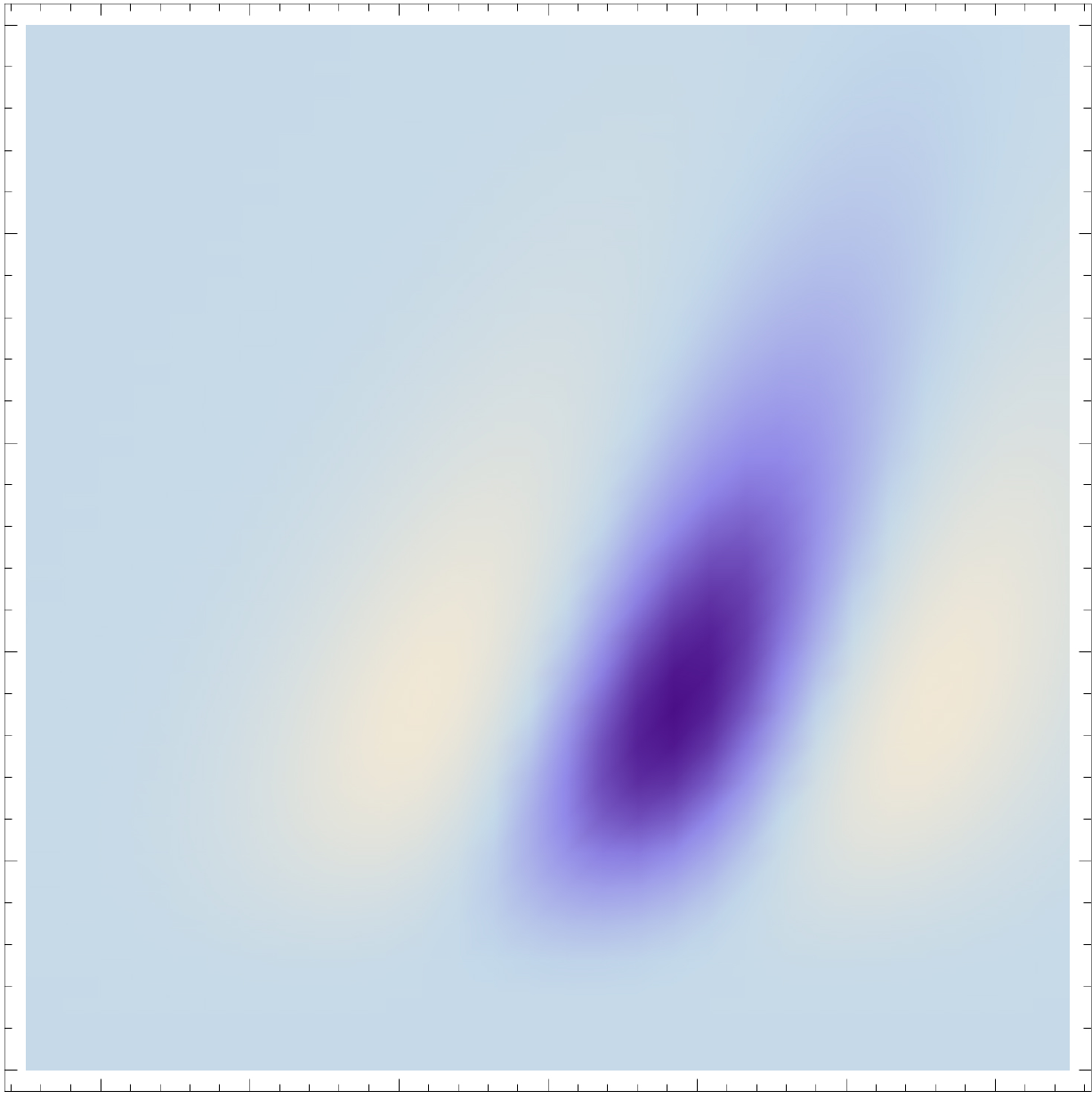} \hspace{-2mm} &
      \hspace{-2mm} \includegraphics[width=0.15\textwidth]{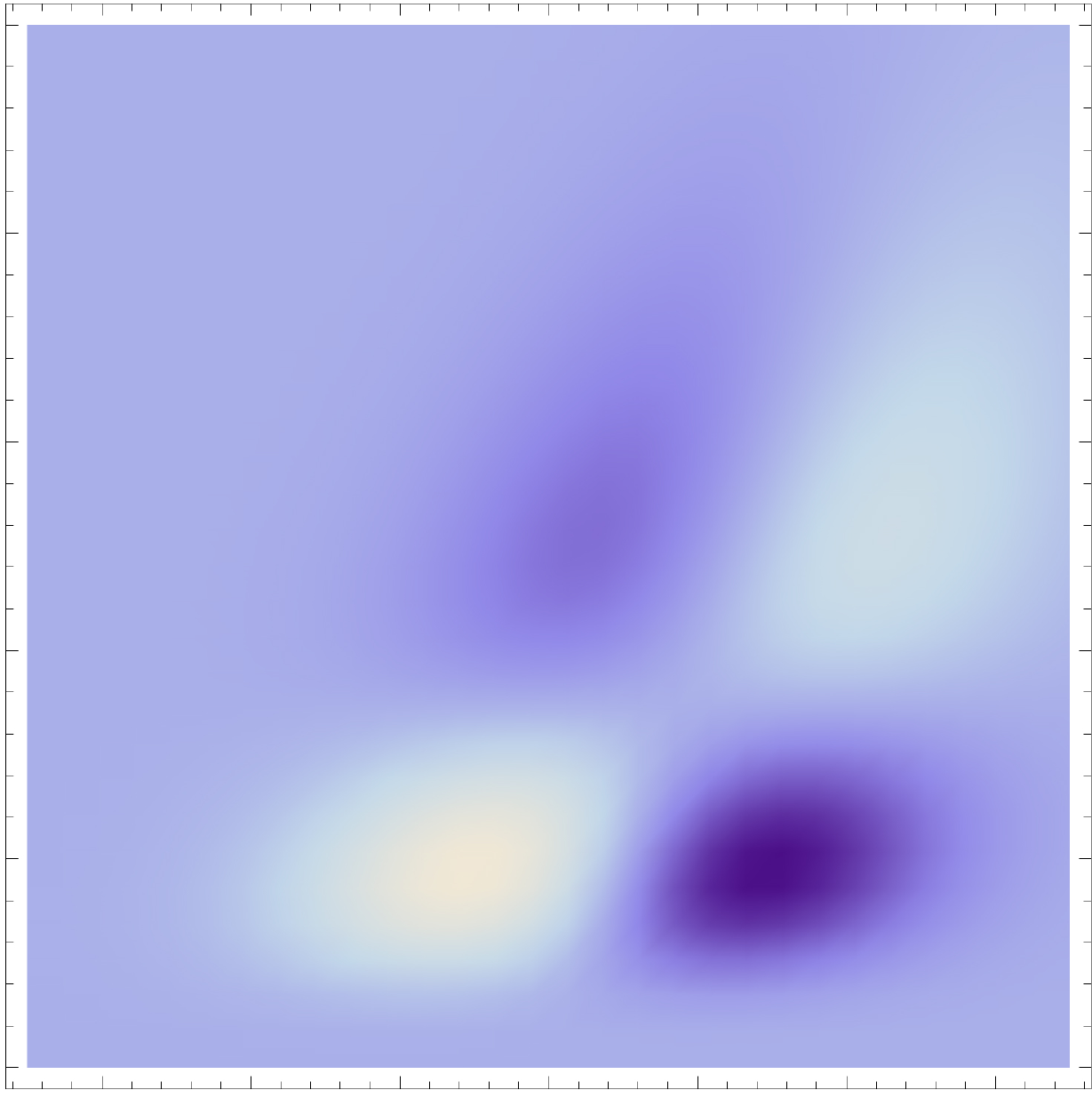} \hspace{-2mm} &
      \hspace{-2mm} \includegraphics[width=0.15\textwidth]{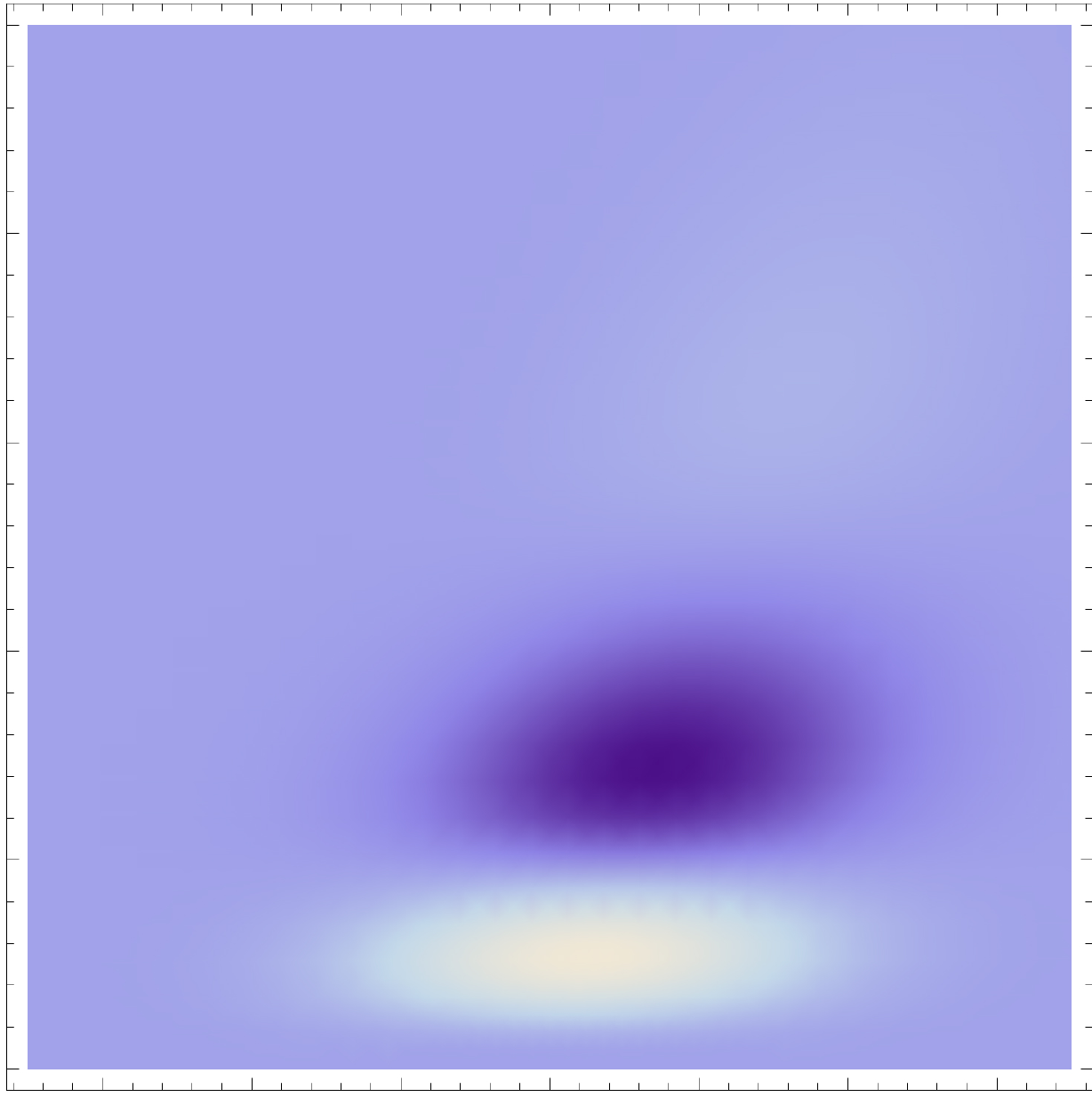} \hspace{-2mm} \\
    \end{tabular} 
  \end{center}
  \caption{{\em Velocity-adapted spatio-temporal kernels\/}
               $T_{x^{m}t^{n}}(x, t;\; s, \tau, v) = \partial_{x^m t^n} (g(x - vt;\; s) \, h(t;\; \tau))$ 
           up to order two obtained as the composition of Gaussian
           kernels over the spatial domain $x$ and a cascade of
           truncated exponential kernels over the temporal domain $t$
           with a logarithmic distribution of the intermediate
           temporal scale levels that approximates the time-causal
           limit kernel ($s = 1$, $\tau = 1$, $K = 7$, $c = \sqrt{2}$, $v = 0.5$).
           In addition to spatial and temporal scaling
           transformations, the corresponding family of receptive
           fields is also closed under Galilean transformations.
           (Horizontal axis: space $x$. Vertical axis: time $t$.)}
  \label{fig-non-caus-vel-adapt-spat-temp-rec-fields}
\end{figure}

Two natural ways of distributing the discrete time constants $\mu_k$
over temporal scales are studied in detail in
\cite{Lin16-JMIV,Lin17-JMIV} corresponding to either a uniform
or a logarithmic distribution in terms of the composed variance
\begin{equation}
  \tau_K = \sum_{k=1}^{K} \mu_k^2.
\end{equation}
Specifically, it is shown in \cite{Lin16-JMIV} that in the case of a
logarithmic distribution of the discrete temporal scale levels, it is
possible to consider an infinite number of temporal scale levels that
cluster infinitely dense near zero temporal scale 
\begin{equation}
  \label{eq-temp-scale-levels-limit-kernel}
  \dots \frac{\tau_0}{c^6}, \frac{\tau_0}{c^4}, \frac{\tau_0}{c^2}, \tau_0,
  c^2 \tau_0, c^4 \tau_0, c^6 \tau_0, \dots 
\end{equation} 
so that a {\em scale-invariant time-causal limit kernel\/} $\Psi(t;\; \tau, c)$ can be defined obeying self-similarity
and scale covariance over temporal scales and with a Fourier transform
of the form
\begin{equation}
  \label{eq-FT-comp-kern-log-distr-limit}
     \hat{\Psi}(\omega;\; \tau, c) 
     = \prod_{k=1}^{\infty} \frac{1}{1 + i \, c^{-k} \sqrt{c^2-1} \sqrt{\tau} \, \omega}.
\end{equation}
Figure~\ref{fig-non-caus-sep-spat-temp-rec-fields} and
Figure~\ref{fig-non-caus-vel-adapt-spat-temp-rec-fields} show
spatio-temporal kernels over a 1+1-dimensional spatio-temporal domain
using approximations of the time-causal limit kernel for temporal smoothing over the
temporal domain and the Gaussian kernel for spatial smoothing over the
spatial domain.
Figure~\ref{fig-non-caus-sep-spat-temp-rec-fields} shows space-time
separable receptive fields corresponding to image velocity $v = 0$,
whereas Figure~\ref{fig-non-caus-vel-adapt-spat-temp-rec-fields} shows
unseparable velocity-adapted receptive fields corresponding to a
non-zero image velocity $v \neq 0$.

The family of space-time separable receptive fields for zero image
velocities is closed under
spatial scaling transformations for arbitrary spatial scaling factors
as well as for temporal scaling transformations with temporal scaling
factors that are integer powers of the distribution parameter $c$ of
the time-causal limit kernel.
The full family of velocity-adapted receptive fields for general
non-zero image velocities is additionally closed under Galilean
transformations, corresponding to variations in the relative motion
between the objects in the world and the observer.
Given that the full families of receptive fields are explicitly
represented in the vision system, this means that it will be possible
to perfectly match receptive field responses computed under the
following types of natural image transformations: 
(i)~objects of different size in the image domain as arising from 
{\em e.g.\/}\ viewing the same object from different distances,
(ii)~spatio-temporal events that occur with different speed, faster or
slower, and
(iii)~objects and spatio-temporal that are viewed with different relative motions between
the objects/event and the visual observer.

If additionally the spatial smoothing is performed over the full
family of spatial covariance matrices $\Sigma$, then receptive field
responses can also be matched (iv)~between different views of the same
smooth local surface patch.

\subsection{Scale normalisation of spatial and spatio-temporal
  receptive fields}

When computing receptive field responses over multiple spatial and
temporal scales, there is an issue about how the receptive field
responses should be normalized so as to enable appropriate comparisons
between receptive field responses at different scales.
Issues of scale normalisation of the derivative based receptive fields
defined from scale-space operations are treated in \cite{Lin97-IJCV,Lin98-IJCV,Lin14-EncCompVis}
regarding spatial receptive fields and in
\cite{Lin16-JMIV,Lin17-JMIV,Lin17-spattempscsel}
regarding spatio-temporal receptive fields.

\paragraph{Scale-normalized spatial receptive fields}

Let $s_{\varphi}$ and $s_{\orth{\varphi}}$ denote the eigenvalues of
the composed affine covariance matrix $s \, \Sigma$ in the spatial receptive field
model (\ref{eq-spat-RF-model}) and let $\partial_{\varphi}$ and
$\partial_{\orth{\varphi}}$ denote directional derivative operators
along the corresponding eigendirections.
Then, the scale-normalized spatial derivative kernel corresponding to
the receptive field model (\ref{eq-spat-RF-model}) is given by
\begin{multline}
  \label{eq-spat-RF-model-der-norm}
   T_{{\varphi}^{m_1} {\orth{\varphi}}^{m_2},norm}(x_1, x_2;\; s, \Sigma) = \\
   s_{\varphi}^{m_1 \gamma_s /2} \, s_{\orth{\varphi}}^{m_2 \gamma_s /2} \,
     \partial_{\varphi}^{m_1} \partial_{\bot \varphi}^{m_2} 
      \left( g(x_1, x_2;\; s \Sigma) \right),
\end{multline}
where $\gamma_s$ denotes the spatial scale normalization parameter of
$\gamma$-normalized derivatives and specifically the choice $\gamma_s
=1$ leads to maximum scale invariance in the sense that the magnitude
response of the spatial receptive field will be covariant under uniform
spatial scaling transformations $(x_1', x_2') = (S_s \, x_1, S_s \, x_2)$,
provided that the spatial scale levels are appropriately matched
$(s_{\varphi}', s_{\orth{\varphi}}') = (S_s^2 \, s_{\varphi}, S_s^2 \, s_{\orth{\varphi}})$.

\paragraph{Scale-normalized spatial receptive fields in the case of a
  non-causal spatio-temporal domain}

For the case of a non-causal spatio-temporal domain, where the temporal
smoothing operation in the spatio-temporal receptive field model is
performed by a non-causal Gaussian temporal kernel
$h(t;\; \tau) = g(t;\; \tau) = 1/\sqrt{2 \pi \tau} \exp(-t^2/2\tau)$,
the scale-normalized spatio-temporal derivative kernel
corresponding to the spatio-temporal receptive field model
(\ref{eq-spat-temp-RF-model-der}) is with corresponding notation
regarding the spatial domain as in (\ref{eq-spat-RF-model-der-norm}) given by
\begin{align}
  \begin{split}
  \label{eq-spat-temp-RF-model-der-norm-non-caus}
    & T_{{\varphi}^{m_1} {\bot \varphi}^{m_2} {\bar t}^n, norm}(x_1, x_2, t;\; s, \tau;\; v, \Sigma)
  \end{split}\nonumber\\
  \begin{split}
    & = s_{\varphi}^{m_1 \gamma_s /2} \, s_{\orth{\varphi}}^{m_2 \gamma_s /2} \, 
          \tau^{n \gamma_{\tau} /2} \, 
  \end{split}\nonumber\\
  \begin{split}
     & \phantom{=} \,\,\,\,
         \partial_{\varphi}^{m_1} \partial_{\bot \varphi}^{m_2} \partial_{\bar t}^n 
          \left( g(x_1 - v_1 t, x_2 - v_2 t;\; s, \Sigma) \, h(t;\; \tau) \right),
  \end{split}
\end{align}
where $\gamma_s$ and $\gamma_{\tau}$ denote the spatial and temporal scale normalization parameters of
$\gamma$-normalized derivatives and specifically the choice 
$\gamma_s =1$ and $\gamma_{\tau} = 1$ leads to maximum scale invariance in the sense that the magnitude
response of the spatio-temporal receptive field will be invariant under
independent scaling transformations of the spatial and the temporal domains
$(x_1', x_2', t') = (S_s \, x_1, S_s \, x_2, S_{\tau} \, t)$,
provided that both the spatial and temporal scale levels are appropriately matched
$(s_{\varphi}', s_{\orth{\varphi}}', \tau') 
 = (S_s^2 \, s_{\varphi}, S_s^2 \, s_{\orth{\varphi}}, S_{\tau}^2 \, \tau)$.

\paragraph{Scale-normalized spatial receptive fields in the case of a
  time-causal spatio-temporal domain}

For the case of a time-causal spatio-temporal domain, where the temporal
smoothing operation in the spatio-temporal receptive field model is
performed by truncated exponential kernels coupled in cascade
$h(t;\; \tau) = h_{composed}(\cdot;\; \mu)$ (\ref{eq-comp-trunc-exp-cascade}),
the corresponding scale-normalized spatio-temporal derivative kernel
corresponding to the spatio-temporal receptive field model
(\ref{eq-spat-temp-RF-model-der}) is given by
\begin{align}
  \begin{split}
  \label{eq-spat-temp-RF-model-der-norm-time-caus}
    & T_{{\varphi}^{m_1} {\bot \varphi}^{m_2} {\bar t}^n, norm}(x_1, x_2, t;\; s, \tau;\; v, \Sigma)
  \end{split}\nonumber\\
  \begin{split}
    & = s_{\varphi}^{m_1 \gamma_s /2} \, s_{\orth{\varphi}}^{m_2 \gamma_s /2} \, 
          \alpha_{n,\gamma_{\tau}}(\tau) \, 
  \end{split}\nonumber\\
  \begin{split}
     & \phantom{=} \,\,\,\,
         \partial_{\varphi}^{m_1} \partial_{\bot \varphi}^{m_2} \partial_{\bar t}^n 
          \left( g(x_1 - v_1 t, x_2 - v_2 t;\; s, \Sigma) \, h(t;\; \tau) \right),
  \end{split}
\end{align}
where $\gamma_s$ and $\gamma_{\tau}$ denote the spatial and and
temporal scale normalization parameters of
$\gamma$-normalized derivatives and $\alpha_{n,\gamma_{\tau}}(\tau)$
is the temporal scale normalization factor, which for the case of
variance-based normalization is given by 
\begin{equation}
   \alpha_{n,\gamma_{\tau}}(\tau) = \tau^{n \gamma_{\tau} /2}
\end{equation}
in agreement with (\ref{eq-spat-temp-RF-model-der-norm-non-caus})
while for the case of $L_p$-normalization it is given by \cite[Equation~(76)]{Lin16-JMIV}
\begin{equation}
   \alpha_{n,\gamma_{\tau}}(\tau) = \frac{G_{n,\gamma_{\tau}}}{\| h_{t^n}(\cdot;\; \tau) \|_p },
\end{equation}
with $G_{n,\gamma_{\tau}}$ denoting the $L_p$-norm of the $n$th order
scale-normalized derivative of a non-causal Gaussian temporal kernel
with scale normalization parameter $\gamma_{\tau}$.
In the specific case when the temporal smoothing is performed using
the scale-invariant limit kernel 
(\ref{eq-FT-comp-kern-log-distr-limit}),
the magnitude response will for the maximally scale invariant choice of scale normalization 
parameters $\gamma_s =1$ and $\gamma_{\tau} = 1$ be invariant under
independent scaling transformations of the spatial and the temporal domains
$(x_1', x_2', t') = (S_s \, x_1, S_s \, x_2, S_{\tau} \, t)$ for
general spatial scaling factors $S_s$ and for temporal
scaling factors $S_{\tau} = c^j$ that are integer powers of the
distribution parameter $c$ of the scale-invariant limit kernel,
provided that both the spatial and temporal scale levels are appropriately matched
$(s_{\varphi}', s_{\orth{\varphi}}', \tau') 
  = (S_s^2 \, s_{\varphi}, S_s^2 \, s_{\orth{\varphi}}, S_{\tau}^2 \, \tau)$.

\subsection{Invariance to local multiplicative illumination
  variations or variations in exposure parameters}
\label{sec-intens-var}

The treatment so far has been concerned with modelling receptive
fields under natural geometric image transformations, modelled as
local scaling transformations, local affine transformations and 
local Galilean transformations representing
the essential dimensions in the variability of a local linearization
of the perspective mapping from a local surface patch in the world to the
tangent plane of the retina.
A complementary issue concerns how to model receptive field responses
under variations in the external illumination and under variations in
the internal exposure mechanisms of the eye that adapt the diameter of
the pupil and the sensitivity of the photoreceptors to the external
illumination.
In this section, we will present a solution for this problem
regarding the subset of intensity transformations that can be modelled as
local multiplicative intensity transformations.

To obtain theoretically well-founded handling of image data under
illumination variations, it is natural to represent the image data on
a logarithmic luminosity scale
\begin{equation}
  f(x_1, x_2, t) \sim \log I(x_1, x_2, t).
\end{equation}
Specifically, receptive field responses that are computed from such a
logarithmic parameterization of the image luminosities can be {\em interpreted physically\/} as a
superposition of relative variations of surface structure and
illumination variations.
Let us assume: (i)~a perspective camera model extended with 
(ii)~a thin circular lens for gathering incoming light from different
directions and 
(iii)~a Lambertian illumination model extended with 
(iv)~a spatially varying albedo factor for modelling the light that is
reflected from surface patterns in the world.
Then, it can be shown \cite[Section~2.3]{Lin13-BICY} that a spatio-temporal receptive field response
\begin{equation}
  L_{{\varphi}^{m_1} {\bot \varphi}^{m_2} {\bar t}^n}(\cdot, \cdot;\; s, \tau) 
  = \partial_{{\varphi}^{m_1} {\bot \varphi}^{m_2} {\bar t}^n} \, {\cal T}_{s,\tau} \, f (\cdot, \cdot)
\end{equation}
of the image data $f$, where ${\cal T}_{s,\tau}$ represents the spatio-temporal
smoothing operator (here corresponding to a spatio-temporal smoothing
kernel of the form (\ref{eq-spat-temp-RF-model})) can be expressed as
\begin{align}
  \begin{split}
    & L_{{\varphi}^{m_1} {\bot \varphi}^{m_2} {\bar t}^n}(x_1, x_2, t;\;
    s, \tau) = \\
 \end{split}\nonumber\\
  \begin{split}
     & = \partial_{{\varphi}^{m_1} {\bot \varphi}^{m_2} {\bar t}^n} \, {\cal T}_{s,\tau} \,
        \left(
            \log \rho(x_1, x_2, t) + \log i(x_1, x_2, t)  \vphantom{\log C_{cam}(\tilde{f}(t))}
        \right.
  \end{split}\nonumber\\
  \begin{split}
     \label{eq-illum-model-spat-rec-field}
     & \phantom{= \partial_{{\varphi}^{m_1} {\bot \varphi}^{m_2} {\bar t}^n} \, {\cal T}_{s,\tau} \,  \left( \right.}
        \left. 
          + \log C_{cam}(\tilde{f}(t)) 
          + V(x_1, x_2)
      \right),
  \end{split}
\end{align}
where 
\begin{itemize}
\item[(i)]
  $\rho(x_1, x_2, t)$ is a spatially dependent {\em albedo factor\/} that reflects
  {\em properties of surfaces of objects\/} in the environment with the implicit
  understanding that this entity may in general refer to points on different
  surfaces in the world depending on the viewing direction 
  and thus the (possibly time-dependent) image position $(x(t), y(t))$,
\item[(ii)]
  $i(x_1, x_2, t)$ denotes a spatially dependent {\em illumination field\/} 
  with the implicit understanding that the 
  amount of incoming light on different surfaces may be different for
  different points in the world as mapped to corresponding image
  coordinates $(x_1, x_2)$ over time $t$,
\item[(iii)]
  $C_{cam}(\tilde{f}(t)) = \frac{\pi}{4} \frac{d}{f}$ represents the
  possibly time-dependent {\em internal camera parameters\/}
  with the ratio $\tilde{f} = f/d$ referred to as the {\em effective
    $f$-number\/}, where $d$ denotes the diameter of the lens and $f$ the
  focal distance, and
\item[(iv)]
  $V(x_1, x_2) = - 2 \log (1 + x_1^2  + x_2^2)$ represents a
  geometric {\em natural vignetting\/} effect corresponding to the factor
  $\log \cos^4(\phi)$ for a planar image plane, with $\phi$ denoting
  the angle between the viewing direction $(x_1, x_2, f)$ and the
  surface normal $(0, 0, 1)$ of the image plane. This vignetting term disappears
  for a spherical camera model.
\end{itemize}
From the structure of Equation~(\ref{eq-illum-model-spat-rec-field})
we can note that for any non-zero order of spatial differentiation 
with at least either $m_1 > 0$ or $m_2 > 0$, the influence of the internal camera parameters in 
$C_{cam}(\tilde{f}(t))$ will disappear because of the spatial
differentiation with respect to $x_1$ or $x_2$, and so will the effects of any
other multiplicative exposure control mechanism.
Furthermore, for any multiplicative illumination variation
$i'(x_1, x_2) = C \, i(x_1, x_2)$, where $C$ is a scalar constant,
the logarithmic luminosity will be transformed as
$\log i'(x_1, x_2) = \log C + \log i(x_1, x_2)$, which implies that
the dependency on $C$ will disappear
after spatial or temporal differentiation.

Thus, given that the image measurements are performed on a logarithmic
brightness scale, the spatio-temporal receptive field responses will
be automatically invariant under local multiplicative illumination
variations as well as under local multiplicative variations in the
exposure parameters of the retina and the eye.

\section{Computational modelling of biological receptive fields}
\label{eq-rel-biol-vision}

In two comprehensive reviews, 
DeAngelis {\em et al.\/}\ \cite{DeAngOhzFre95-TINS,deAngAnz04-VisNeuroSci} present overviews of 
spatial and temporal response properties of (classical) receptive fields 
in the central visual pathways.
Specifically, the authors point out the limitations of defining
receptive fields in the spatial domain only
and emphasize the need to characterize receptive fields in the
{\em joint\/} space-time domain,
to describe how a neuron processes the visual image.
Conway and Livingstone \cite{ConLiv06-JNeurSci} and 
Johnson {\em et al.\/}\ \cite{JohHawSha08-JNeuroSci}
 show results of corresponding
investigations concerning spatio-chromatic receptive fields.

In the following, we will describe how the above derived idealized
functional models of linear receptive fields can be used for modelling the 
spatial, spatio-chromatic and spatio-temporal response properties of biological
receptive fields.
Indeed, it will be shown that the derived idealized functional models 
lead to predictions of receptive field profiles that are
qualitatively very similar to {\em all\/} the receptive field types
presented in (DeAngelis {\em et al.\/}\ \cite{DeAngOhzFre95-TINS,deAngAnz04-VisNeuroSci}) and schematic simplifications
of most of the receptive
fields shown in (Conway and Livingstone \cite{ConLiv06-JNeurSci}) and
(Johnson {\em et al.\/}\ \cite{JohHawSha08-JNeuroSci}).

\begin{figure}[hbtp]
   \begin{center}
     \begin{tabular}{c}
        \hspace{-1mm} \includegraphics[width=0.48\textwidth]{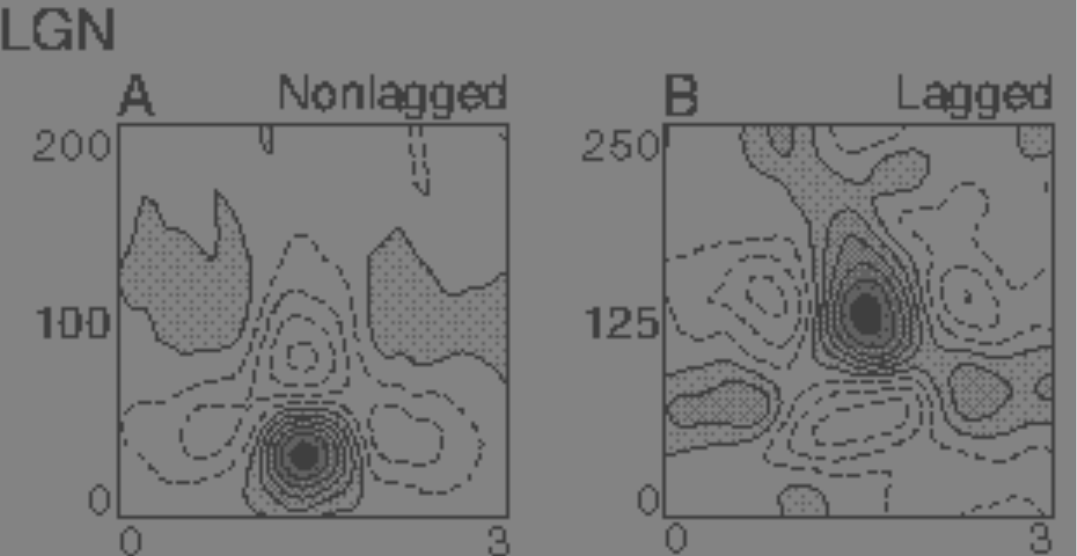}
    \end{tabular}
  \end{center}
  \begin{center}
    \begin{tabular}{cc}
      \hspace{2mm} {\footnotesize $h_{xxt}(x, t;\; s, \tau)$}
      & \hspace{-2mm} {\footnotesize $-h_{xxtt}(x, t;\; s, \tau)$} \\
\hspace{2mm} \includegraphics[width=0.22\textwidth]{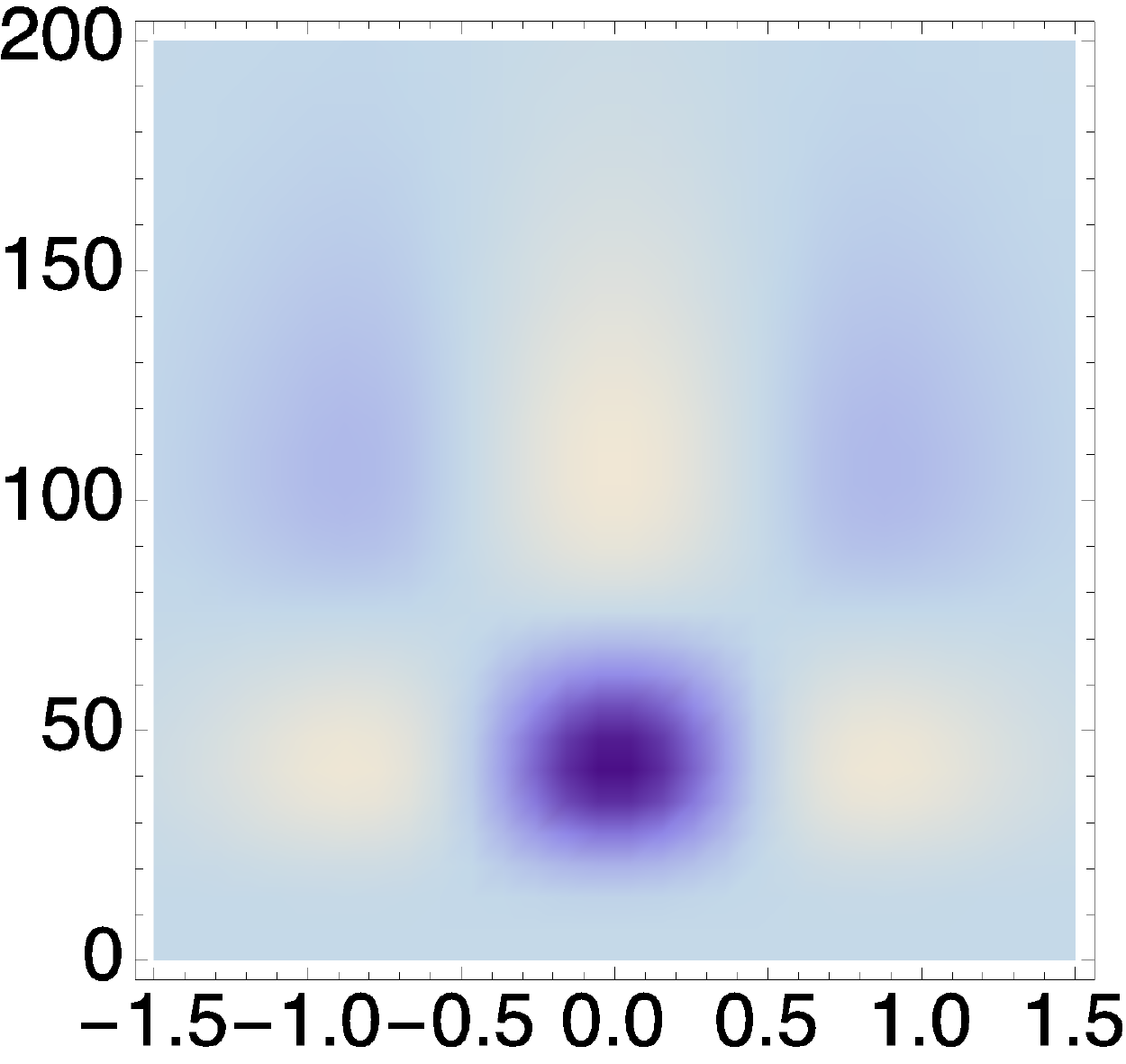}
      &
\hspace{-2mm}  \includegraphics[width=0.22\textwidth]{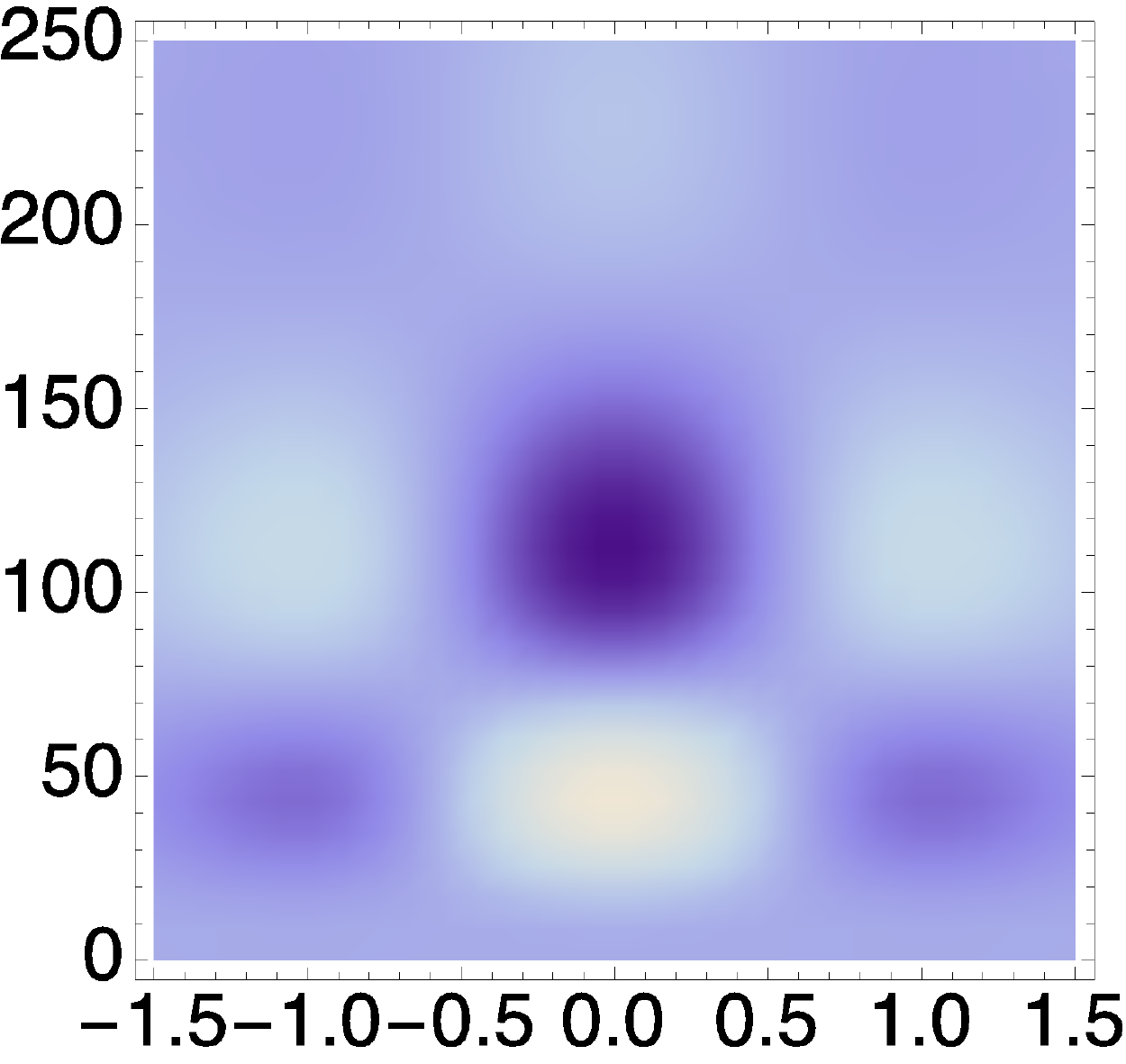} 
    \end{tabular}
  \end{center}
\caption{Computational modelling of space-time separable receptive
  field profiles in the lateral geniculate nucleus (LGN)
           as reported by DeAngelis {\em et al.\/}\ \protect\cite{DeAngOhzFre95-TINS} using
    idealized spatio-temporal receptive fields of the form
     $T(x, t;\; s, \tau) = \partial_{x^m} \partial_{t^n} (g(;\; s) \, h(t;\; \tau))$
   according to Equation~(\protect\ref{eq-spat-temp-RF-model-der}) with
    the temporal smoothing function $h(t;\; \tau)$
    modelled as a cascade of first-order integrators/truncated exponential
    kernels of the form (\protect\ref{eq-comp-trunc-exp-cascade}):
    (left) a ``non-lagged cell'' modelled using first-order temporal derivatives,
    (right) a ``lagged cell'' modelled using second-order temporal
    derivatives.
    Parameter values with $\sigma_x = \sqrt{s}$ and $\sigma_t = \sqrt{\tau}$: 
    (a) $h_{xxt}$: $\sigma_x = 0.5$~degrees, $\sigma_t = 40$~ms.
    (b) $h_{xxtt}$: $\sigma_x = 0.6$~degrees, $\sigma_t = 60$~ms.
           (Horizontal dimension: space $x$. Vertical dimension: time~$t$.)}
  \label{fig-deang-tins-temp-resp-prof-lagged-nonlagged}

  \bigskip

  \begin{center}
     \begin{tabular}{cc}
        & {\small $\nabla^2 g(x, y;\; s)$} \\
       \hspace{-4mm} \includegraphics[height=0.10\textheight]{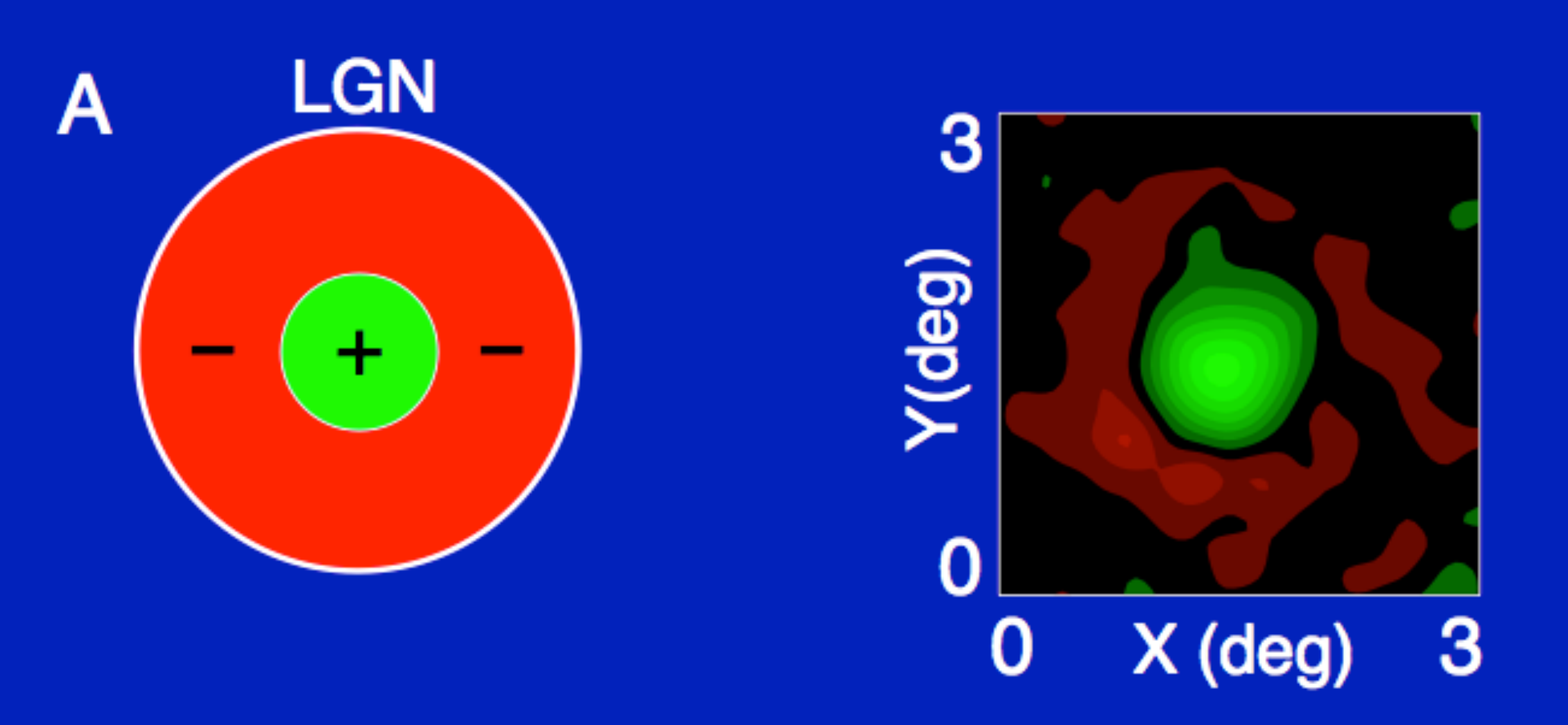}
       & \includegraphics[height=0.10\textheight]{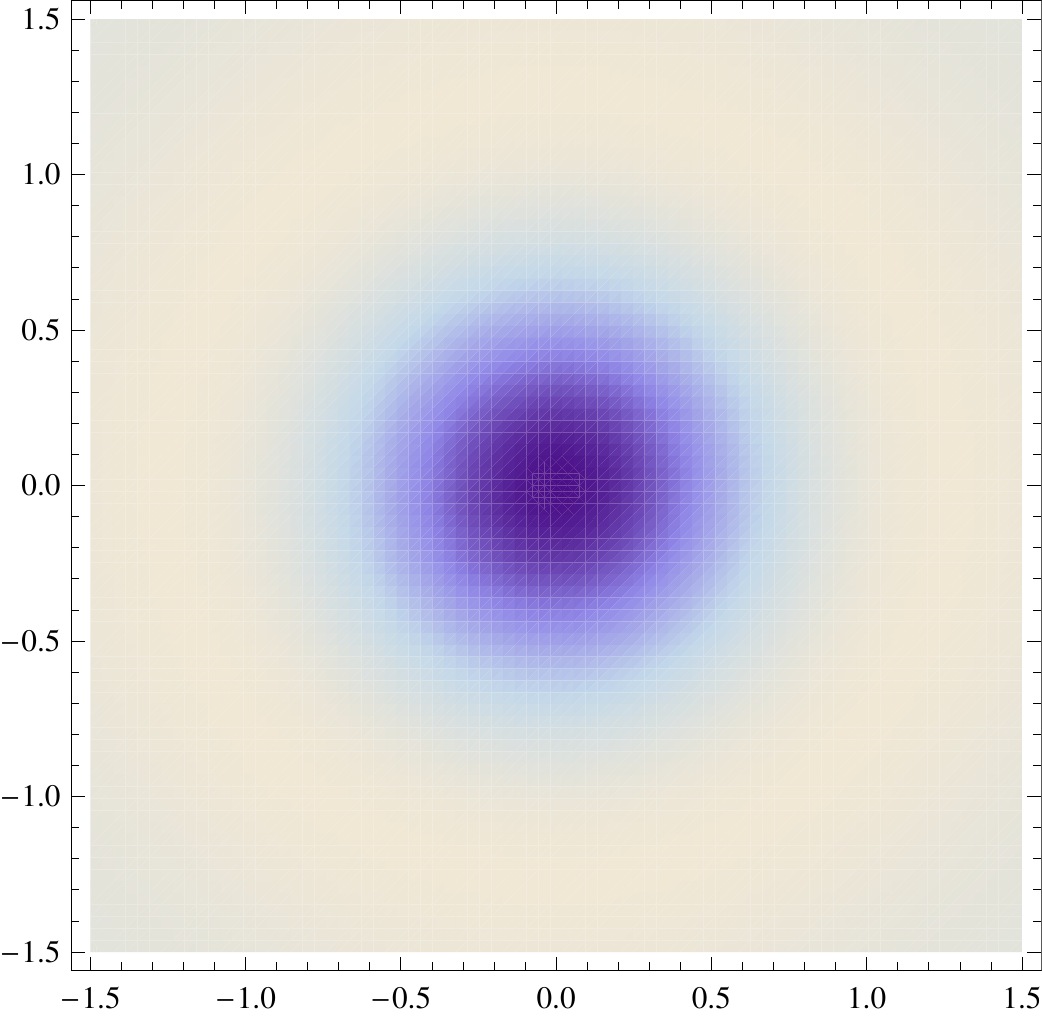}
     \end{tabular}
   \end{center}

  \caption{Computational modelling the spatial component of receptive fields in the
    LGN using the Laplacian of the Gaussian:
                (left) Receptive fields in the LGN have approximately
               circular center-surround responses in the spatial
               domain, as reported by DeAngelis {\em et al.\/}\ \protect\cite{DeAngOhzFre95-TINS}.
               (right)
               In terms of Gaussian derivatives, this spatial response 
               profile can be modelled by the Laplacian of the
               Gaussian $\nabla^2 g(x, y;\; s) 
            = (x^2 + y^2 - 2s)/(2 \pi s^3) \exp(-(x^2+y^2)/2s)$,
               here with $\sigma_s = \sqrt{s} = 0.6$ in units of degrees of visual
               angle.}
  \label{fig-deang-tins-LGN-spat}
\end{figure}

\begin{figure}[!htp]

  \begin{center}
    \begin{tabular}{c}
      \includegraphics[scale=0.45]{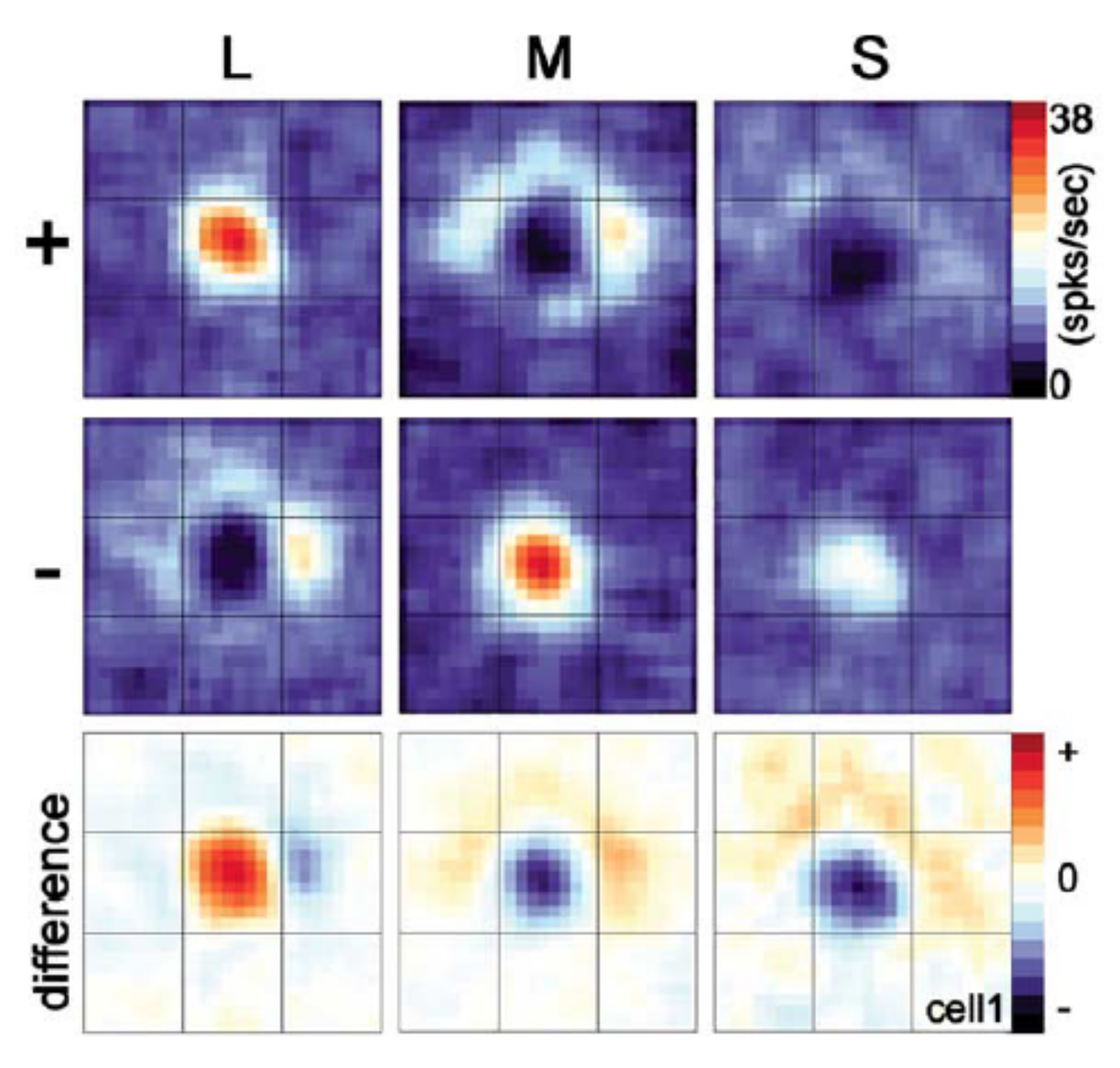} \\
    \end{tabular} 
  \end{center}
 \vspace{-3mm}
  \caption{Spatio-chromatic receptive field response of a
    {\em double-opponent neuron \/}as reported by Conway and Livingstone
    \protect\cite[Figure~2, Page~10831]{ConLiv06-JNeurSci},
    with the colour channels L, M and S
    essentially corresponding to red, green and blue, respectively. 
    (From these L, M and S colour channels, corresponding red/green and 
    yellow/blue colour-opponent channels can be formed from the
    differences between L to M and between L+M to S.)}
  \label{fig-col-opp-neuron}

\bigskip

  \begin{center}
    \begin{tabular}{cc}
      \includegraphics[scale=0.30]{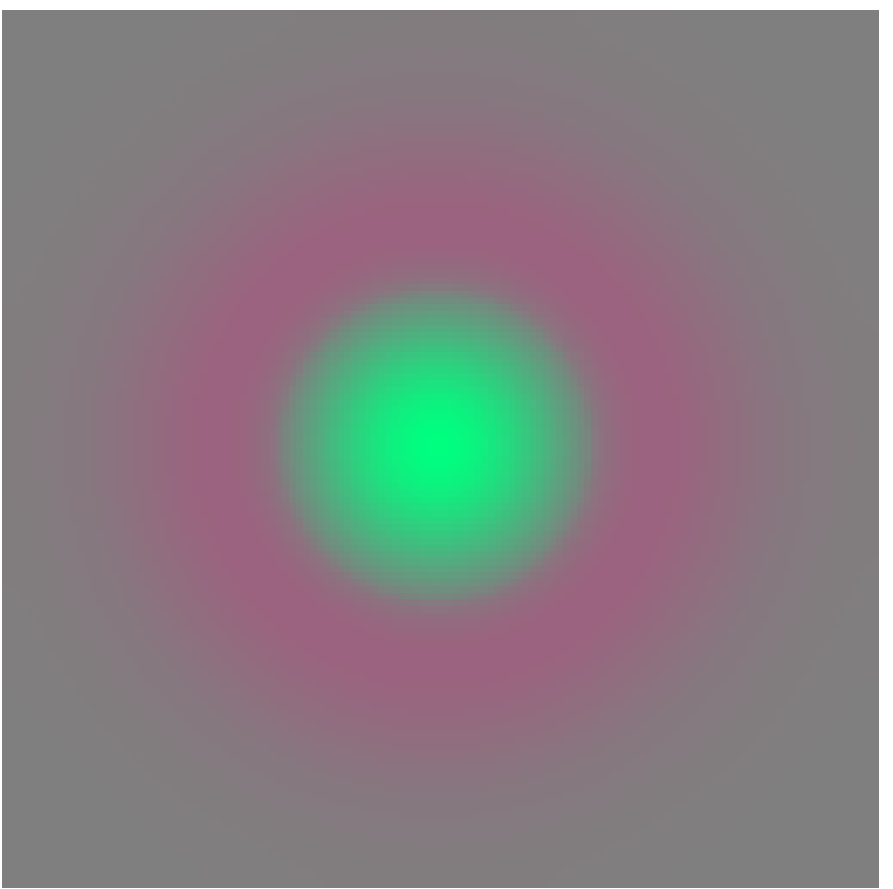} &
      \includegraphics[scale=0.30]{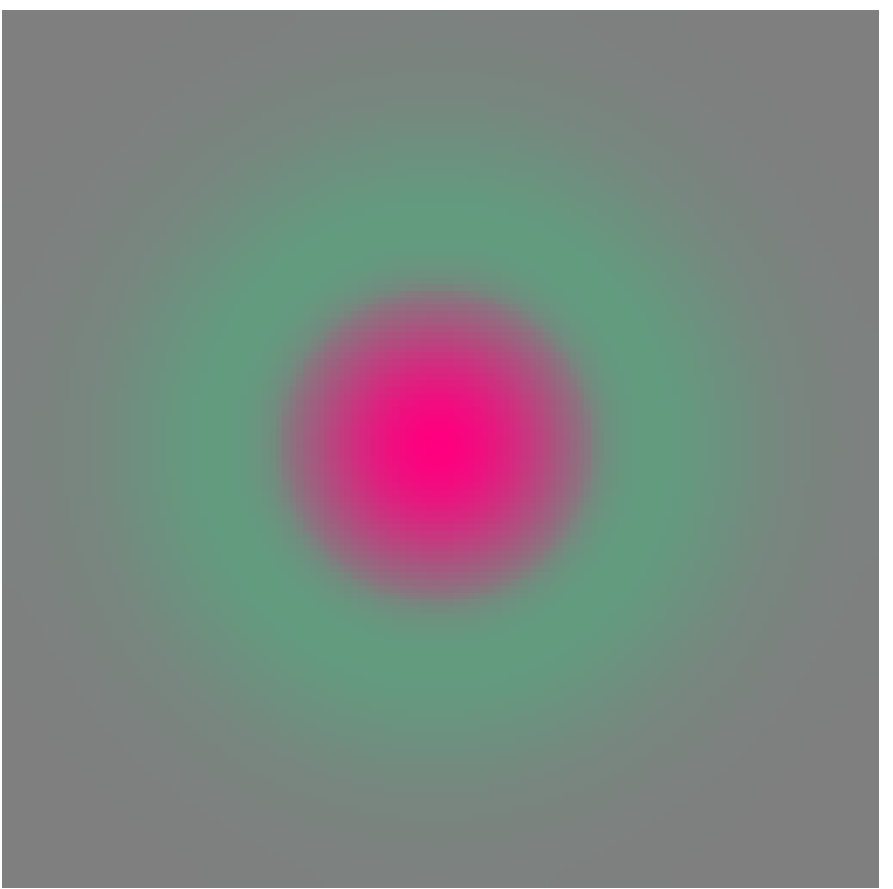}  \\
    \end{tabular} 
  \end{center}
  \vspace{-2mm}
  \begin{center}
    \begin{tabular}{cc}
      \includegraphics[scale=0.30]{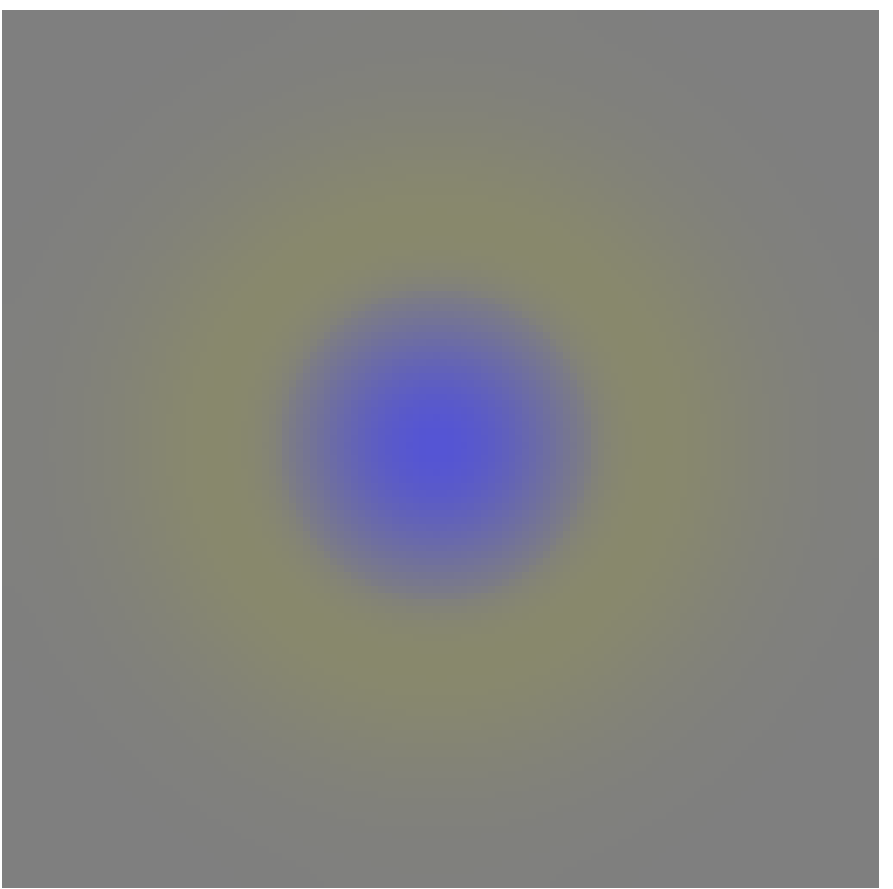}  &
      \includegraphics[scale=0.30]{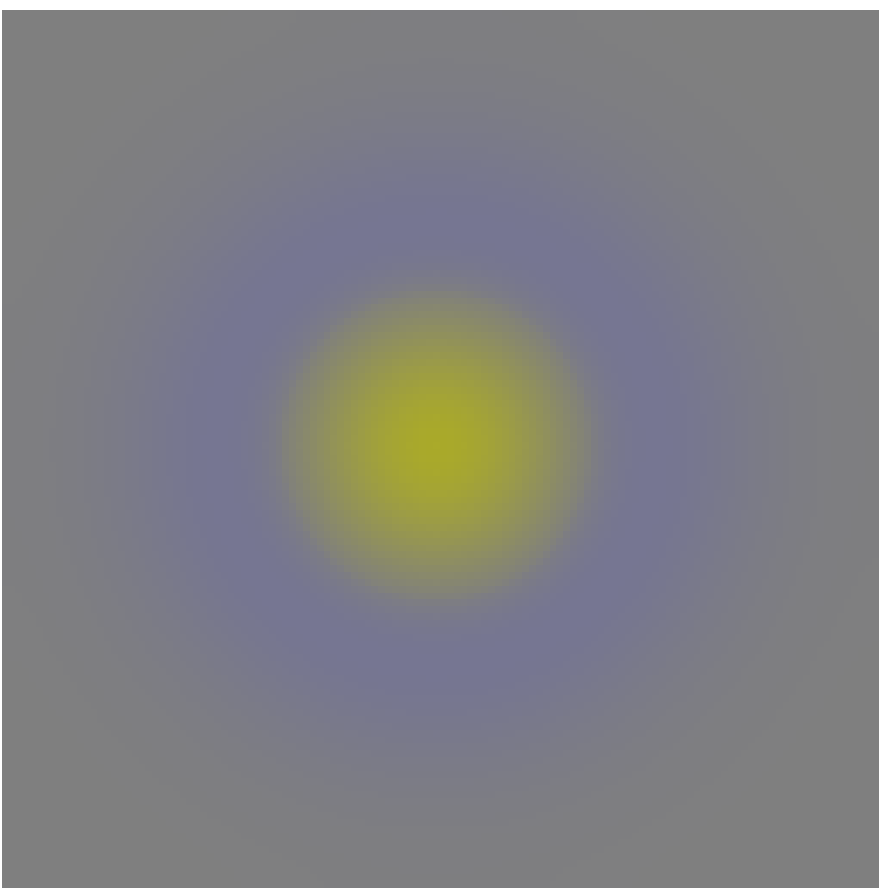} \\
    \end{tabular} 
  \end{center}
 \vspace{-3mm}
  \caption{Idealized spatio-chromatic receptive fields over
    the spatial domain corresponding to the
    application of the Laplacian operator to positive and negative
    red/green and yellow/blue colour opponent channels, respectively.
    These receptive fields can be seen as idealized models of the
    spatial component of double-opponent spatio-chromatic receptive
    fields in the LGN.}
  \label{fig-col-gauss-Laplace-ders}
\end{figure}

\subsection{Spatial and spatio-temporal receptive fields in the LGN}

Regarding visual receptive fields in the lateral geniculate nucleus (LGN), 
DeAngelis {\em et al.\/}\ \cite{DeAngOhzFre95-TINS,deAngAnz04-VisNeuroSci}
report that most neurons
(i)~have approximately circular center-surround organization in the
 spatial domain and that 
(ii)~most of the receptive fields are separable in space-time.
There are two main classes of temporal responses for such cells:
(i)~a ``non-lagged cell'' is defined as a cell for which the first temporal lobe is the largest one
  (Figure~\ref{fig-deang-tins-temp-resp-prof-lagged-nonlagged}(left)), whereas
(ii)~a ``lagged cell'' is defined as a cell for which the second lobe dominates
(Figure~\ref{fig-deang-tins-temp-resp-prof-lagged-nonlagged}(right)).

When using a time-causal temporal smoothing kernel, the first peak of
a first-order temporal derivative will be strongest, whereas the second
peak of a second-order temporal derivative will be strongest
(see \cite[Figure~2]{Lin16-JMIV}).
Thus, according to this theory, non-lagged LGN cells can be seen as
corresponding to first-order time-causal temporal derivatives, whereas lagged LGN
cells can be seen as corresponding to second-order time-causal temporal derivatives.

The spatial response, on the other hand, shows a high similarity
to a Laplacian of a Gaussian, leading to an idealized receptive field
model of the form \cite[Equation~(108)]{Lin13-BICY}
\begin{equation}
  \label{eq-lgn-model-1}
  h_{LGN}(x, y, t;\; s, \tau) 
  = \pm (\partial_{xx} + \partial_{yy}) \, g(x, y;\; s) \, \partial_{t^n} \, h(t;\; \tau).
\end{equation}
Figure~\ref{fig-deang-tins-LGN-spat} shows a comparison between 
the spatial component of a receptive field in the LGN with a Laplacian of the Gaussian.
This model can also be used for modelling spatial on-center/off-surround and
off-center/on-surround receptive fields in the retina.
Figure~\ref{fig-deang-tins-temp-resp-prof-lagged-nonlagged} shows
results of modelling space-time separable receptive fields in the LGN in this way,
using a cascade of truncated exponential kernels of the form
(\protect\ref{eq-comp-trunc-exp-cascade}) for temporal smoothing over
the temporal domain.

Regarding the spatial domain, the model in terms of spatial
Laplacians of Gaussians
$(\partial_{xx} + \partial_{yy}) \, g(x, y;\; s)$ 
is closely related to
differences of Gaussians, which have previously been shown to
constitute a good approximation of the spatial variation of receptive fields in
the retina and the LGN (Rodieck \cite{Rod65-VisRes}).
This property follows from the fact that the
rotationally symmetric Gaussian satisfies the isotropic diffusion
equation
\begin{align}
  \begin{split}
    \frac{1}{2} \nabla^2 L(x, y;\; s) 
    & = \partial_s L(x, y;\; s) 
  \end{split}\nonumber\\
  \begin{split}
    & \approx 
        \frac{L(x, y;\; s + \Delta s) - L(x, y;\; s)}{\Delta s}
  \end{split}\nonumber\\
  \begin{split}
    \label{eq-expl-DoG-Laplace}
    & = \frac{DOG(x, y;\; s, \Delta s)}{\Delta s},
  \end{split}
\end{align}
which implies that differences of Gaussians can be interpreted
as approximations of derivatives over scale and hence to Laplacian
responses.
Conceptually, this implies very good agreement with the spatial
component of the LGN model (\ref{eq-lgn-model-1}) in terms of Laplacians
of Gaussians.
More recently, Bonin {\em et al.\/}\ \cite{BonManCar05-JNeuroSci} have found that LGN
responses in cats are well described by difference-of-Gaussians and temporal smoothing
complemented by a non-linear contrast gain control mechanism (not
modelled here).

\begin{figure}[hbtp]
   \begin{center}
     \begin{tabular}{cc}
        & {\small$\partial_x g(x, y;\; \Sigma)$} \\
       \hspace{-3mm} \includegraphics[height=0.094\textheight]{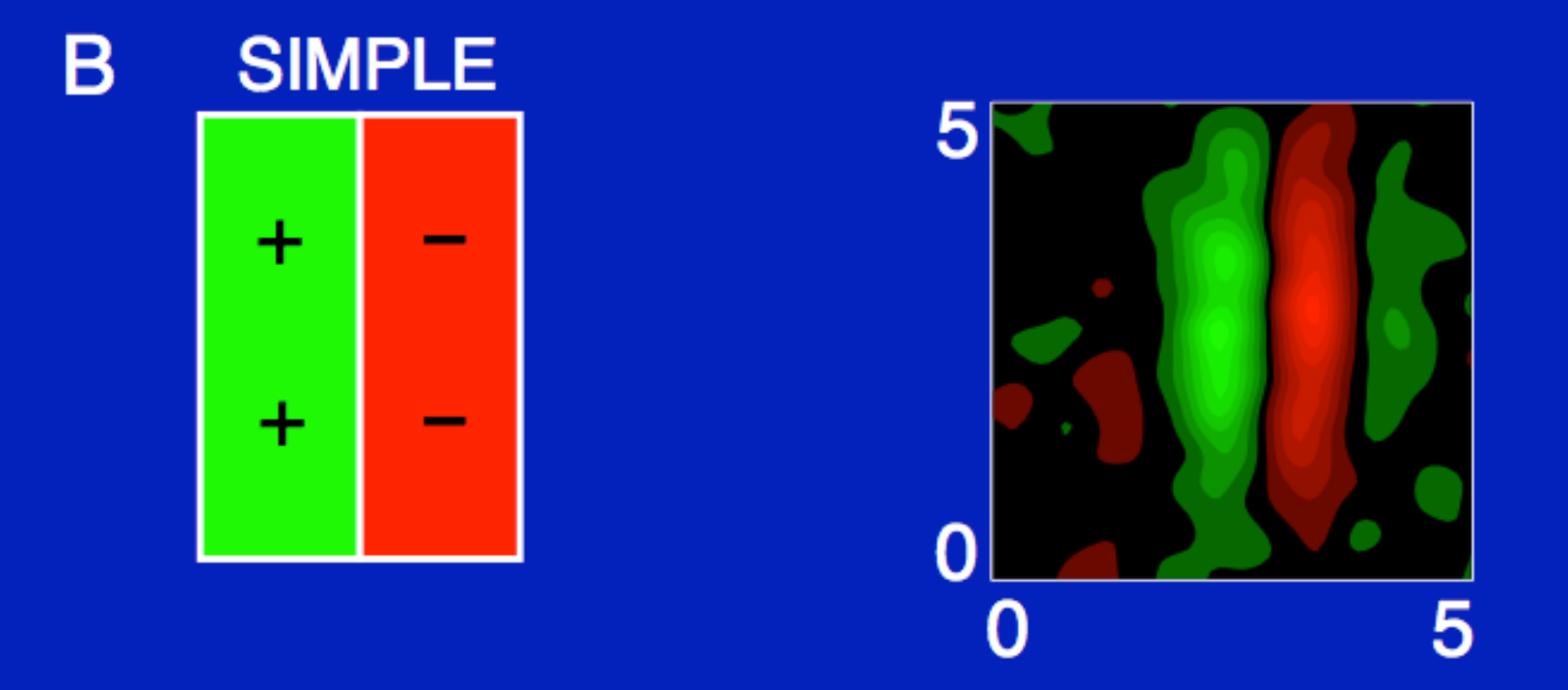}
       & \includegraphics[height=0.094\textheight]{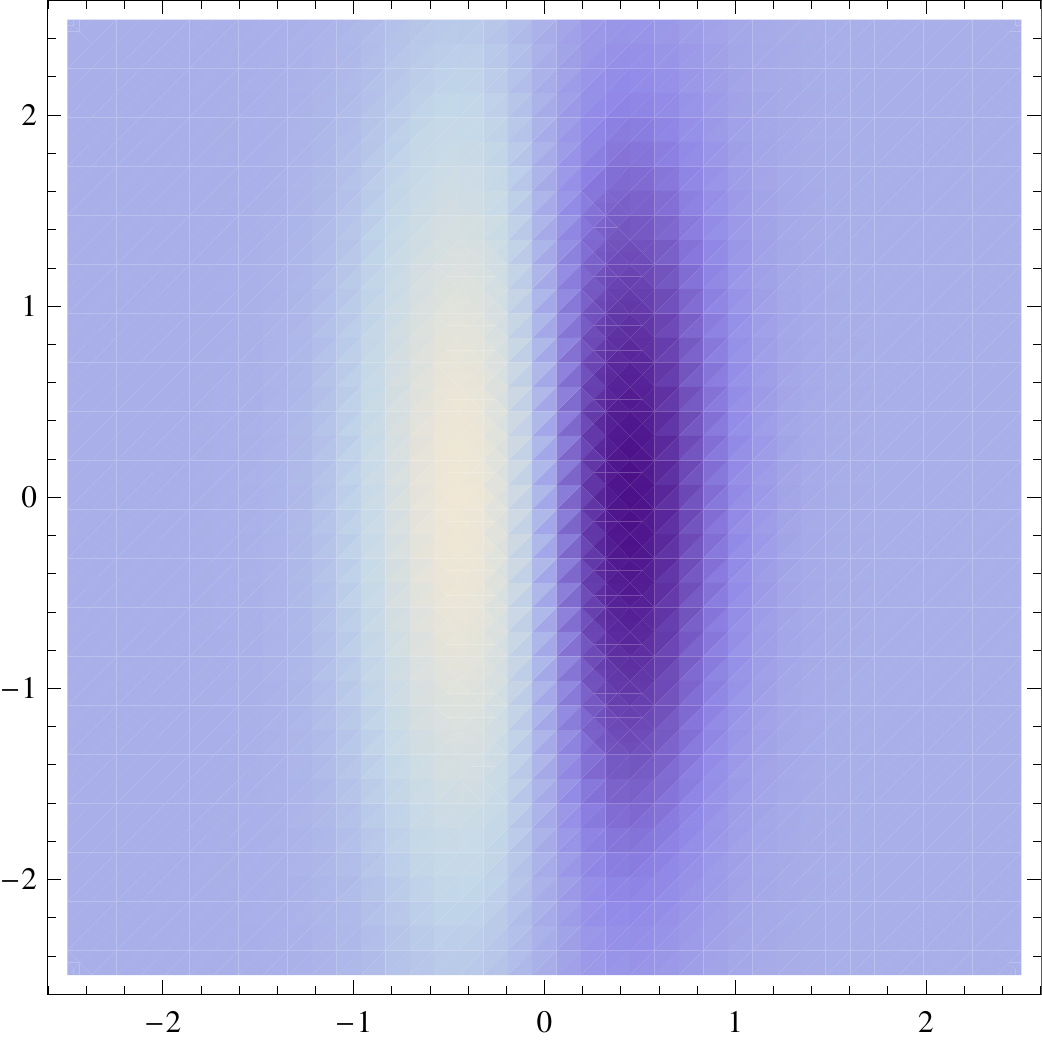}
     \end{tabular}
   \end{center}
  \caption{Computational modelling of a receptive field profile over the spatial domain
           in the primary visual cortex (V1) as reported by DeAngelis {\em et al.\/}\
           \cite{DeAngOhzFre95-TINS,deAngAnz04-VisNeuroSci} using
           affine Gaussian derivatives:
          (middle) Receptive field profile of a simple cell over image
          intensities as reconstructed from cell
          recordings, with positive weights 
          represented as green and negative weights by red. (left) Stylized simplification of the receptive
          field shape. (right) Idealized model of the receptive field
          from a first-order directional derivative of an affine
          Gaussian kernel $\partial_x g(x, y;\; \Sigma) = 
            \partial_x g(x, y;\; \lambda_x, \lambda_y)$ according to (\ref{eq-aff-gauss}),
           here with $\sigma_x = \sqrt{\lambda_x} = 0.45$ and 
           $\sigma_y = \sqrt{\lambda_y} = 1.4$ in units of
           degrees of visual angle, and with positive weights with
           respect to image intensities represented by white and
           negative values by violet.}
  \label{fig-simple-cell-aff-gauss-model}

  \medskip

  \begin{center}
     \begin{tabular}{ccc}
        & & {\small $\partial_{\orth \varphi}g(x, y;\; \Sigma)$} \\
       \hspace{-3mm} \includegraphics[height=0.096\textheight]{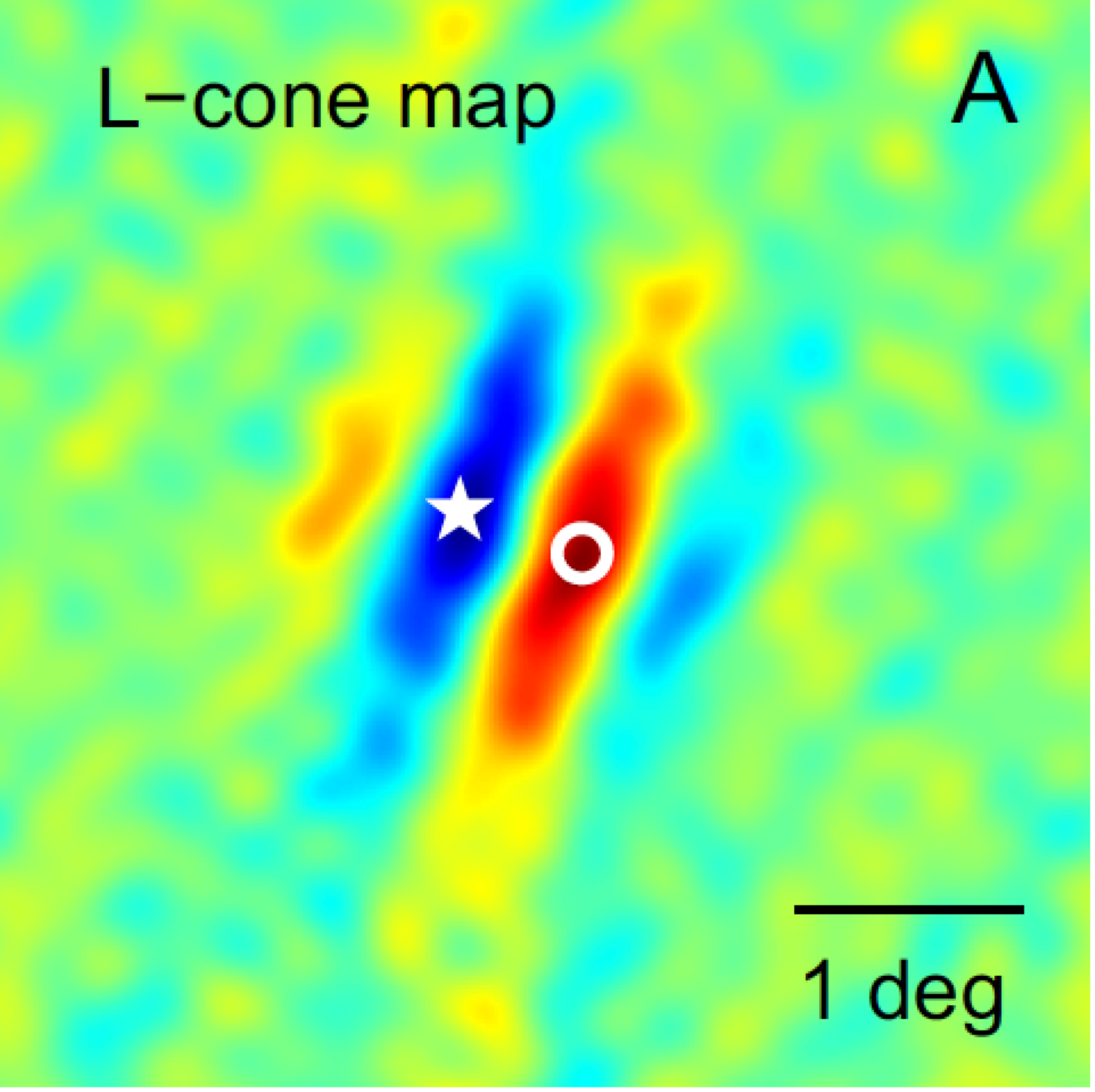}
       & \includegraphics[height=0.096\textheight]{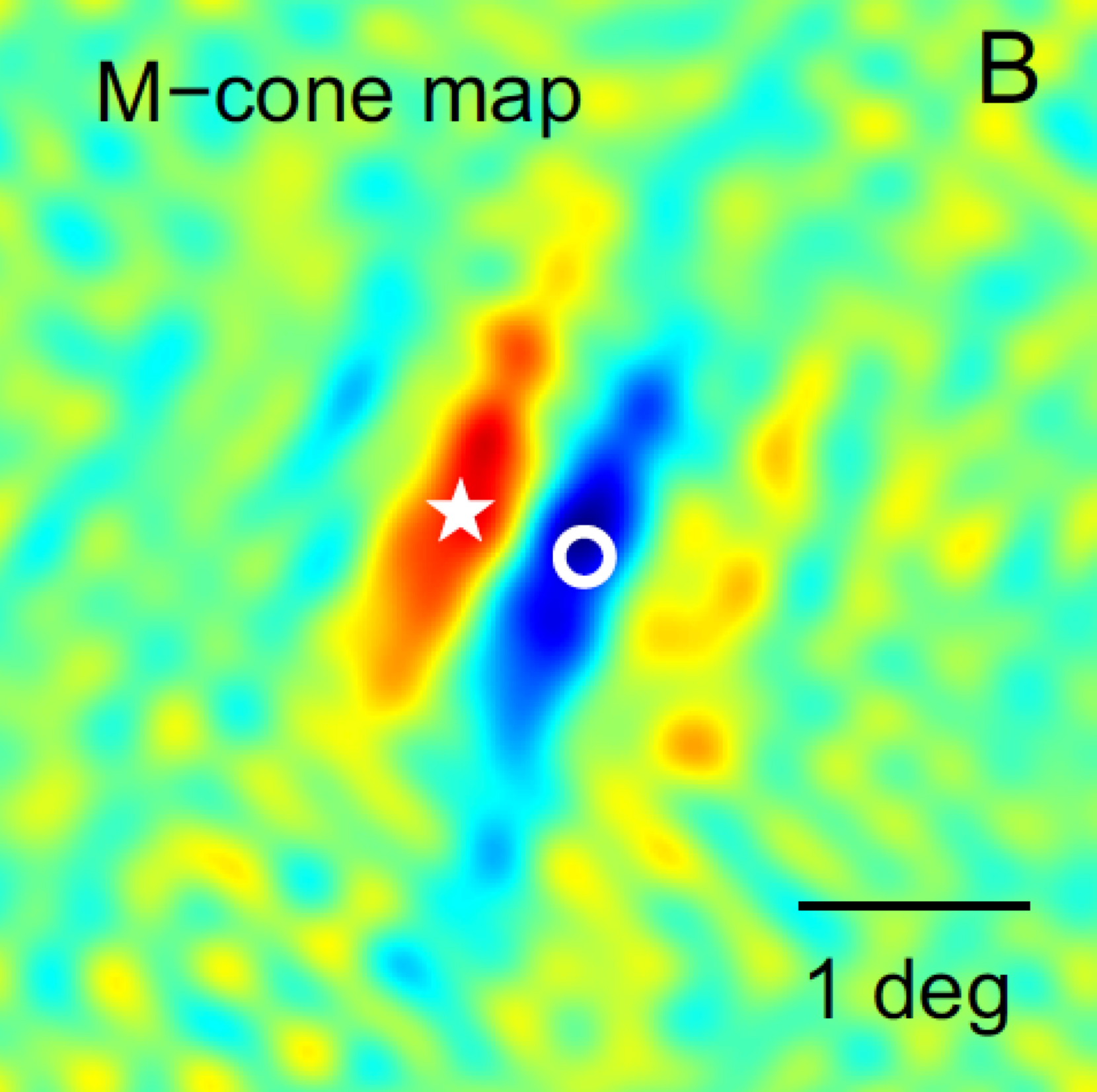}
       & \includegraphics[height=0.096\textheight]{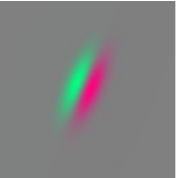}
     \end{tabular}
   \end{center}
  \caption{Computational modelling of a spatio-chromatic receptive field profile over the spatial domain
           for a double-opponent simple cell in the primary visual
           cortex (V1) as measured by Johnson {\em et al.\/}\
           \cite{JohHawSha08-JNeuroSci} using affine Gaussian
           derivatives over a colour-opponent channel:
          (left) Responses to L-cones corresponding to long wavelength
          red cones, with positive weights
          represented by red and negative weights by blue. 
          (middle) Responses to M-cones corresponding to medium wavelength
          green cones, with positive weights
          represented by red and negative weights by blue. 
          (right) Idealized model of the receptive field
          from a first-order directional derivative of an affine
          Gaussian kernel $\partial_{\orth \varphi}g(x, y;\; \Sigma)$ 
          according to (\ref{eq-aff-gauss}) over a red-green
          colour-opponent channel for $\sigma_1 = \sqrt{\lambda_1} = 0.6$ and
         $\sigma_2 = \sqrt{\lambda_2} = 0.2$ in units of
           degrees of visual angle, $\alpha = 67~\mbox{degrees}$ and with positive
           weights for the red-green colour-opponent channel represented by red and
           negative values by green.}
  \label{fig-simple-cell-aff-gauss-model-col-opp}
\end{figure}

\begin{figure*}[hbtp]
  \begin{center}
     \begin{tabular}{c}
        \includegraphics[width=0.78\textwidth]{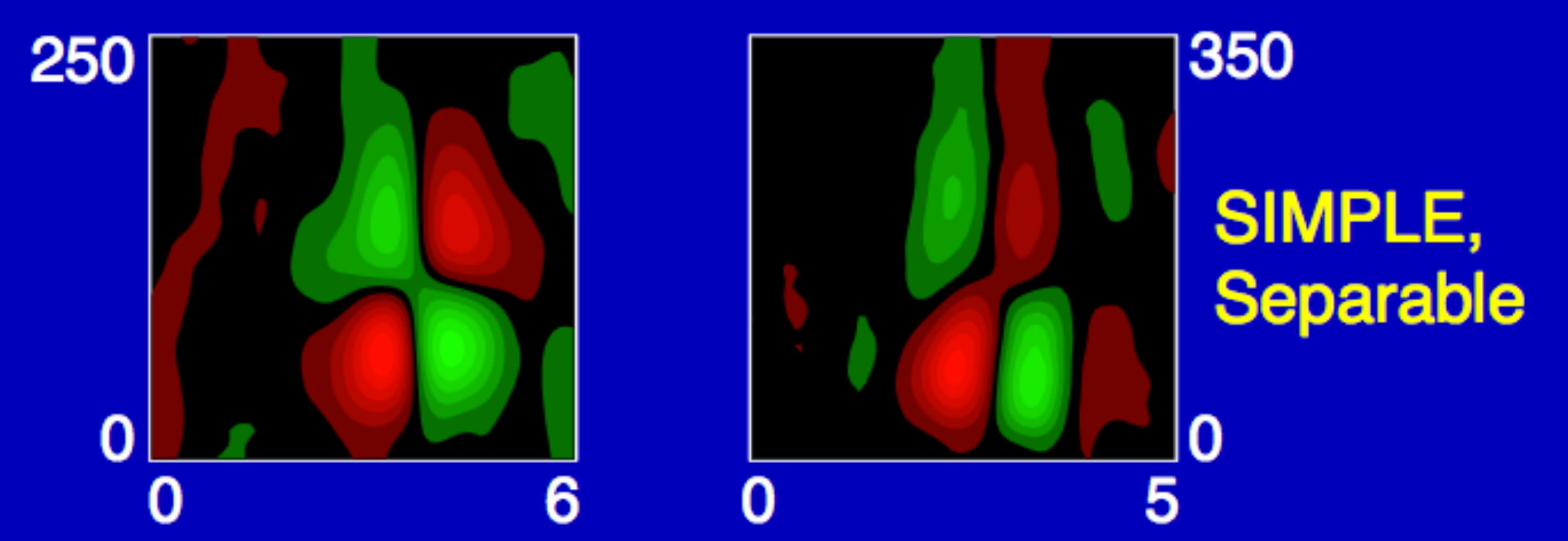} 
    \end{tabular}
  \end{center}
  \begin{center}
    \begin{tabular}{cc}
      \hspace{-26mm} {\footnotesize $h_{xt}(x, t;\; s, \tau)$}
      & \hspace{4mm} {\footnotesize $-h_{xxt}(x, t;\; s, \tau)$} \\
\hspace{-26mm} \includegraphics[width=0.24\textwidth]{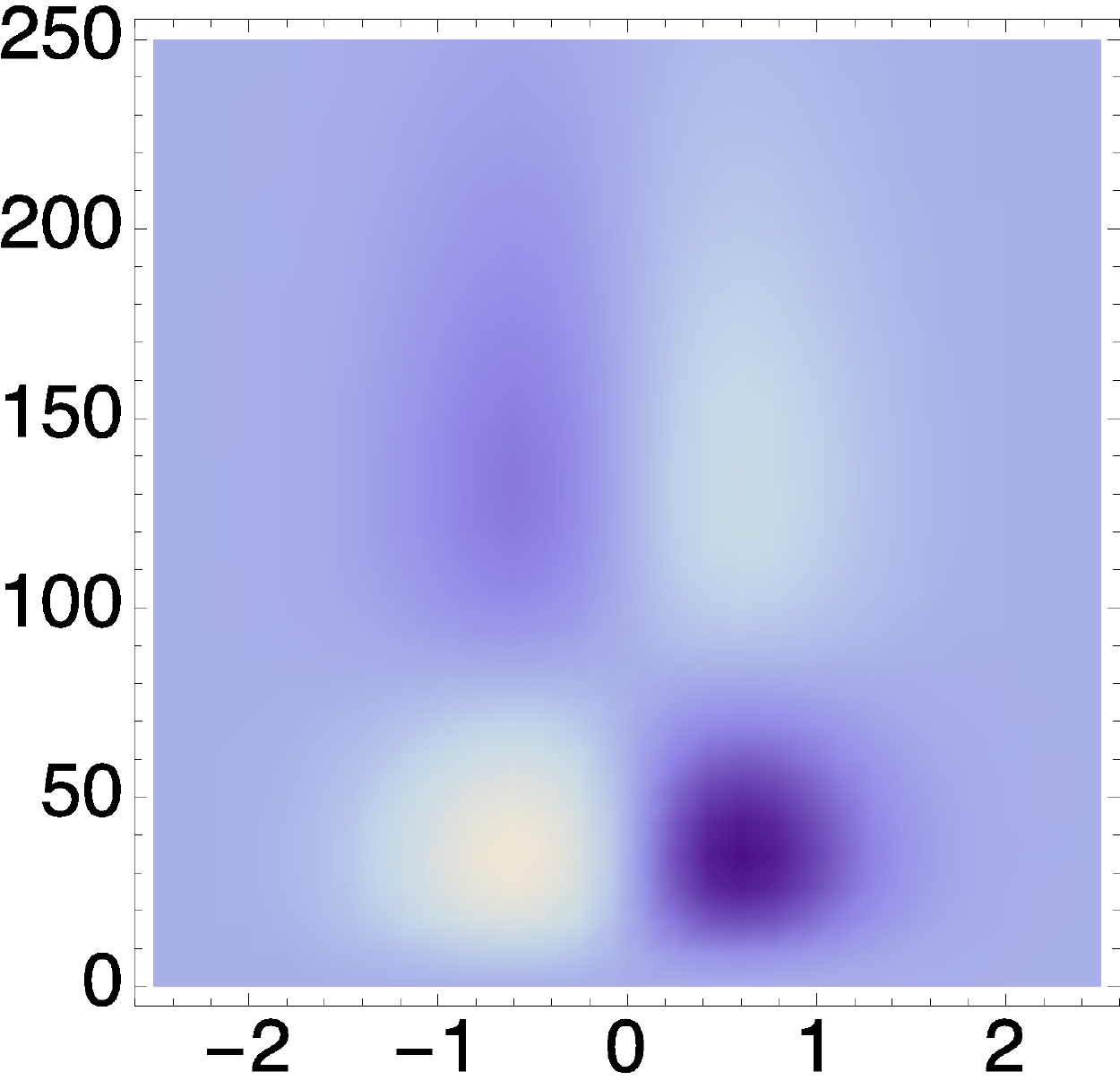}
      &
\hspace{4mm}  \includegraphics[width=0.24\textwidth]{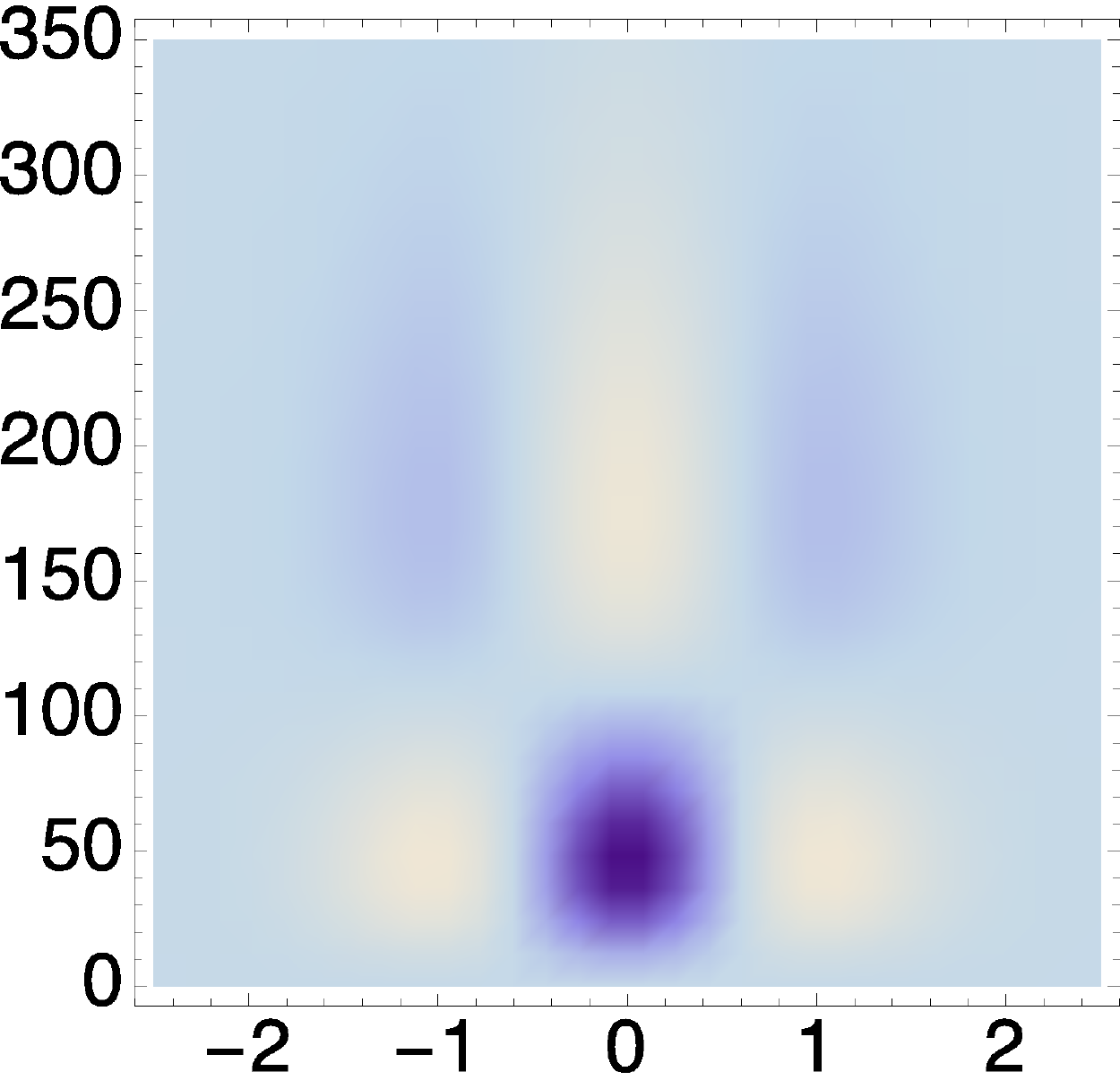} 
    \end{tabular}
  \end{center}

\smallskip

  \begin{center}
     \begin{tabular}{c}
        \includegraphics[width=0.78\textwidth]{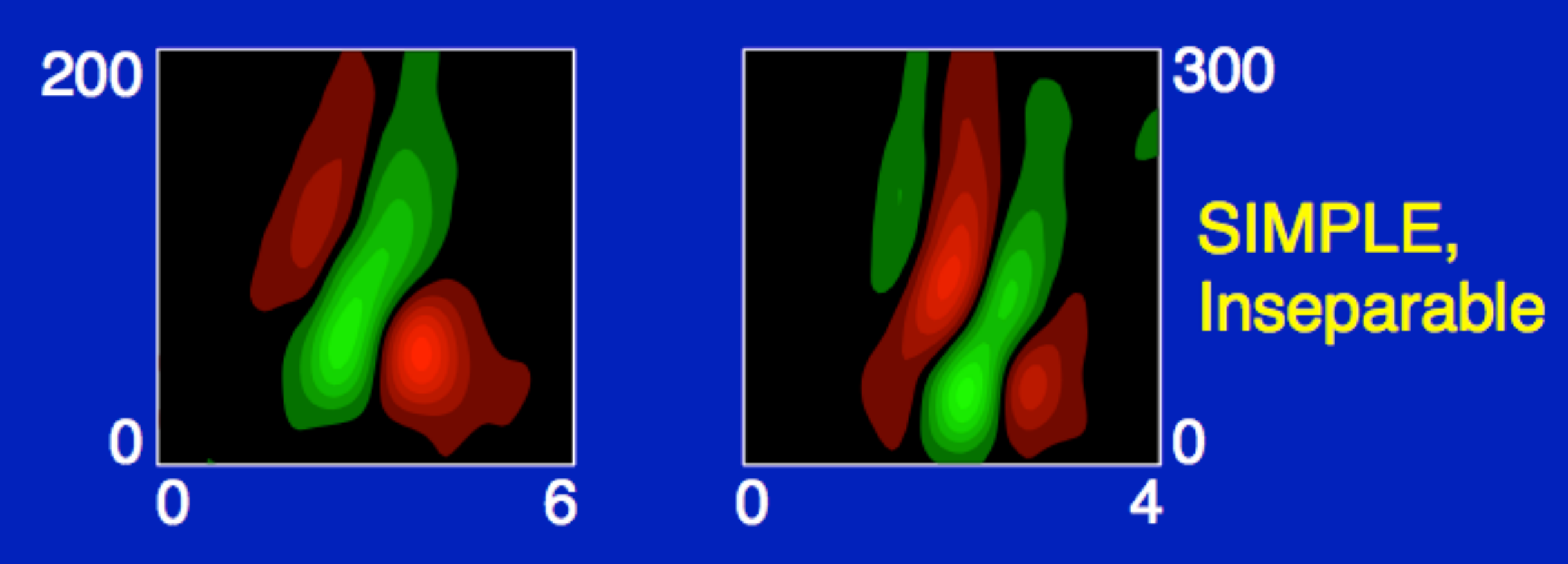}
    \end{tabular}
  \end{center}
  \begin{center}
    \begin{tabular}{cc}
      \hspace{-26mm} {\footnotesize $h_{xx}(x, t;\; s, \tau, v)$}
      & \hspace{4mm} {\footnotesize $-h_{xxx}(x, t;\; s, \tau, v)$} \\
\hspace{-26mm}
\includegraphics[width=0.24\textwidth]{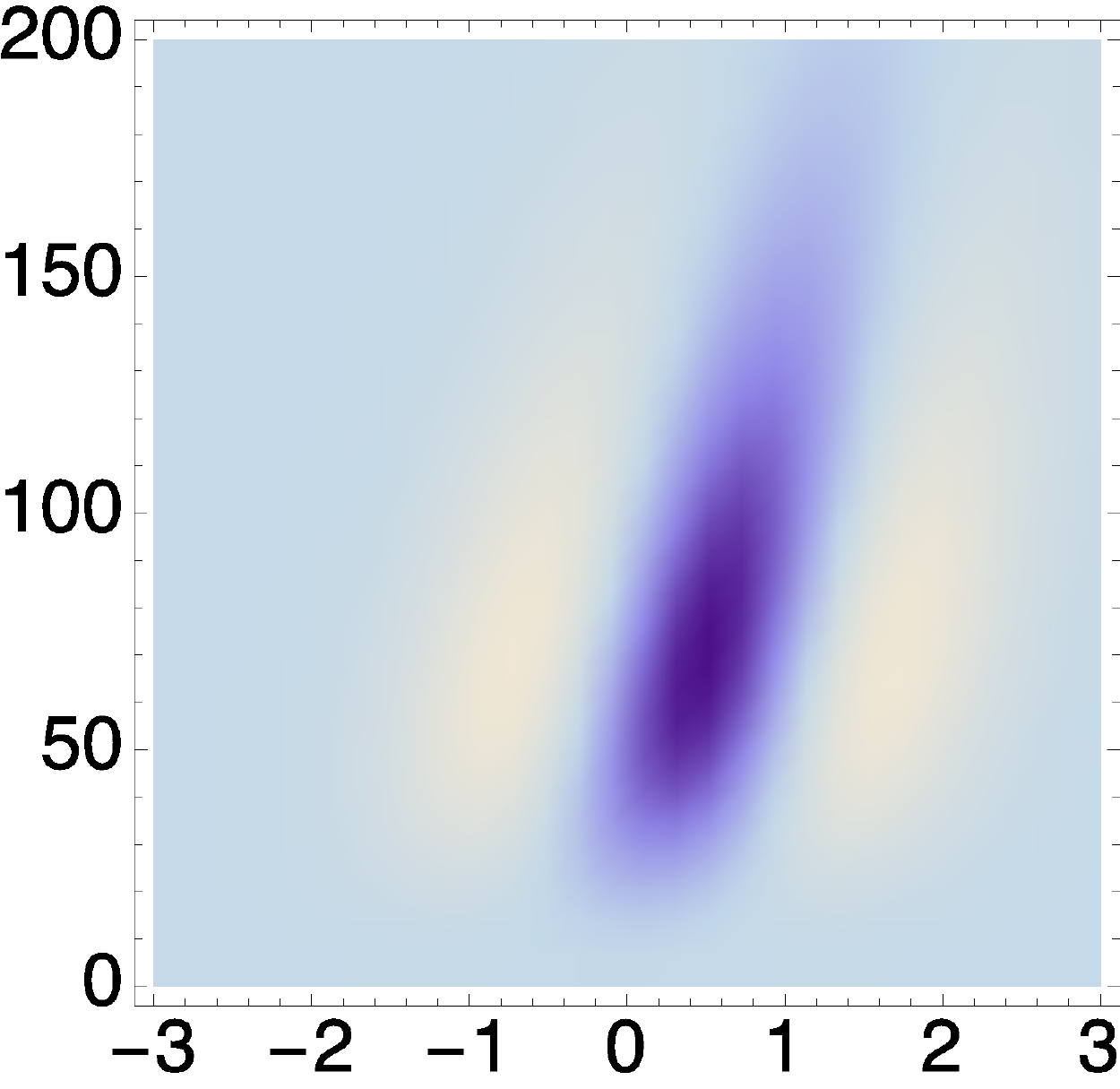}
      &
\hspace{4mm}  \includegraphics[width=0.24\textwidth]{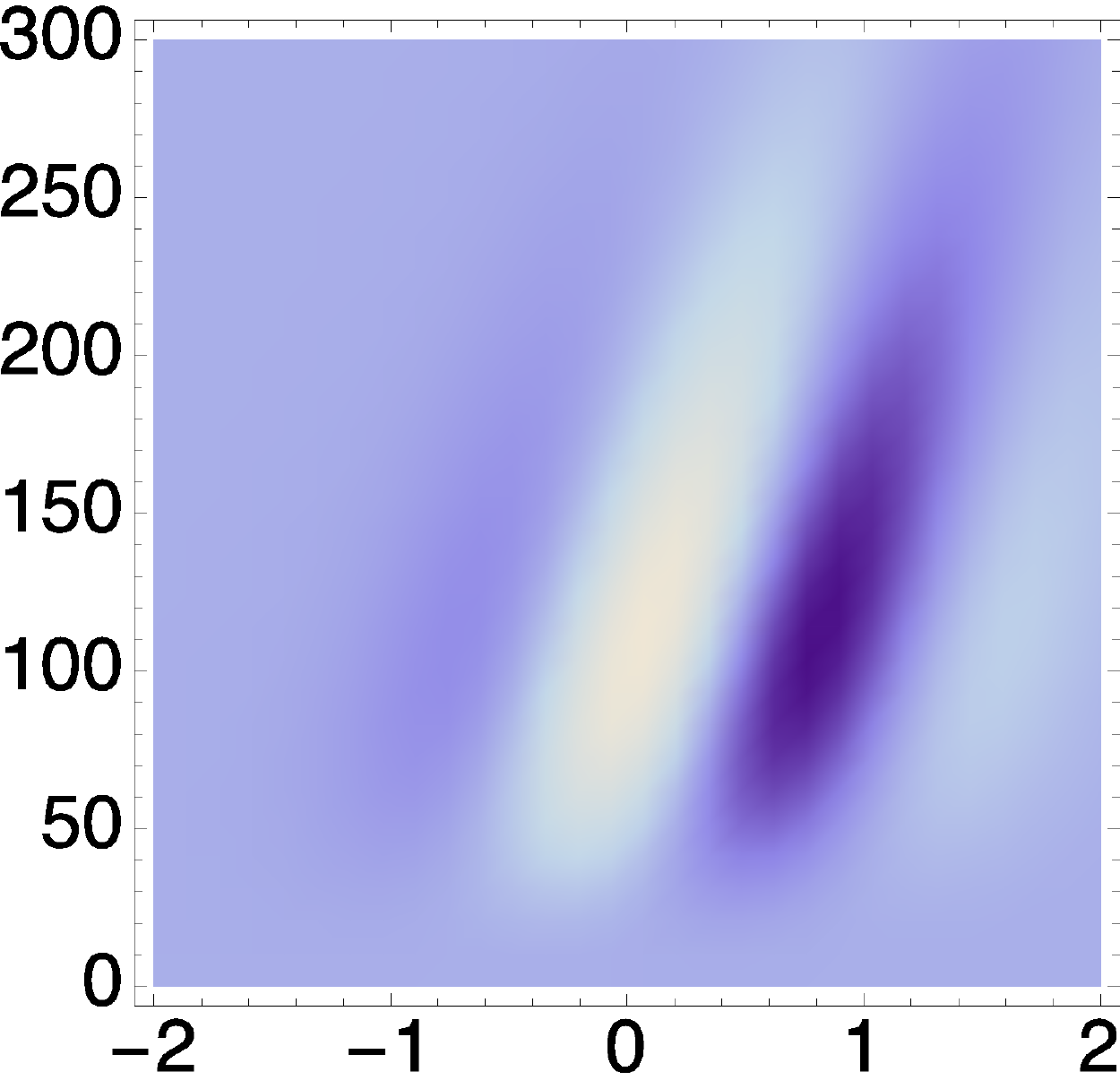}
    \end{tabular}
  \end{center}

  \caption{Computational modelling of simple cells in the primary
    visual cortex (V1) based on neural cell recordings reported
    by DeAngelis {\em et al.\/}\ \cite{DeAngOhzFre95-TINS} and using
    idealized spatio-temporal receptive fields of the form
    $T(x, t;\; s, \tau, v)
           = \partial_{x^m} \partial_{t^n} (g(x - v t;\; s) \, h(t;\; \tau))$
    according to Equation~(\protect\ref{eq-spat-temp-RF-model-der}) and
    with the temporal smoothing function $h(t;\; \tau)$
    modelled as a cascade of first-order integrators/truncated exponential
    kernels of the form (\protect\ref{eq-comp-trunc-exp-cascade}):
    (upper part) Separable receptive fields corresponding to mixed
    derivatives of first- or second-order derivatives over space with
    first-order derivatives over time.
    (lower part) Inseparable velocity-adapted receptive fields
    corresponding to second- or third-order derivatives over space.
    Parameter values with $\sigma_x = \sqrt{s}$ and $\sigma_t = \sqrt{\tau}$: 
    (a) $h_{xt}$: $\sigma_x = 0.6$~degrees, $\sigma_t = 60$~ms.
    (b) $h_{xxt}$: $\sigma_x = 0.6$~degrees, $\sigma_t = 80$~ms.
    (c) $h_{xx}$: $\sigma_x = 0.7$~degrees, $\sigma_t = 50$~ms, $v = 0.007$~degrees/ms.
    (d) $h_{xxx}$: $\sigma_x = 0.5$~degrees, $\sigma_t = 80$~ms, $v = 0.004$~degrees/ms.
    (Horizontal axis: Space $x$ in degrees of visual angle. Vertical axis: Time $t$ in ms.)}
   \label{fig-biol-model-simple-cells-rec-filters-over-time}
\end{figure*}

\subsection{Double-opponent spatio-chromatic receptive fields in the LGN}

In a study of spatio-chromatic response properties of V1 neurons in
the alert macaque monkey, Conway and Livingstone \cite{ConLiv06-JNeurSci} describe 
receptive fields with approximately circular red/green and 
yellow/blue colour-opponent response properties over 
the spatio-chromatic domain, see Figure~\ref{fig-col-opp-neuron}.
Such cells are referred to as {\em double-opponent cells\/},
since they simultaneously compute both spatial and chromatic opponency.
According to Conway and Livingstone \cite{ConLiv06-JNeurSci}, this cell type can be regarded
as the first layer of spatially opponent colour computations.

If we, motivated by the previous application of Laplacian of Gaussian functions to model
rotationally symmetric on-center/off-surround and
off-center/on-surround receptive fields in the LGN
(\ref{eq-lgn-model-1}),
apply the Laplacian of the Gaussian operator to 
red/green and yellow/blue colour-opponent channels, 
\begin{equation}
  \label{eq-col-opp-space-uv-from-rgb}
  \left(
    \begin{array}{c}
      f\\
      u\\
      v
    \end{array}
   \right)
    =
   \left(
     \begin{array}{ccc}
       \tfrac{1}{3} &   \tfrac{1}{3} & \tfrac{1}{3} \\
       \tfrac{1}{2} & - \tfrac{1}{2} & 0 \\
       \tfrac{1}{2} &   \tfrac{1}{2} & -1 \\
    \end{array}
   \right)
   \left(
     \begin{array}{c}
       R \\
       G\\
       B
     \end{array}
   \right),
\end{equation}
respectively,
we obtain equivalent spatio-chromatic receptive fields 
corresponding to red-center/green-surround,
green-center/red-surround,
yellow-center/blue-surround or
blue-center/yellow-surround, respectively,
as shown in Figure~\ref{fig-col-gauss-Laplace-ders} and corresponding
to the following spatio-chromatic receptive field model applied to the RGB channels
\begin{multline}
  \label{eq-double-opp-spat-model}
  h_{double-opponent}(x, y;\; s) = \\
  \pm (\partial_{xx} + \partial_{yy}) \, g(x, y;\; s) 
   \left(
     \begin{array}{ccc}
       \tfrac{1}{2} & - \tfrac{1}{2} & 0 \\
       \tfrac{1}{2} &   \tfrac{1}{2} & -1 \\
    \end{array}
   \right).
\end{multline}
In this respect, these spatio-chromatic receptive fields can be used as an
idealized model for the spatio-chromatic response properties for
double-opponent cells.

\subsection{Spatial, spatio-chromatic and spatio-temporal receptive fields in V1}

Concerning the neurons in the primary visual cortex (V1), 
DeAngelis {\em et al.\/}\
\cite{DeAngOhzFre95-TINS,deAngAnz04-VisNeuroSci}
describe that their receptive fields are generally different 
from the receptive fields in the LGN in the sense that they are
(i)~oriented in the spatial domain and 
(ii)~sensitive to specific stimulus velocities.
Cells (iii)~for which there are precisely localized ``on'' and ``off''
subregions with (iv)~spatial summation within each subregion,
(v)~spatial antagonism between on- and off-subregions and
(vi)~whose visual responses to stationary or moving spots can be
predicted from the spatial subregions are referred to as 
  {\em simple cells\/} (Hubel and Wiesel \cite{HubWie59-Phys,HubWie62-Phys,HubWie05-book}).

Figure~\ref{fig-simple-cell-aff-gauss-model} shows an example of the
spatial dependency of a simple cell, that can be well modelled by a
first-order affine Gaussian derivative over image intensities. 
Figure~\ref{fig-simple-cell-aff-gauss-model-col-opp} shows
corresponding results for a color-opponent receptive field of a simple
cell in V1, that can be modelled as a first-order affine Gaussian spatio-chromatic
derivative over an R-G colour-opponent channel.

Figure~\ref{fig-biol-model-simple-cells-rec-filters-over-time} shows
the result of modelling the spatio-temporal receptive fields of simple
cells in V1 in this way, using the general idealized model of
spatio-temporal receptive fields in Equation~(\protect\ref{eq-spat-temp-RF-model-der})
in combination with a temporal smoothing kernel obtained by coupling a
set of truncated exponential kernels in cascade (\ref{eq-comp-trunc-exp-cascade}). 
The results in the upper part show space-time separable spatio-temporal
receptive fields corresponding to zero image velocity $v = 0$,
and corresponding to either first- or second-order spatial derivatives over
image space in combination with first-order temporal derivatives over time.
The results in the lower part show inseparable spatio-temporal receptive
fields corresponding to non-zero image velocities and based on either
second- or third-order spatial derivatives over image space.

As can be seen from these figures,
the proposed idealized receptive field models do quite well 
reproduce the qualitative shape of the neurophysiologically 
recorded biological receptive fields.

\section{Relations to previous work}
\label{sec-rel-prev-work}

Young \cite{You87-SV} has shown how spatial receptive fields
in cats and monkeys can be well modelled by Gaussian derivatives up to
order four. 
Young {\em et al.\/}\ \cite{YouLesMey01-SV,YouLes01-SV} have also shown how spatio-temporal
receptive fields can be modelled by Gaussian derivatives over a
spatio-temporal domain, corresponding to the Gaussian spatio-temporal concept
described here, although with a different type of parameterization;
see also \cite{Lin97-ICSSTCV,CVAP257} for our closely
related earlier work.
The normative theory for visual receptive fields presented 
in \cite{Lin10-JMIV,Lin13-BICY,Lin13-PONE,Lin16-JMIV}
and here does first of all provide additional 
theoretical foundation for Young's spatial modelling work based on Koenderink
and van Doorn's theory \cite{Koe84,KoeDoo87-BC} and does additionally provide a conceptual
extension to a time-causal spatio-temporal domain that takes into explicit
account the fact that the future cannot be accessed.
Additionally, our model provides
a better parameterization of the spatio-temporal receptive field model over a
non-causal spatio-temporal domain based on the Gaussian
spatio-temporal scale-space concept.

This model or earlier versions of it has in turn been exploited for modelling of biological
receptive fields by 
Lowe \cite{Low00-BIO},
May and Georgeson \cite{MayGeo05-VisRes},
Hesse and Georgeson \cite{HesGeo05-VisRes},
Georgeson  {\em et al.\/}\ \cite{GeoMayFreHes07-JVis},
Wallis and Georgeson \cite{WalGeo09-VisRes},
Hansen and Neumann \cite{HanNeu09-JVis},
Wang and Spratling \cite{WanSpra16-CognComp}, 
Mahmoodi \cite{Mah16-JMIV,Mah17-JMIV} and 
Pei {\em et al.\/}\ \cite{PeiGaoHaoQiaAi16-NeurRegen}.

\subsection{Relations to modelling by Gabor functions}

Gabor functions \cite{Gab46}
\begin{equation}
  G(x;\; s, \omega) = e^{-i\omega x} \, g(x;\; s),
\end{equation}
have been frequently used for modelling spatial
receptive fields 
(Mar\v{c}elja \cite{Mar80-JOSA}; Jones and Palmer \cite{JonPal87a,JonPal87b};
 Ringach \cite{Rin01-JNeuroPhys,Rin04-JPhys})
motivated by their property of minimizing the uncertainty relation.
This motivation can, however, be questioned on both theoretical and
empirical grounds. 
Stork and Wilson \cite{StoWil90-JOSA} argue that (i)~only complex-valued Gabor
functions that cannot describe single receptive field minimize the
uncertainty relation, (ii)~the real functions that minimize this
relation are Gaussian derivatives rather than Gabor functions and
(iii)~comparisons among Gabor and alternative fits to both
psychophysical and physiological data have shown that in many cases
other functions (including Gaussian derivatives) provide better fits
than Gabor functions do.

Conceptually, the ripples of the Gabor functions, which are given by
complex sine waves, are related to the ripples of Gaussian
derivatives, which are given by Hermite functions.
A Gabor function, however, requires the specification of a scale parameter and a
frequency, whereas a Gaussian derivative requires a scale parameter
and the order of differentiation (per spatial and temporal dimension).
With the Gaussian derivative model, receptive fields of different
orders can be mutually related by derivative operations, 
and be computed from each other by nearest-neighbour operations. 
The zero-order receptive fields as well as the derivative based
receptive fields can be modelled by diffusion equations, 
and can therefore be implemented by computations between 
neighbouring computational units.

In relation to invariance properties, the family of affine Gaussian
kernels is closed under affine image deformations, whereas the family
of Gabor functions obtained by multiplying rotationally symmetric
Gaussians with sine and cosine waves is not closed under affine image
deformations. This means that it is not possible to compute truly
affine invariant image representations from such Gabor functions.
Instead, given a pair of images that are related by a non-uniform
image deformation, the lack of affine covariance implies that there 
will be a systematic bias in image representations derived from such
Gabor functions, corresponding to the difference between the
backprojected Gabor functions in the two image domains.
If using receptive profiles defined from directional derivatives of
affine Gaussian kernels, it will on the other hand be possible to
compute affine invariant image representations \cite{Lin13-PONE}, in turn providing
better internal consistency between receptive field responses computed
from different views of objects in the world.

With regard to invariance to multiplicative illumination variations,
the even cosine component of a Gabor function does in general not have
its integral equal to zero, which means that the illumination
invariant properties under multiplicative illumination variations or
exposure control mechanisms described in Section~\ref{sec-intens-var} do not hold for Gabor functions.

In this respect, the Gaussian derivative model is simpler, it can be
related to image measurements by differential geometry, be derived
axiomatically from symmetry principles, be computed from a minimal
set of connections and allows for provable invariance properties under
locally linearized image deformations (affine transformations)
as well as local multiplicative illumination variations and exposure
control mechanisms.

\subsection{Relations to approaches for learning receptive fields from natural
  image statistics}

\begin{figure*}[!t]
  \begin{center}
     \begin{tabular}{c}
        \includegraphics[width=0.78\textwidth]{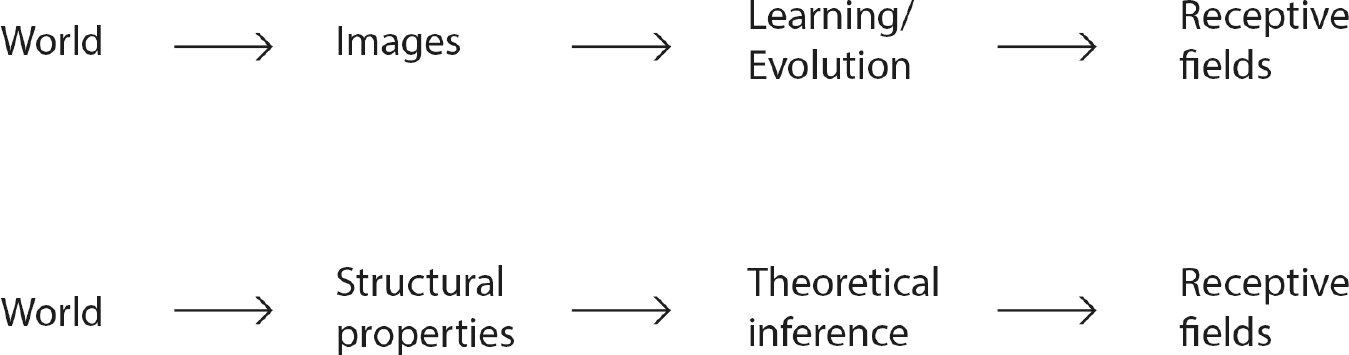}
    \end{tabular}
  \end{center}

  \bigskip

  \caption{Two structurally different ways of deriving receptive field shapes
    for a vision system intended to infer properties of the world by
    either biological or artificial visual perception.
    (top row) A traditional model for learning receptive fields
    shapes consists of collecting real-world image data from the
    environment, and then applying learning algorithms possibly in
    combination with evolution over multiple generations of the
    organism that the vision system is a part of.
    (bottom row) With the normative theory for receptive fields
    presented in this paper, a short-cut is made in the sense that the
  derivation of receptive field shapes starts from structural
  properties of the world (corresponding to symmetry properties in
  theoretical physics) from which receptive field shapes are
  constrained by theoretical mathematical inference.}
  \label{fig-infer-RF-from-struct-props}
\end{figure*}

Work has also been performed on learning receptive field properties 
and visual models from the statistics of natural image data 
(Field \cite{Fie87-JOSA};
 van~der~Schaaf and van~Hateren \cite{SchHat96-VisRes};
 Olshausen and Field \cite{OlsFie96-Nature};
 Rao and Ballard \cite{RaoBal98-CompNeurSyst};
 Simoncelli and Olshausen \cite{SimOls01-AnnRevNeurSci};
 Geisler \cite{Wil08-AnnRevPsychol};
 Hyv{\"a}rinen {\em et al.\/}\ \cite{HyvHurHoy09-NatImgStat};
 L{\"o}rincz \cite{LoePalSzi12-PLOS-CB})
and been shown to lead to the formation of similar receptive fields as
found in biological vision.
The proposed theory of receptive fields can be seen as
describing basic physical constraints under which a learning based
method for the development of receptive fields will operate and the
solutions to which an optimal adaptive system may converge to,
if exposed to a sufficiently large and representative set of natural
image data  (see Figure~\ref{fig-infer-RF-from-struct-props}).

Field \cite{Fie87-JOSA} as well as Doi and Lewicki \cite{DoiLew05-JapCogSci} have
described how ''natural images are not random, instead they exhibit
statistical regularities'' and have used such statistical regularities for
constraining the properties of receptive fields. 
The theory presented in
this paper can be seen as a theory at a higher level of abstraction,
in terms of basic principles that reflect properties of the environment
that in turn determine properties of the image data, without need for
explicitly constructing specific statistical models for the image
statistics. Specifically, the proposed theory can be used for
explaining why the above mentioned statistical models lead to
qualitatively similar types of receptive fields as the idealized
functional models of receptive fields obtained from our theory.

An interesting observation that can be made from the similarities 
between the receptive field families derived by necessity from the
assumptions and the receptive profiles found by cell recordings in
biological vision, is that receptive fields in the
retina, LGN and V1 of higher mammals are very close to {\em ideal\/} in 
view of the stated structural requirements/symmetry properties.
In this sense, biological vision can be seen as having adapted very
well to the transformation properties of the outside world and the
transformations that occur when a three-dimensional world is projected
to a two-dimensional image domain.

\subsection{Logarithmic brightness scale}

The notion of a {\em logarithmic brightness scale\/} goes back to the Greek astronomer
Hipparchus, who constructed a subjective scale for the brightness of
stars in six steps labelled ``1 \dots 6'', where the brightest stars
were said to be of the first magnitude ($m = 1$), while the faintest
stars near the limits of human perception were of the sixth magnitude.
Later, when quantitative physical measurements were made possible of the
intensities of different stars, it was noted that Hipparchus
subjective scale did indeed correspond to a logarithmic scale.
In astronomy today, the {\em apparent brightness\/} of stars is still
measured on a logarithmic scale, although extended over a much wider
span of intensity values.
A logarithmic transformation of image intensities is also used in the
retinex theory (Land \cite{Lan74-RoyInst,Lan86-VR}).

In psychophysics, the {\em Weber-Fechner law\/} attempts to describe
the relationship between the physical magnitude and the
perceived intensity of stimuli.
This law states that the ratio of an increment threshold $\Delta I$
for a just noticeable difference in relation to the background
intensity $I$ is constant over large ranges of magnitude variations \cite[Pages~671--672]{Pal99-Book}
\begin{equation}
  \label{eq-Weber-ratio}
  \frac{\Delta I}{I} = k,
\end{equation}
where the constant $k$ is referred to as the Weber ratio.
The theoretical analysis of invariance properties of a logarithmic
brightness scale under multiplicative transformations of the
illumination field as well as multiplicative exposure control
mechanisms in Section~\ref{sec-intens-var} is in excellent agreement with these psychophysical
findings.
If one considers an adaptive image exposure mechanism in the retina
that adapts the diameter of the pupil and the sensitivity of the
photopigments such that relative range variability in the signal
divided by the mean illumination is held constant
(\ref{eq-Weber-ratio}) (see {\em e.g.\/}\ Peli \cite{Pel90-JOSA}),
then such an adaptation mechanism can be seen as implementing an
approximation of the derivative of a logarithmic transformation
\begin{equation}
  d(\log z) = \frac{dz}{z}.
\end{equation}
For a strictly positive entity $z$, there are also information
theoretic arguments to regard $\log z$ as a default parameterization
(Jaynes \cite{Jay68-SMC}).
This property is essentially related to the fact that the ratio $dz/z$ then
becomes a dimensionless integration measure.
A general recommendation of care should, however, be taken when
using such reasoning based on dimensionality arguments, since
important phenomena could be missed, {\em e.g.\/}, in the presence of
hidden variables.
The physical modelling of the effect of illumination variations on
receptive field measurements in Section~\ref{sec-intens-var} provides 
a formal justification for using a logarithmic brightness scale in
this context as well as an additional contribution of showing how the
receptive field measurements can be related to inherent physical
properties of object surfaces in the environment.

\section{Summary}
\label{sec-summ}

Neurophysiological cell recordings have shown that 
mammalian vision has developed receptive fields that are tuned to different
sizes and orientations in the image domain as well as to different
image velocities in space-time.
A main message of this article has been to show that it is possible
to derive such families of receptive field profiles {\em by necessity\/}, 
given a set of structural requirements on the first stages of visual
processing as formalized into the notion of an {\em idealized vision
  system\/},
and whose functionality is determined by set of mathematical and
physical assumptions
(see Figure~\ref{fig-infer-RF-from-struct-props}).

These structural requirements reflect {\em structural properties of the world\/}
for the receptive fields to be compatible with natural image
transformations including:
(i)~variations in the sizes of objects in the world,
(ii)~variations in the viewing distance,
(iii)~variations in the viewing direction,
(iv)~variations in the relative motion between objects in the world and the observer,
(v)~variations in the speed by which temporal events occur and
(vi)~local multiplicative illumination variations or multiplicative exposure control mechanisms,
which are natural to {\em adapt to\/} for a vision system that is to
{\em interact with the world\/} in a successful manner.
In a competition between different organisms, adaptation to 
these properties may constitute an {\em evolutionary advantage\/}.

The presented theoretical model provides a {\em normative theory\/} for
deriving {\em functional models of linear receptive fields\/} based on
Gaussian derivatives and closely related operators.
Specifically, the proposed theory can {\em explain\/} the different
shapes of receptive field profiles that are found in biological vision
from a requirement that the visual system should be able to compute
covariant receptive field responses under the natural types of image
transformations that occur in the environment, to enable the
computation of invariant representations for perception at higher
levels \cite{Lin13-PONE}.


The presented theory leads to a computational framework for 
defining spatial and spatio-temporal receptive fields from visual data
with the attractive properties that:
(i)~the receptive field profiles can be derived {\em by necessity\/} from
first principles and 
(ii)~it leads to {\em predictions\/} about receptive field profiles in good agreement
with receptive fields found by cell recordings in biological vision.
Specifically, idealized functional models have been presented for 
space-time separable receptive fields in the retina and the LGN 
and for both space-time separable and non-separable simple cells 
in the primary visual cortex (V1).

The qualitatively very good agreement between the predicted receptive
field profiles from the normative axiomatic theory with the receptive
field profiles found by neurophysiological measurements indicates that the earliest receptive
fields in higher mammal vision have reached a
state that can be seen as very close to {\em ideal\/} in view of the stated
structural requirements/symmetry properties.
In this sense, biological vision can be seen as having adapted very well to the
transformation properties of the outside world and to the 
transformations that occur when a three-dimensional world
is projected onto a two-dimensional image domain.

Compared to more common approaches of learning receptive field
profiles from natural image statistics, the proposed framework 
makes it possible to derive the shapes
of idealized receptive fields without any need for training data. 
The proposed framework for invariance and covariance properties
also adds explanatory value by showing that the families of receptive profiles 
tuned to different orientations in space and image velocities in
space-time that can be observed in biological vision can be {\em explained\/}
from the requirement that the receptive fields should be covariant
under basic image transformations to enable true invariance
properties.
If the underlying receptive fields would not be covariant, then there
would be a systematic bias in the visual operations, corresponding to
the amount of mismatch between the backprojected receptive fields.

Corresponding types of arguments applied to the area of hearing, lead to
the formulation of a normative theory of auditory receptive fields 
(Lindeberg and Friberg \cite{LinFri15-PONE,LinFri15-SSVM}).

\section*{Acknowledgments}

The support from the Swedish Research Council 
(Grant Number 2014-4083) is gratefully acknowledged.

\bibliographystyle{SSBAtrans}
\bibliography{bib/defs,bib/tlmac}

\end{document}